\documentclass[11pt]{revtex4}
\usepackage{epsfig, amssymb, latexsym, amsfonts, amsmath, amsthm}

\DeclareRobustCommand{\ted}{			
  \ifmmode \else \leavevmode\unskip\penalty9999 \hbox{}\nobreak\hfill
  \fi\quad\hbox{$\triangledown$}}

\DeclareRobustCommand{\Ted}{			
  \ifmmode \else \leavevmode\unskip\penalty9999 \hbox{}\nobreak\hfill
  \fi\quad\hbox{$\blacktriangledown$}}

\theoremstyle{plain}

\newcommand{\OP}{ {\mathcal{O}} }

\newcommand{\OPT}{ {\mathcal{O}}^{\circ} }
\newcommand{\QU}{ {\mathcal{Q}} }

\newcommand{\SP}{ {\mathcal{S}} }

\newcommand{\m}{ {\mathfrak{m}} }
\newcommand{\TRD}{ {\mathfrak{Trd}} }
\newcommand{\SPD}{ {\mathfrak{Spd}} }
\newcommand{\SPO}{ {\mathfrak{S}} }
\newcommand{\SO}{ {\mathfrak{S}} }

\newcommand{\HP}[4]{ H_{ #1_1\dots#1_{#3} | #4 }^{ #2_1\dots#2_{#3} } }

\newcommand{\HPT}[4]{ \tilde H_{ #1_1\dots#1_{#3} | #4 }^{ #2_1\dots#2_{#3} } }
\newcommand{\HPTP}[4]{ \tilde H_{ #1_1\dots#1_{#3} | #4 }^{ ' #2_1\dots#2_{#3} } }

\newcommand{\tw}{ {\text{tw}} }

\newcommand{\PRa}{ \mathcal{P} }

\newcommand{\Y}{ \mathcal{Y} }

\newcommand{\YH}{ \hat{\mathcal{Y}} }

\newcommand{\X}{ \mathbf{X} }
\newcommand{\XS}[2]{ X^{ #1_1 \dots #1_{#2} } }
\newcommand{\x}{ \mathbf{x} }
\newcommand{\D}{ \mathbf{d} }
\newcommand{\DA}{ \overset{\lra}{D} }
\newcommand{\TDA}{ \overset{\lra}{\text{D}} }
\newcommand{\pd}{ \partial }

\newcommand{\EYPN}[4]{ \YH^{ \kln{#3}| #2_1 \dots #2_k }_{ #1_1 \dots #1_k | #4 } }

\def \tl#1{\overset{\kern 2pt\circ}{#1}}
\def \TL#1{\overset{\kern -28pt \circ}{#1}}

\newcommand {\kln}[1]{\left( #1 \right)}
\newcommand {\Kln}[1]{\bigl( #1 \bigr)}

\newcommand {\KLno}[1]{\Bigl( #1 \Bigr.}
\newcommand {\KLNo}[1]{\biggl( #1 \biggr.}
\newcommand {\KLNNo}[1]{\Biggl( #1 \Biggr.}

\newcommand {\oKLn}[1]{\Bigl. #1 \Bigr)}
\newcommand {\oKLN}[1]{\biggl. #1 \biggr)}
\newcommand {\oKLNN}[1]{\Biggl. #1 \Biggr)}

\newcommand {\kls}[1]{\left\{ #1 \right\}}
\newcommand {\Kls}[1]{\bigl\{ #1 \bigr\}}
\newcommand {\KLs}[1]{\Bigl\{ #1 \Bigr\}}

\newcommand {\KLSo}[1]{\biggl\{ #1 \biggr.}
\newcommand {\KLSSo}[1]{\Biggl\{ #1 \Biggr.}

\newcommand {\oKLS}[1]{\biggl. #1 \biggr\}}
\newcommand {\oKLSS}[1]{\Biggl. #1 \Biggr\}}

\newcommand {\kle}[1]{\left[ #1 \right]}
\newcommand {\Kle}[1]{\bigl[ #1 \bigr]}
\newcommand {\KLe}[1]{\Bigl[ #1 \Bigr]}
\newcommand {\KLE}[1]{\biggl[ #1 \biggr]}

\newcommand {\KLEo}[1]{\biggl[ #1 \biggr.}

\newcommand {\oKLE}[1]{\biggl. #1 \biggr]}

\newcommand {\matel}[3]{\left< #1 \left|\; #2\, \right| #3 \right>}

\newcommand {\PIpe}[1]{\Bigl|_{#1} \Bigr.}
\newcommand {\PIPe}[1]{\biggl|_{#1} \biggr.}

\newcommand {\C} {{\mathbb{C}}}
\newcommand {\R} {{\mathbb{R}}}

\newcommand {\Id} {{\mathbb{I}}}
\newcommand {\im}{{\text{i}}}
\newcommand {\e}{{\text{e}}}

\newcommand {\ix} {\int {\mathrm{d}}^4 \! x}

\newcommand {\lra} {\leftrightarrow}

\newcommand {\la} {\leftarrow}

\newcommand {\Ra} {\longrightarrow}

\newcommand {\ra} {\rightarrow}

\newcommand {\tx} {\tilde x}

\newcommand {\fo}{\forall\,}

\newcommand {\les}{\leqslant}	
\newcommand {\ges}{\geqslant}	

\begin{document}

\title{Geometric twist decomposition off the light--cone for nonlocal QCD operators}

\author{ J\"org Eilers }
\email{ eilers@itp.uni-leipzig.de }
\affiliation{ Center for Theoretical Studies and Institute of Theoretical Physics, Leipzig University, Augustusplatz~10, D-04109~Leipzig, Germany}

\date{\today}

\begin{abstract}

\vspace*{0.1cm}

\noindent A general procedure is introduced allowing for the infinite decomposition of nonlocal operators off the light--cone into operators of definite geometric twist.

\vspace*{0.25cm}

\noindent PACS: 24.85.+p, 13.88.+e, 11.30.Cp\\ Keywords: Twist decomposition, Nonlocal off--cone operators, Tensor harmonic polynomials, Target mass corrections

\end{abstract}

\maketitle

\section{Introduction}
\label{Introduction}

Nonperturbative parton distributions together with appropriate kinematical factors parametrize the matrix elements of quark--antiquark and gluon operators. These QCD operators enter hadronic processes via the nonlocal light~cone expansion~\cite{AZ78,ZAV}. In fact, many different scattering processes are governed by one and the same set of light cone operators. For deeply virtual Compton scattering, for example, the so--called double distributions parametrize the corresponding nonforward matrix elements. For deep inelastic forward scattering and Drell--Yan processes the parton distributions parametrize forward matrix elements of light~cone operators. It is therefore convincing to focus on the nonlocal operators. Here, Gross and Treiman in Ref.~\cite{GT71} have shown using Wilson short--distance expansion \cite{W69} that in the kinematic regime where factorization holds, i.e. for large space--like virtualities, the dominant contributions to the physical scattering amplitudes come from leading twist operators. The notion of (geometric) twist is defined as twist = (canonical) dimension $-$ (Lorentz) spin and makes use of the irreducible tensor representations of the orthochronous Lorentz group. These representations are characterized by their Young symmetry which has to be realized on the space of traceless tensors. A classification of nonlocal QCD operators according to their (geometric) twist naturally leads to an expansion into dominant and sub--dominant light~cone contributions. The leading twist contributions thereby correspond to the dominant part of the scattering process where the quark and nucleon mass can be neglected. Higher twist effects describe contributions proportional to the quark and nucleon mass but also yield information about the quark--gluon correlations inside the hadrons. Beyond leading order one also has to take care of radiative corrections. Experimental precision allows the determination of nonleading contributions to physical scattering amplitudes and therefore a reliable treatment of the various sub--dominant effects is required.

Higher twist effects being related to target--mass effects again split up into two parts. First, there are higher twist contributions stemming from the subtraction of the traces in leading twist operators. Here we address these contributions as geometrical twist effects which are sometimes also called kinematical twist effects in the literature~\cite{RW00}. Second, there are higher twist contributions including total derivatives of operators of lower geometrical twist. These contributions are present in all exclusive processes but are absent in forward--like situations; see Refs.~\cite{BB99,BBK03} and~\cite{BL01} for a treatment of this problem. We also remark that the separation of geometrical and dynamical contributions is not translation invariant, since the total derivative acts on the reference point chosen for the light~cone expansion. One therefore always needs both parts for a decent treatment of power corrections in off--forward situations. Using a purely group theoretical procedure the (finite) twist decomposition of nonlocal light~cone operators in configuration space, as far as they are relevant for light~cone dominated hadronic processes, has been performed in Ref.~\cite{GLR99,GLR00} for QCD operators up to second rank. However, if one wants to calculate target--mass corrections for scattering processes one is forced to consider the (infinite) twist decomposition off the light~cone thereby taking into account all trace terms which, after Fourier transformation into momentum space, lead to contributions suppressed by powers of the relevant variable $M^2/Q^2$, where $M$ is the nucleon/target mass.

The target--mass corrections resulting from leading and, eventually, next--to--leading twist contributions have first been discussed by Nachtmann~\cite{Nachtmann73} in unpolarized deep--inelastic scattering by expanding into a series of Gegenbauer polynomials. Later on, this method has been applied to polarized deep--inelastic scattering in Refs.~\cite{BE76,BE76_b,W77,MU80,KU95}. Another method for the determination of target--mass effects was first given by Georgi and Politzer~\cite{GP} for unpolarized deep--inelastic scattering and then extended to polarized scattering and to general electro--weak couplings by Refs.~\cite{PR98,BLTA1}. Ref.~\cite{BM01} treats twist--2 target--mass corrections in deeply virtual Compton scattering.

However, all the above mentioned articles treated only the leading twist contributions since, up to now, a complete off--cone decomposition for nonlocal operators carrying free tensor indices in $x$--space has been possible only when their $n$th moments are totally symmetric allowing for the application of the group theoretical results of Bargmann and Todorov~\cite{BT77}. Especially, this holds for all scalar operators; see Ref.~\cite{GLR01} for the infinite twist decomposition of such objects. But once free indices are involved nontrivial Young patterns (including antisymmetries) occur. In this paper we will develop a general algorithm which allows one to perform a complete twist decomposition of, in principle, any local and nonlocal operator off as well as on the light~cone. We thereby extend previous work on the twist decomposition of nonlocal operators in Ref.~\cite{GLR99,GLR00,GLR01,GL01,L02}.

This paper is organized as follows. In Section~\ref{Compton} we give certain nonlocal QCD operators appearing in the Compton amplitude for 
nonforward virtual Compton scattering via the nonlocal light~cone expansion and explain the concept of geometric twist. The process of twist decomposition is sketched in three steps. The following two Sections~\ref{TracePart} and~\ref{SymmetryPart} explain in detail how a local tensor is decomposed into $SO(2h;\C)$--irreducible components. Section~\ref{TracePart} treats all problems related to the tracelessness of these tensors, whereas Section~\ref{SymmetryPart} treats the problem of Young symmetry. In Section~\ref{ApplicationQCD} we apply our results to various local QCD operators.

\section{Compton amplitude and the concept of geometric twist}
\label{Compton}

Let us now discuss Compton scattering of a virtual photon off a hadron,
\begin{equation}
\label{ohne_meson}
 {\gamma^*}(q_1) + {\mathrm H}(P_1) \Ra {\gamma^*}(q_2) +{\mathrm H}(P_2) \;  ,
\end{equation}
which is an important process in Quantum Chromodynamics. This general process covers a series of different reactions through which a variety of inclusive informations on the short--distance structure of nucleons become accessible at large space--like virtualities. It is also closely connected to the spin problem of the nucleon. The case of forward scattering $P_1 = P_2 = P$ describes deep inelastic scattering (DIS) off unpolarized or polarized targets which is widely discussed in the literature, see e.g.~\cite{DIS1,DIS2,MUTA,IZ}, and $P_1 \neq P_2$ corresponds to the generic nonforward virtual Compton scattering, see References~\cite{MRGHD,CSJ1,CSJ2,CSJ3}.

The Compton amplitude for the process (\ref{ohne_meson}) is given by
\begin{equation}
\label{Compton_amp}
 T_{\mu\nu}\kln{P_+,P_-;q} = \im \ix \; \e^{\im qx} \, \matel{P_2,S_2}{R \, T\kle{J_\mu\kln{\frac{x}{2}} J_\nu\kln{-\frac{x}{2}}\mathcal{S}} }{P_1,S_1}
\end{equation}
where
\begin{eqnarray}
\label{DefPPM}
 P_\pm = P_2 \pm P_1, \quad  q = \hbox{$\frac{1}{2}$} \kln{q_1 + q_2}  \,  ,
\end{eqnarray}
are chosen as independent kinematic variables. As usual, $P_1\kln{P_2}$ and $q_1\kln{q_2}$ denote the four--momenta of the incoming (outgoing) nucleons and photons, respectively, where $S_1$ and $S_2$ are the spins of these nucleons.

The renormalized, time--ordered product of two electromagnetic currents $R \, T \kle{J_\mu\kln{\frac{x}{2}} J_\nu\kln{-\frac{x}{2}} \mathcal{S} }$ contained in the Compton amplitude (\ref{Compton_amp}) can be decomposed via the nonlocal light~cone expansion which we give here including a reference point $y$
\begin{eqnarray}
\label{int_lce_y}
&&
R \, T \KLE{J_\mu\kln{y+\frac{x}{2}} \, J_\nu\kln{y-\frac{x}{2}} \mathcal{S} }
\\
\nonumber
&&
\quad
= \int_{-1}^1 \text{d} \kappa_1 \, \int_{-1}^1 \text{d} \kappa_2 \;\,
C_{\mu\nu}^\Gamma \! \kln{x^2,\kappa_1 \, \tx,\kappa_2 \, \tx;\mu^2} \; R \, T
\Kle{O_\Gamma\!\kln{y + \kappa_1 \, \tx,y + \kappa_2 \, \tx}\mathcal{S}}
\\
\nonumber
&&
 \quad \quad
 + \quad \text{higher order terms} \; .
\end{eqnarray}
Here, $\Gamma$ denotes the tensor structure as well as the flavor structure of the nonlocal QCD operators
$O_\Gamma\! \kln{y + \kappa_1 \, \tx,y + \kappa_2 \, \tx}$. Examples for nonlocal QCD operators appearing in the expansion
(\ref{int_lce_y}) are the operators $O_{\alpha_1}^a \kln{\kappa_1 \, \tx, \kappa_2 \, \tx}$ and $O_{\alpha_1}^{5\,a} \kln{\kappa_1 \, \tx, \kappa_2 \, \tx} $ as well as $M^a_{[\alpha_1\alpha_2]} \kln{\kappa_1 \, \tx, \kappa_2 \, \tx}$ and $N^a \kln{\kappa_1 \, \tx, \kappa_2 \, \tx}$.

In the following we will give these operators in explicit form.   
\begin{align}
\label{str_oprab}
 O^a_{\alpha_1}\kln{\kappa_1 \, \tx,\kappa_2 \, \tx}
&=
\bar\psi\kln{\kappa_1 \, \tx}
\, \gamma_{\alpha_1} \, \lambda^a_f \, U\kln{\kappa_1 \, \tx,\kappa_2 \, \tx} \, \psi\kln{\kappa_2 \, \tx}
\\
\nonumber
&
\qquad \quad 
- \bar\psi\kln{\kappa_2 \, \tx}
\, \gamma_{\alpha_1} \, \lambda^a_f \, U\kln{\kappa_2 \, \tx,\kappa_1 \, \tx} \, \psi\kln{\kappa_1 \, \tx} 
\\
\nonumber
&
\\
\label{str_oprbb}
 O^{5\,a}_{\alpha_1}\kln{\kappa_1 \, \tx,\kappa_2 \, \tx}
&=
\bar\psi\kln{\kappa_1 \, \tx}
\,\gamma^5 \gamma_{\alpha_1} \, \lambda^a_f \, U\kln{\kappa_1 \, \tx,\kappa_2 \, \tx} \,
\psi\kln{\kappa_2 \, \tx} 
\\
\nonumber
&
\qquad \quad 
+ \bar\psi\kln{\kappa_2 \, \tx} \,\gamma^5
\gamma_{\alpha_1} \, \lambda^a_f \, U\kln{\kappa_2 \, \tx,\kappa_1 \, \tx} \, \psi\kln{\kappa_1 \, \tx} \; .
\\
\intertext{In the nonsinglet case $\lambda^a_f \neq 1$, it is sufficient to consider the above vector
operators $O^a_{\alpha_1}\kln{\kappa_1 \, \tx,\kappa_2 \, \tx}$ and $O^{5\,a}_{\alpha_1}\kln{\kappa_1 \, \tx,\kappa_2 \, \tx}  $.
In the flavor singlet case $\lambda^0_f = 1$ however, these operators mix under renormalization with the gluon operators}
\label{str_oprab_gluon}
 G_{\alpha_1}\kln{\kappa_1 \, \tx,\kappa_2 \, \tx}
&=
\quad
 \delta_{\alpha_1}^{(\beta_1} \tx^{\beta_2)} \; \KLno{} F^{a\,\rho}_{\beta_1}\kln{\kappa_1 \, \tx}
\, U^{ab}\kln{\kappa_1 \, \tx,\kappa_2 \, \tx} \, F^{b}_{\rho\beta_2}\kln{\kappa_2 \, \tx}
\\
\nonumber
&
\qquad \qquad \qquad \quad
+ F^{a\,\rho}_{\beta_1} \kln{\kappa_2 \, \tx}
\, U^{ab}\kln{\kappa_2 \, \tx,\kappa_1 \, \tx} \, 
F^{b}_{\rho\beta_2}\kln{\kappa_1 \, \tx} \oKLn{}
\\
\nonumber
&
\\
\label{str_oprbb_gluon}
 G^{5}_{\alpha_1}\kln{\kappa_1 \, \tx,\kappa_2 \, \tx}
&=
\quad
 \delta_{\alpha_1}^{(\beta_1} \tx^{\beta_2)} \; \KLno{} F^{a\,\rho}_{\beta_1}\kln{\kappa_1 \, \tx}
 \, U^{ab}\kln{\kappa_1 \, \tx,\kappa_2 \, \tx} \, \tilde F^{b}_{\rho\beta_2}\kln{\kappa_2 \, \tx}
\\
\nonumber
&
\qquad \qquad \qquad \quad
- F^{a\,\rho}_{\beta_1}\kln{\kappa_2 \, \tx} 
\, U^{ab}\kln{\kappa_2 \, \tx,\kappa_1 \, \tx} \, \tilde F^{b}_{\rho\beta_2}\kln{\kappa_1 \, \tx} \oKLn{} \; ,
\end{align}
where
\begin{equation}
F_{\mu\nu} = F_{\mu\nu}^a \, t^a \qquad \text{and} \qquad \tilde F_{\mu\nu} = \frac{1}{2} \; \varepsilon_{\mu\nu\alpha\beta} \; F^{\alpha\beta}
\end{equation}
is the gluon field strength and the dual field strength, respectively. The leading quark--mass contributions are determined by the operators $M^a_{[\alpha_1\alpha_2]} \kln{\kappa_1 \, \tx, \kappa_2 \, \tx}$ and $N^{a}\kln{\kappa_1 \, \tx,\kappa_2 \, \tx}$ which we now give for general values of $\kappa_1$ and $\kappa_2$ including the phase factor $U\kln{\kappa_1 \, \tx, \kappa_2 \, \tx}$
\begin{align}
\label{str_oprab_M}
 M^a_{[\alpha_1\alpha_2]}\kln{\kappa_1 \, \tx,\kappa_2 \, \tx}
&=
\bar\psi\kln{\kappa_1 \, \tx}
\, \sigma_{[\alpha_1\alpha_2]} \, \lambda^a_f \, U\kln{\kappa_1 \, \tx,\kappa_2 \, \tx}\,\psi\kln{\kappa_2 \, \tx}
\\
\nonumber
&
\qquad \quad 
- \bar\psi\kln{\kappa_2 \, \tx}
\, \sigma_{[\alpha_1\alpha_2]} \,\lambda^a_f \, U\kln{\kappa_2 \, \tx,\kappa_1 \, \tx} \,  
\psi\kln{\kappa_1 \, \tx}
\\
\nonumber
&
\\
\label{str_oprbb_N}
 N^{a}\kln{\kappa_1 \, \tx,\kappa_2 \, \tx}
&=
\bar\psi\kln{\kappa_1 \, \tx}
\,\lambda^a_f \, U\kln{\kappa_1 \, \tx,\kappa_2 \, \tx} \,
\psi\kln{\kappa_2 \, \tx}
\\
\nonumber
&
\qquad \quad 
+ \bar\psi\kln{\kappa_2 \, \tx}
\,\lambda^a_f \, U\kln{\kappa_2 \, \tx,\kappa_1 \, \tx} \, \psi\kln{\kappa_1 \, \tx} \; .
\\
\intertext{In the flavor singlet case, these two operators mix with the gluon operators}
\label{str_oprab_gluon_M}
 G^5_{[\alpha_1\alpha_2]}\kln{\kappa_1 \, \tx,\kappa_2 \, \tx}
&=
{F^{a\,\rho}}_{[\alpha_1} \kln{\kappa_1 \, \tx}
\, U^{ab}\kln{\kappa_1 \, \tx,\kappa_2 \, \tx} \, \tilde F^{b}_{\alpha_2]\rho}\kln{\kappa_2 \, \tx}
\\
\nonumber
&
\qquad \quad
+ {F^{a\,\rho}}_{[\alpha_1} \kln{\kappa_2 \, \tx}
\, U^{ab}\kln{\kappa_2 \, \tx,\kappa_1 \, \tx} \,
\tilde F^{b}_{\alpha_2]\rho}\kln{\kappa_1 \, \tx}
\\
\nonumber
&
\\
\label{str_oprbb_gluon_N}
 G\kln{\kappa_1 \, \tx,\kappa_2 \, \tx}
&=
F^{a\,\sigma\rho}\kln{\kappa_1 \, \tx}
 \, U^{ab}\kln{\kappa_1 \, \tx,\kappa_2 \, \tx} \,  
F^{b}_{\sigma\rho}\kln{\kappa_2 \, \tx}
\\
\nonumber
&
\qquad \quad
- F^{a\,\sigma\rho}\kln{\kappa_2 \, \tx} 
\, U^{ab}\kln{\kappa_2 \, \tx,\kappa_1 \, \tx} \, F^{b}_{\sigma\rho}\kln{\kappa_1 \, \tx} \; .
\end{align}

\subsection{The concept of geometric twist}

Now, we will discuss a method for the decomposition of local QCD operators into operators with definite \textit{geometric twist}. The method is essentially based upon the application of representation theory of the Lorentz group to an infinite tower of local operators $\OP_{\Gamma|n}\kln{x}$. Thereby, we decompose these local operators into components carrying an irreducible tensor representation of the Lorentz group. In four space--time dimensions these irreducible tensors carry an integer Lorentz spin $j_n$ which allows us to define the geometric twist according to Ref.~\cite{GT71} as
\begin{equation}
\label{DefinitionOfTwist}
\text{twist} \, \kln{\tau} \; = \; \text{canonical dimension} \, \kln{d_n} - \text{Lorentz spin} \, \kln{j_n} \; .
\end{equation}
Since the twist $\tau$ is not $n$--dependent, we can sum the infinite tower of local operators $\OP^{\text{tw}\kln{\tau}}_{\Gamma|n}\kln{x}$ of definite geometric twist to nonlocal operators of (definite) geometric twist which yields two advantages. First, these nonlocal twist operators are Lorentz covariant tensors and second, the twist decomposition is process-- and model--independent since the geometric twist is defined for the nonlocal \textit{operators} and not for their matrix elements.

Here, it is important to demarcate the notion of \textit{geometric twist} from the notion of \textit{dynamical twist} which has been proposed by Jaffe and Ji~\cite{JJ92}. This dynamical twist $\kln{t}$ counts powers $Q^{2-t}$ of the energy--momentum transfer $Q$ and is directly related to the power by which the corresponding distribution amplitudes contribute to the scattering amplitude. Since distribution amplitudes can only be defined for matrix elements of nonlocal operators, the notion of dynamical twist is not independent of the specific process and also not Lorentz covariant. Therefore, we are using the concept of geometric twist.  

To explain the procedure of twist decomposition we generically denote the nonlocal QCD operators by  
\begin{equation}
\label{GenericOperators}
\OP_\Gamma \kln{\kappa_1 \, x,\kappa_2 \, x}
=
\Phi' \kln{ \kappa_1 \, x} \Gamma \; U\kln{\kappa_1 \, x,\kappa_2 \, x}
\Phi \kln{\kappa_2 \, x} \; ,
\end{equation}
where we understand the operators $\OP_\Gamma\kln{\kappa_1 \, x,\kappa_2 \, x}$ to be unrenormalized quantities. $\Phi$ generically denotes the different local fields contained in the bilocal operators and $\Gamma$ like in (\ref{int_lce_y}) denotes the free indices of the nonlocal operators.

For $d=2h$ space--time dimensions these fields are the scalar field $\phi$ with canonical dimension $d_\phi = h-1$, the Dirac spinor $\psi$ with canonical dimension $d_\psi = h-1/2$ as well as the gauge field strength $F_{[\mu\nu]}$ with canonical dimension $d_F = h$. The covariant derivative $D_\mu = \pd_\mu + \im \, g \, A_\mu$ has a dimension of one in any space--time dimension $2h$ since $d_g = 2-h$ and $d_A = h-1$.

The \textit{complete (infinite) twist decomposition} of the nonlocal operators (\ref{GenericOperators})
\textit{off the light~cone} basically proceeds in three steps.

\subsubsection*{First step}

We take the operators $\OP_\Gamma \kln{\kappa_1 \, x,\kappa_2 \, x}$ off the light~cone and perform a \textit{Taylor expansion} for arbitrary values of $x$ at the expansion point $y=0$:
\begin{equation}
\label{TaylorExpansionZero}
\OP_\Gamma \kln{\kappa_1 \, x,\kappa_2 \, x}
= \sum_{n=0}^{\infty} \frac{1}{n!} \; x^{\zeta_1}\cdots x^{\zeta_n} \; \KLE{ \Phi'(y) \, \Gamma \; \TDA_{\zeta_1} \! \kln{\kappa_1, \kappa_2} \cdots \TDA_{\zeta_n} \! \kln{ \kappa_1, \kappa_2} \Phi(y) }_{y=0}
\end{equation}
with the generalized covariant derivatives
\begin{align}
\TDA_{\mu}\kln{\kappa_1, \kappa_2}
&=
\kappa_1 \, \overset{\la}{D}_\mu + \kappa_2 \, \overset{\ra}{D}_\mu
\\
\overset{\la}{D}_{\mu}
&=
\overset{\la}{\pd^y_\mu} - \im \, g \, A_\mu\kln{y}
\\
\overset{\ra}{D}_{\mu}
&=
\overset{\ra}{\pd^y_\mu} + \im \, g \, A_\mu\kln{y} \; .
\\
\intertext{By defining the variables}
\kappa_+ 
&:=
\frac{1}{2} \, \kln{\kappa_2 + \kappa_1}
\\
\kappa_-
&:=
\frac{1}{2} \, \kln{\kappa_2 - \kappa_1}
\\
\intertext{and the derivatives}
\pd_\mu^+
&:=
\overset{\ra}{D}_{\mu} + \overset{\la}{D}_{\mu}
=  \overset{\ra}{\pd^y_\mu} + \overset{\la}{\pd^y_\mu}
\\
\label{DefCovariantGene}
\DA_\mu
&:=
\overset{\ra}{D}_{\mu} - \overset{\la}{D}_{\mu}
=
\overset{\ra}{\pd^y_\mu} - \overset{\la}{\pd^y_\mu} + 2 \, \im \, g \, A_\mu\kln{y}
\\
\intertext{we can re--express $\TDA_{\mu}\kln{\kappa_1, \kappa_2}$ by}
\TDA_{\mu}\kln{\kappa_1, \kappa_2}
&=
\kappa_- \, \DA_\mu + \kappa_+ \, \pd_\mu^+ \; .
\end{align}
If we now restrict our considerations to the centered bilocal operators $\OP_\Gamma\kln{-\kappa \, x, \kappa  \, x}$ with $\kappa = \kappa_-$ and $\kappa_+ = 0$, we obtain a much simpler expression for the Taylor expansion (\ref{TaylorExpansionZero})\footnote{
The consideration of the centered operator $\OP_\Gamma\kln{-\kappa \, x, \kappa  \, x}$ is no restriction of generality since 
a Taylor expansion around the expansion point $y=\kappa_+ \, x$ yields the same local operators (\ref{GenericLocalOperators}).},
\begin{equation}
\label{SummationToNonlocalOperators}
\OP_\Gamma\kln{-\kappa \, x, \kappa  \, x}
=
\sum_{n=0}^\infty \frac{\kappa^n}{n!} \; \OP_{\Gamma|n}\kln{x}
\end{equation}
with the local operators
\begin{eqnarray}
\label{GenericLocalOperators}
\OP_{\Gamma|n}\kln{x}
&=&
x^{\zeta_1} \cdots x^{\zeta_n} \; \OP_{\Gamma(\zeta_1\dots\zeta_n)}\kln{y} \PIpe{y=0}
\\
&=&
X^{\zeta_1\dots\zeta_n} \; \Phi'\kln{y} \; \Gamma \; \DA_{(\zeta_1} \cdots \DA_{\zeta_n)} \Phi\kln{y} \PIpe{y=0}
\\
&=&
\Phi'\kln{y} \; \Gamma \, \Kln{x\DA}^n \, \Phi\kln{y} \PIpe{y=0} \; .
\end{eqnarray}
Obviously, the operators $\OP_{\Gamma|n}\kln{x}$ are homogeneous polynomials of degree $n$ in $x \in \mathbb{M}^{2h}$ and constitute
an infinite tower of local operators. Here and in the following, we will denote the symmetric tensor $x^{\zeta_1} \cdots x^{\zeta_n}$ by $X^{\zeta_1\dots\zeta_n}$.

\subsubsection*{Second step}

This tower of local operators $\OP_{\Gamma|n}\kln{x}$ is now decomposed into tensors which carry an irreducible tensor representation of the $2h$--dimensional Lorentz group $SO(1,2h-1;\R)$. From (\ref{GenericLocalOperators}) we see that $\OP_{\Gamma|n}\kln{x}$ is irreducible if and only if
$\OP_{\Gamma(\zeta_1\dots\zeta_n)}\kln{y}$ is irreducible and we know from representation theory of classical Lie groups (see Ref.~\cite{BR77} Chapter~8 and  Ref.~\cite{VK93} Chapter~16) that the irreducible (finite dimensional) tensor representations of $SO(1,2h-1;\R)$ are related to the irreducible (finite dimensional) tensor representations of $SO(2h;\C)$ by complex extension. Any complex--analytic irreducible representation $\mathcal{T}_{G_\C}$ of a semi--simple Lie group $G_\C$ subduces a real  irreducible representation $\mathcal{T}_{\tilde G_\R}$ of a real form\footnote{A group $\tilde G_\R$ is called a real form of the group $G_\C$ if its (unique) complex extension coincides with $G_\C$.} $\tilde G_\R$ of $G_\C$. On the other hand, any irreducible representation $\mathcal{T}_{\tilde G_\R}$ of a real semi--simple Lie group induces a complex irreducible representation $\mathcal{T}_{G_\C}$ of the complex extension $G_\C$ of $\tilde G_\R$. Therefore, in order to investigate the irreducible tensor representations of the Lorentz group $SO(1,2h-1;\R)$ we may consider equally well the irreducible tensor representations of the complex group $SO(2h;\C)$.

These representations are realized on the space of traceless tensors of rank $m=n+k$ ($k$ denotes the number of indices of $\Gamma$), whose symmetry class is determined by (orthogonal) Young operators $\YH_{[\m]}=f_{(\m)}/m! \; q_{[\m]} \, \QU_{[\m]} \, \PRa_{[\m]}$. Here, $(\m)=(\m_1,\dots,\m_r)$ with
\begin{equation}
\m_1 \ges \m_2 \ges \dots \ges \m_r \qquad \text{and} \qquad \sum_{i=1}^r = m
\end{equation}
denotes the corresponding Young pattern and $[\m]$ denotes a standard Young tableau obtained from the pattern $(\m)$ by inserting the $m$ indices in lexicographic order. $\PRa_{[\m]}$ and $\QU_{[\m]}$ denote symmetrization and antisymmetrization with respect to the tableau $[\m]$, whereas the operation $q_{[\m]}$ ensures the orthogonality of the Young operators~$\YH_{[\m]}$; see Section~\ref{SymmetryPart} for details about the construction of $\YH_{[\m]}$.

Since the indices $\zeta_1$ to $\zeta_n$ of the operators $\OP_{\Gamma(\zeta_1\dots\zeta_n)}\kln{y}$ are symmetric, they must necessarily by inserted into the first row of the Young pattern $(\m)$. If we first restrict our considerations to a generic nonlocal second rank tensor, i.e. $k=2$, we find four possible patterns
\begin{enumerate}
\item[(I)]{
\begin{picture}(5,2)
\unitlength0.5cm
\linethickness{0.075mm}
\put(1.000,0){\framebox(1,1){}}
\put(2.015,0){\framebox(1,1){}}
\put(3.030,0){\framebox(1,1){}}
\put(4.045,0){\framebox(1,1){}}
\put(5.060,0){\framebox(3,1){$\cdots$}}
\put(8.075,0){\framebox(1,1){}}
\put(9.09,0){\framebox(1,1){}}
\put(10.105,0){\framebox(1,1){}}
\put(11.120,0){\framebox(1,1){}}
\end{picture}\\}

\item[(II)]{
\begin{picture}(5,2)
\unitlength0.5cm
\linethickness{0.075mm}
\put(1.000,0){\framebox(1,1){}}
\put(2.015,0){\framebox(1,1){}}
\put(3.030,0){\framebox(1,1){}}
\put(4.045,0){\framebox(1,1){}}
\put(5.060,0){\framebox(3,1){$\cdots$}}
\put(8.075,0){\framebox(1,1){}}
\put(9.09,0){\framebox(1,1){}}
\put(10.105,0){\framebox(1,1){}}
\put(1.000,-1.015){\framebox(1,1){}}
\end{picture}\\ \\}

\item[(III)]{
\begin{picture}(5,2)
\unitlength0.5cm
\linethickness{0.075mm}
\put(1.000,0){\framebox(1,1){}}
\put(2.015,0){\framebox(1,1){}}
\put(3.030,0){\framebox(1,1){}}
\put(4.045,0){\framebox(1,1){}}
\put(5.060,0){\framebox(3,1){$\cdots$}}
\put(8.075,0){\framebox(1,1){}}
\put(9.090,0){\framebox(1,1){}}
\put(1.000,-1.015){\framebox(1,1){}}
\put(1.000,-2.030){\framebox(1,1){}}
\end{picture}\\ \\ \\}

\item[(IV)]{
\begin{picture}(5,2)
\unitlength0.5cm
\linethickness{0.075mm}
\put(1.000,0){\framebox(1,1){}}
\put(2.015,0){\framebox(1,1){}}
\put(3.030,0){\framebox(1,1){}}
\put(4.045,0){\framebox(1,1){}}
\put(5.060,0){\framebox(3,1){$\cdots$}}
\put(8.075,0){\framebox(1,1){}}
\put(9.090,0){\framebox(1,1){}}
\put(1.000,-1.015){\framebox(1,1){}}
\put(2.015,-1.015){\framebox(1,1){}}
\end{picture}\\}
\end{enumerate}
Since all the above Young symmetries are to be realized on traceless tensors, all patterns must fulfill the relation $\ell_1 + \ell_2 \les d=2h$ or will vanish. Here, $\ell_i$ is the length of the $i$th column of the pattern $(\m)$. This has a major impact on the complete decomposition into irreducible tensors. For $d=2$, for example, only the complete symmetric Young pattern (I) contributes. The most important case from the point of view of twist decomposition is of course $d=4$. Here, the Young patterns (I) to (IV) contribute to irreducible tensors with definite geometric twist.

In four dimensions we have the possibility to label the irreducible representations
of the Lorentz group by the Lorentz spin $j$. For the contributing Young patterns (I) to (IV) we find the spins
\begin{eqnarray}
\label{SpinsI}
\text{(I)}
&&
j = m,m-2,m-4,\dots
\\
\label{SpinsII}
\text{(II)}
&&
j = m-1,m-2,m-3,\dots
\\
\label{SpinsIII}
\text{(III)}
&&
j = m-2,m-3,m-4,\dots
\\
\label{SpinsIV}
\text{(IV)}
&&
j = m-2,m-3,m-4,\dots 
\end{eqnarray}
The highest spins correspond to irreducible (traceless) tensors of the respective Young symmetry $(\m)$ while the lower spins are related to trace terms containing irreducible tensors which do not necessarily carry the symmetry $(\m)$. This quite complicated interplay between trace terms and irreducible tensors will become clear in the Sections~\ref{TracePart} and~\ref{SymmetryPart} where we perform the complete decomposition of generic local operators into irreducible tensors. In fact, it is the major challenge in this decomposition to find all irreducible contributions inside the trace terms.

For the local QCD operator $\OP_{\Gamma(\zeta_1\dots\zeta_n)}\kln{y}$ this problem is solvable only due to the fact that the $n$ indices
$\zeta_1$ to $\zeta_n$ are totally symmetric because of their multiplication with the symmetric tensor~$X^{\zeta_1\dots\zeta_n}$.
We emphasize that it would be impossible to find a closed form for the complete decomposition into irreducible tensors without this
symmetrization, i.e. for the (unsymmetric) local operator $\OP_{\Gamma\zeta_1\dots\zeta_n}\kln{y}$ with unrestricted $n$.
For the (generic) local QCD operators $\OP_{\Gamma|n}\kln{x}$ given by (\ref{GenericLocalOperators}) the tensor
$X^{\zeta_1\dots\zeta_n}$ is naturally contained due to the Taylor~expansion (\ref{TaylorExpansionZero}) and
it is compelling to use this tensor as a tool to obtain the desired decompositions. Thereby, we express, loosely speaking,
all operations performed on the local operators $\OP_{\Gamma(\zeta_1\dots\zeta_n)}\kln{y}$\footnote{
For example, the subtraction of trace terms and the applications of Young projectors;
see Sections~\ref{TracePart} and~\ref{SymmetryPart} for the details.} 
by differential operators acting on $X^{\zeta_1\dots\zeta_n}$. We will refer to this technique as \textit{polynomial technique}.

We have already remarked that the tensor $X^{\zeta_1\dots\zeta_n}$ is naturally generated for local QCD operators. But on the other hand, this tensor can as well be introduced by hand as an auxiliary device if a decomposition of the uncontracted tensor $\OP_{\Gamma(\zeta_1\dots\zeta_n)}\kln{y}$ shall be obtained. Via the polynomial technique we first obtain the complete
decomposition for $\OP_{\Gamma|n}\kln{x;y} = X^{\zeta_1\dots\zeta_n} \, \OP_{\Gamma(\zeta_1\dots\zeta_n)}\kln{y}$ and then remove all vectors $x^{\zeta_i}$ in a second step by the simple replacements $x^2 \ra \delta_{\zeta_i\zeta_j}$ and $x_{\alpha_i} \ra \delta_{\alpha_i\zeta_k}$\footnote{Here, $\alpha_i$ indicates an index contained in $\Gamma$.}. A final symmetrization of $\zeta_1$ to $\zeta_n$ then yields the decomposition of $\OP_{\Gamma(\zeta_1\dots\zeta_n)}\kln{y}$.

A generic form for the complete decomposition of $\OP_{\Gamma|n}\kln{x}$ into irreducible tensors is given by
\begin{equation}
\label{FormalExpansionDimD}
\OP_{\Gamma|n}\kln{x}
= \bigoplus_{[\m]} \; \, \bigoplus_{i} \; \, \bigoplus_{t} \;
c_{\Gamma^t|n^t}^{i|[\m]}\kln{\delta,x,n,d} \cdot \OP_{\Gamma \backslash \Gamma^t|n-n^t}^{i|[\m]}\kln{x} \; ,
\end{equation}
where $\Gamma \backslash \Gamma^t$ is the complement of $\Gamma^t$ in $\Gamma$. First, we have to sum over all contributing Young symmetries $[\m]$ which define the irreducible local operators $\OP_{\Gamma'|n'}^{i|[\m]}\kln{x}$. The second direct sum over $i$ denotes that one and the same Young symmetry is in general realized by different (independent) irreducible operators with a different "internal" structure. Furthermore, these operators can appear in different types of trace terms (for example $x^2 \, \delta_{\alpha_i\alpha_j}$ and $x_{\alpha_i} x_{\alpha_j}$) which we indicate by the third sum over $t$.

The coefficient functions\footnote{The coefficients $c_{\Gamma^t|n^t}^{i|[\m]}\kln{x,n,d}$ are denoted with the
same number $i$ and the same Young symmetry $[m]$ which, of course, does not mean that these coefficients are irreducible in any
sense.} $c_{\Gamma^t|n^t}^{i|[\m]}\kln{\delta,x,n,d}$ can depend on the traces $\delta_{\alpha_i\alpha_j}$, $x_{\alpha_i}$ and $x^2$ and  are proportional to fractions of polynomials in $n$ and $h$. These weights thereby depend upon the order $n$ of the local operator $\OP_{\Gamma|n}\kln{x}$, the space--time dimension $d=2h$ as well as the Young symmetry $[\m]$. If the coefficient $c_{\Gamma^t|n^t}^{i|[\m]}\kln{\delta,x,n,d}$ depends on $\delta$ or $x$ the respective term is a trace term.

After the restriction of the operators $\OP_{\Gamma'|n'}^{i|[\m]}\kln{x}$ to the real Lorentz group $SO(1,3;\R)$ the sum over the contributing Young tableaux $[\m]$ can be replaced by a sum over the respective twist of the irreducible local operators $\OP_{\Gamma'|n'}^{i|[\m]}\kln{x}$ and we obtain for $d=4$
\begin{equation}
\label{FormalExpansionDim4}
\OP_{\Gamma|n}\kln{x}
= \bigoplus_{\tau=\tau_{\text{min}}}^{\tau_{\text{max}}} \bigoplus_{j} \; \, \bigoplus_{t} \;
c_{\Gamma^t|n^t}^{\,j|\tau}\kln{\delta,x,n} \cdot \OP_{\Gamma \backslash \Gamma^t|n-n^t}^{j|\tw\kln{\tau}}\kln{x} \; .
\end{equation}
Thereby, we also have to enlarge the sum over $i$ since two irreducible tensor operators which carry different inequivalent Young symmetries $[\m]$ and $[\m']$ can, in general, have the same spin and twist and we therefore replaced the sum over $i$ by a (in general larger) sum over $j$. $\tau_{\text{max}}$ is restricted by the order $n$ of the local operator $\OP_{\Gamma|n}\kln{x}$ and it holds
\begin{equation}
\label{restrictionTwist}
\tau_{\text{max}} \les \text{dim}\Kln{\Phi} + \text{dim}\Kln{\Phi'} + n \; .
\end{equation}
The content of the formal decomposition (\ref{FormalExpansionDimD}) will become clearer in the Sections~\ref{TracePart} and~\ref{SymmetryPart} where we discuss explicit examples. Further examples for the generic local twist decomposition~(\ref{FormalExpansionDim4}) will be given in Section~\ref{ApplicationQCD}.

\subsubsection*{Third step}

As a third and last step, one has to perform a \textit{resummation} of the local operators belonging to the same symmetry class and thereby having the same twist according to
\begin{equation}
\label{NonLocalSummation}
\OP_{\Gamma}^{i|\tw\kln{\tau}} \kln{-\kappa \, x, \kappa \, x}
= \bigoplus_{n=0}^\infty \frac{\kappa^n}{n!} \; \OP_{\Gamma|n}^{i|\tw\kln{\tau}}\kln{x} \; .
\end{equation}
At this point, an important remark is in order. Since the nonlocal operators $\OP_{\Gamma}^{i|\tw\kln{\tau}} \kln{-\kappa \, x, \kappa \, x}$ of definite geometric twist are an infinite sum of different local operators $\OP_{\Gamma|n}^{i|\tw\kln{\tau}}\kln{x}$ whose Lorentz spins $j_n$ are proportional to the order $n$ of the local operator (see the spin content of the Young patterns (I) to (IV) given by (\ref{SpinsI}) to (\ref{SpinsIV})), the nonlocal operator does not carry a definite Lorentz spin and is not irreducible under $SO(1,3;\R)$. Therefore, the notion of twist is much weaker in the nonlocal setting then in the local setting. However, only the geometric twist $\tau$ defined according to (\ref{DefinitionOfTwist}) is invariant under the summation (\ref{NonLocalSummation}) and can be used for the classification of nonlocal operators.

After the summation~(\ref{NonLocalSummation}) to nonlocal operators, the local version (\ref{FormalExpansionDim4}) of the twist decomposition obtains the form
\begin{equation}
\label{NonLocalFormalExpansion}
\OP_{\Gamma} \kln{-\kappa \, x, \kappa \, x}
= \bigoplus_{\tau=\tau_{\text{min}}}^{\infty} \bigoplus_{j} \;\, \bigoplus_{t} \;
\int_0^1 \text{d} \lambda \;\, c_{\Gamma^t}^{\,j|\tau} \kln{\delta,x,\pd,\lambda,\kappa} \cdot
\OP_{\Gamma \backslash \Gamma^t}^{j|\tw\kln{\tau}}\kln{- \lambda \kappa \, x, \lambda \kappa \, x} \; .
\end{equation}
In contrast to the local coefficients $c_{\Gamma^t|n^t}^{j|\tau}\kln{\delta,x,n}$, the coefficients $c_{\Gamma^t}^{j|\tau} \kln{\delta,x,\pd,\lambda,\kappa}$ are differential operators acting on the nonlocal operators $\OP_{\Gamma \backslash \Gamma^t}^{j|\tw\kln{\tau}}\kln{- \lambda \kappa \, x, \lambda \kappa \, x}$ of definite geometric twist. The $\lambda$--integration in (\ref{NonLocalFormalExpansion}) stems from the nonlocal representation of the denominators of the local coefficients $c_{\Gamma^t|n^t}^{\,j|\tau}\kln{\delta,x,n}$ whereas the differentiations in (\ref{NonLocalFormalExpansion}) are related to nonlocal representations of the respective numerators. Since $n$ is no longer restricted, the complete local twist decomposition~(\ref{FormalExpansionDim4}) turns into an infinite twist decomposition after the nonlocal summation; see relation~(\ref{restrictionTwist}).

Unfortunately, this third and last step in the process of twist decomposition is very problematic for two reasons.
\begin{itemize}
\item{Although the infinite twist decomposition for each nonlocal operator is of the form~(\ref{NonLocalFormalExpansion}) there is no generic nonlocal twist decomposition covering all operator of one and the same rank. Off the light~cone such a generic decomposition exists only in local form.}  
\item{The single integral representation for the coefficient functions $c_{\Gamma^t}^{\,j|\tau} \kln{\delta,x,\pd,\lambda,\kappa}$ yields hypergeometric functions $F_{[p,q]}\Kln{[a_1,\dots,a_p],[b_1,\dots,b_q],z}$, which can only be avoided by choosing a multiple integral representation for the local coefficients $c_{\Gamma^t|n^t}^{\,j|\tau}\kln{\delta,x,n}$. This multiple integral representation however is no longer unique. }
\end{itemize}
To avoid these problems we will not perform summations to nonlocal operators in this paper. The generic decompositions into $SO(2h;\C)$--irreducible tensors given in Section~\ref{spindecompositions} are given in local form only. This also holds for the twist decompositions of various QCD operators in Section~\ref{ApplicationQCD}. If a decomposition is given in nonlocal form in the literature we will refer to the respective references.

\newpage

\section{Trace Part}
\label{TracePart}

Now that we have generalized the problem of twist decomposition of nonlocal QCD operators to the problem of the complete decomposition of a general tensor $\OP_{\alpha_1\dots\alpha_k\kln{\zeta_1\dots\zeta_n}}$ into irreducible $SO(2h,\C)$--tensors we will treat all problems related to tracelessness of these irreducible tensors in this Section. The $k$ free indices contained in $\Gamma$ are now denoted by $\alpha_1$ to $\alpha_k$. In principle, two problems have to be treated.
\begin{itemize}
\item{First, we have to describe a method that can render any tensor $\OP_{\alpha_1\dots\alpha_k\kln{\zeta_1\dots\zeta_n}}$ traceless; this problem is treated in Subsection~\ref{SubtractionOfAllTraces}.}
\item{Second, we have to separate all trace terms from all traceless contributions in these tensors, i.e. we have to construct a complete trace decomposition. This is done in Subsection~\ref{CompleteTraceDecomposition}.}
\end{itemize}
The remaining  problems related to Young symmetry including the construction of spin~projectors and the final spin~decompositions are treated in Section~\ref{SymmetryPart}.

\subsection{The subtraction of all traces of $\OP_{\alpha_1\dots\alpha_k|n}$}
\label{SubtractionOfAllTraces}

Since an irreducible $SO(2h,\C)$--tensor must be traceless, the construction of a projector $\HP{\alpha}{\alpha'}{k}{n}$ which subtracts all traces of the partly contracted tensor
\begin{align}
\OP_{\alpha_1\dots\alpha_k|n} 
&:= \XS{\zeta}{n} \; \OP_{\alpha_1\dots\alpha_k\kln{\zeta_1\dots\zeta_n}}
\intertext{is a natural starting point for the discussion. The resulting traceless tensor will be denoted by $\OPT_{\alpha_1\dots\alpha_k|n}$ and is given by}
\OPT_{\alpha_1\dots\alpha_k|n}
&:= \HP{\alpha}{\alpha'}{k}{n} \, \OP_{\alpha'_1\dots\alpha'_k|n} \; .
\end{align}
Here, we postulate the existence of such a projector $\HP{\alpha}{\alpha'}{k}{n}$ and will now deduce it properties. If we join both sets of indices $\kls{\alpha}$ and $\kls{\zeta}$ to
\begin{equation}
\label{JoinSets}
\kls{\xi} := \kls{\zeta} \cup \kls{\alpha} \qquad \text{and} \qquad m := n + k
\end{equation}
we can, of course, formulate the condition of tracelessness for $\OPT_{\alpha_1\dots\alpha_k\kln{\zeta_1\dots\zeta_n}}$ by
\begin{equation}
\label{generalTracelessCondition}
\delta^{\xi_i \xi_j} \; \OPT_{\alpha_1\dots\alpha_k\kln{\zeta_1\dots\zeta_n}} = 0 \qquad \fo \quad 1 \les i < j \les m \; ,
\end{equation}
which then implies the tracelessness of $\OPT_{\alpha_1\dots\alpha_k|n}$. Here, $\OPT_{\alpha_1\dots\alpha_k\kln{\zeta_1\dots\zeta_n}}$ in analogy to $\OPT_{\alpha_1\dots\alpha_k|n}$ is made traceless by a constant projector $\HP{\xi}{\xi'}{m}{0}$
\begin{equation}
\OPT_{\alpha_1\dots\alpha_k\kln{\zeta_1\dots\zeta_n}} = \HP{\xi}{\xi'}{m}{0} \, \OP_{\alpha'_1\dots\alpha'_k\kln{\zeta'_1\dots\zeta'_n}} \; .
\end{equation}
If we contract $\OPT_{\alpha_1\dots\alpha_k\kln{\zeta_1\dots\zeta_n}}$ with $\XS{\zeta}{n}$ we arrive at $\OPT_{\alpha_1\dots\alpha_k|n}$, which means that the projectors $\HP{\xi}{\xi'}{m}{0}$ and $\HP{\alpha}{\alpha'}{k}{n}$ must obey the relation
\begin{equation}
\label{RealationHHxi}
\XS{\zeta}{n} \; \HP{\xi}{\xi'}{m}{0} = \HP{\alpha}{\alpha'}{k}{n} \; \XS{\zeta'}{n} \; .
\end{equation}
This relation expresses the fact that the projector $\HP{\alpha}{\alpha'}{k}{n}$ must be a differential operator which reproduces the partly contracted projector $\XS{\zeta}{n} \, \HP{\xi}{\xi'}{m}{0}$ by acting on the symmetric tensor $\XS{\zeta'}{n}$. Furthermore we can deduce an important consistency condition for $\HP{\alpha}{\alpha'}{k}{n}$ from the latter relation~(\ref{RealationHHxi}), namely it holds
\begin{equation}
\label{generalConsistencyCondition}
x^{\alpha_i} \, \HP{\alpha}{\alpha'}{k}{n}
= H_{\alpha_1 \dots \alpha_{i-1} \alpha_{i+1} \dots \alpha_k | n+1}^{\alpha'_1\dots\alpha'_{i-1}\alpha'_{i+1}\dots\alpha'_k} \; x^{\alpha'_i} \qquad \fo \quad  1 \les i \les k \; .
\end{equation}
This relation however does not yet encode the condition of tracelessness (\ref{generalTracelessCondition}). To find suitable conditions for the contracted traceless tensor $\OPT_{\alpha_1\dots\alpha_k|n}$ let us formulate condition~(\ref{generalTracelessCondition}) for the respective projector
\begin{align}
\label{generalTracelessConditionProj}
\delta^{\xi_i \xi_j} \; \HP{\xi}{\xi'}{m}{0}
&= 0 \qquad \fo \quad 1 \les i < j \les m \; .
\intertext{Since any trace contained in $\OP_{\alpha_1\dots\alpha_k\kln{\zeta_1\dots\zeta_n}}$ is proportional to some $\delta_{\xi'_i \xi'_j}$, the relation~(\ref{generalTracelessConditionProj}) can analogously be formulated as}
\label{generalTracelessConditionProjB}
\HP{\xi}{\xi'}{m}{0} \; \delta_{\xi'_i \xi'_j} 
&= 0 \qquad \fo \quad 1 \les i < j \les  m \; .
\end{align}
The latter relation~(\ref{generalTracelessConditionProjB}) expresses the simple fact that $\HP{\xi}{\xi'}{m}{0}$ projects out all trace terms.

To transform these two general conditions of tracelessness~(\ref{generalTracelessConditionProj}) and (\ref{generalTracelessConditionProjB}) into a set of conditions for the differential operator $\HP{\alpha}{\alpha'}{k}{n}$ we have to distinguish three different cases for the indices $\xi_i$ and $\xi_j$. In the following, $\OP_n$ denotes some fully contracted tensor of order $n$, i.e. a homogeneous polynomial.
\begin{enumerate}
\item{$\xi_i,\xi_j \in \kls{\alpha}$: In this case the general condition (\ref{generalTracelessConditionProj})
is expressed by the same condition $\delta^{\alpha_i\alpha_j} \, \HP{\alpha}{\alpha'}{k}{n} = 0$ for $1 \les i < j \les k$ and for all $n$. \\
For primed indices $\xi'_i,\xi'_j \in \kls{\alpha'}$ we get, in analogy to (\ref{generalTracelessConditionProjB}), $\HP{\alpha}{\alpha'}{k}{n} \; \delta_{\alpha'_i \alpha'_j} = 0$.}
\item{$\xi_i,\xi_j \in \kls{\zeta}$: Since $\Delta \, x^{\zeta_i} x^{\zeta_j} = 2 \; \delta^{\zeta_i\zeta_j}$ holds,
condition (\ref{generalTracelessConditionProj}) transforms into the condition $\Delta \, \HP{\alpha}{\alpha'}{k}{n} \, \OP_n = 0$ for all $n$ in this case. For $n$ indices $\zeta_i$ the Laplacian generates all possible contractions in $\kls{\zeta}$.
\\
For primed indices $\xi'_i,\xi'_j \in \kls{\zeta'}$ we get, according to the relations (\ref{RealationHHxi}) and (\ref{generalTracelessConditionProjB}), the analogous condition $\HP{\alpha}{\alpha'}{k}{n} \; x^2 \, \OP_{n-2} = 0$ for $n \ges 2$ since $x^{\zeta'_i} x^{\zeta'_j} \; \delta_{\zeta'_i \zeta'_j} = x^2$.}
\item{$\xi_i \in \kls{\alpha}$ and $\xi_j \in \kls{\zeta}$: Now, the general condition of tracelessness~(\ref{generalTracelessConditionProj}) is realized through $\pd^{\alpha_l} \, \HP{\alpha}{\alpha'}{k}{n} \, \OP_n = 0$ for $1 \les l \les k$ and all $n$ since $\partial^{\alpha_l} \, x^{\zeta_i} x^{\zeta_j} = \delta^{\alpha_l\zeta_i} x^{\zeta_j} + x^{\zeta_i} \delta^{\alpha_l\zeta_j}$ holds.
For $n$ indices $\zeta_i$ the derivative $\pd^{\alpha_l}$ of course generates all possible contractions of $\alpha_l$ and indices contained in $\kls{\zeta}$.
\\
For primed indices $\xi'_i \in \kls{\alpha'}$ and $\xi'_j \in \kls{\zeta'}$ we get an analogous relation $\HP{\alpha}{\alpha'}{k}{n} \; x_{\alpha'_l} \, \OP_{n-1} = 0$ for $n \ges 1$ since $x^{\zeta'_j} \, \delta_{\alpha'_l\zeta'_j} = x_{\alpha'_l}$. Again, this holds because of relations (\ref{RealationHHxi}) and (\ref{generalTracelessConditionProjB}).}
\end{enumerate}
Let us collect the six derived relations into two blocks. The first three relations encode condition~(\ref{generalTracelessConditionProj}) after the contraction with $\XS{\zeta}{n}$
\begin{align}
\label{operatorConditionDelta}
\delta^{\alpha_i\alpha_j} \, \HP{\alpha}{\alpha'}{k}{n} \; \OP_n
&=
0 \qquad \fo \quad  1 \les i < j \les k
\\
\label{operatorConditionLaplace}
\Delta \, \HP{\alpha}{\alpha'}{k}{n} \; \OP_n
&=
0
\\
\label{operatorConditionDivergence}
\pd^{\alpha_i} \, \HP{\alpha}{\alpha'}{k}{n} \; \OP_n
&=
0 \qquad \fo \quad  1 \les i \les k \; .
\intertext{The following three relations encode condition~(\ref{generalTracelessConditionProjB}) and describe the action of $\HP{\alpha}{\alpha'}{k}{n}$ on pure trace terms}
\label{operatorConditionDeltaSecondVersion}
\HP{\alpha}{\alpha'}{k}{n} \, \delta_{\alpha'_i\alpha'_j} \; \OP_n
&=
0 \qquad \fo \quad  1 \les i < j \les k, \quad n \ges 0
\\
\label{operatorConditionXQ}
\HP{\alpha}{\alpha'}{k}{n} \, x^2 \; \OP_{n-2}
&=
0 \qquad \fo \quad  n \ges 2
\\
\label{operatorConditionDivergenceXQ}
\HP{\alpha}{\alpha'}{k}{n} \, x_{ \alpha'_i} \; \OP_{n-1  }
&=
0 \qquad \fo \quad  1 \les i \les k,  \quad n \ges 1 \; .
\end{align}
The latter relations (\ref{operatorConditionDeltaSecondVersion}) to (\ref{operatorConditionDivergenceXQ}) are equivalent to the relations (\ref{operatorConditionDelta}), (\ref{operatorConditionLaplace}) and (\ref{operatorConditionDivergence}).

With this \textit{polynomial technique} we have been able to transform the conditions (\ref{generalTracelessConditionProj}) and (\ref{generalTracelessConditionProjB}) which we cannot solve for arbitrary $n$ into the six conditions of tracelessness (\ref{operatorConditionDelta}) to (\ref{operatorConditionDivergenceXQ}) which we can solve for a fixed number of free indices $k$.

\subsubsection*{Connection to the complex~cone}

Relation~(\ref{operatorConditionXQ}) reveals an important view onto the limit of $\HP{\alpha}{\rho}{k}{n}$\footnote{From now on we will denote the indices $\alpha_i'$ by $\rho_i$.} onto the complex~cone. To follow this path let us first define the complex~cone $\mathbb{K}^{2h}\kln{\C} $ by
\begin{equation}
\mathbb{K}^{2h}\kln{\C} =
\kls{\tx \in \mathbb{\C}^{2h} \PIpe{} \tx^2 = \tx_1^2 + \tx_2^2 + \dots + \tx_{2h}^2 = 0 }
\end{equation}
and a related graded algebra $P\Kln{\mathbb{K}^{2h}}$ by
\begin{equation}
P\Kln{\mathbb{K}^{2h}} := \bigoplus_{n=0}^{\infty} \, K^{2h}_n \; ,
\end{equation}
where $K^{2h}_n$ denote the spaces of homogeneous polynomials $f_n \! \kln{\tx}$ of degree $n$ on the complex~cone $\mathbb{K}^{2h}\kln{\C}$. According to Ref.~\cite{BT77} a differential operator $\tilde Q$ is called an interior differential operator on the cone if it satisfies
\begin{align}
\label{ConditionOnCone}
\tilde Q \; \Kle{x^2 \, f_n \! \kln{x}} \PIpe{x=\tx}
&= 0 \; .
\intertext{If we restrict relation~(\ref{operatorConditionXQ}) to the complex~cone we get}
\label{ConditionOnConeH}
\HPT{\alpha}{\rho}{k}{n} \; \Kle{x^2 \, \tilde \OP_{n-2} } \PIpe{x=\tx }
&= 0
\end{align}
which of course means that $\HPT{\alpha}{\rho}{k}{n}$ must be an interior differential operator on $\mathbb{K}^{2h}\kln{\C}$ since $\tilde \OP_{n-2}$ is a homogeneous polynomial. This observation gives us the building blocks for the on--cone projector $\HPT{\alpha}{\rho}{k}{n}$ because Ref.~\cite{BT77} establishes a representation of the conformal Lie algebra $\mathfrak{so}\kln{2,2h}$ in terms of interior differential operators. A basis of this representation then also serves as a basis for the interior operators $\tilde Q$ and this basis of interior differential operators reads
\begin{eqnarray}
\tx_\alpha 
&=&
x_\alpha \, \PIpe{x=\tx}
\\
\tilde \D_\alpha
&=&
\X \, \pd_\alpha - \frac{1}{2} \, x_\alpha \, \Delta \, \PIPe{x=\tx}
\\
\tilde \X
&=&
x \pd + h - 1 \, \PIpe{x=\tx}
\\
\tilde \X_{\kle{\alpha\beta}}
&=&
- x_{\alpha} \, \pd_{\beta} + x_{\beta} \, \pd_{\alpha} \, \PIpe{x=\tx}
\\
\nonumber
\text{and} \qquad \delta_{\alpha\beta} \; .
\end{eqnarray}
One can easily see that these operators are interior operators on the complex cone since the only non vanishing commutators with $x^2$ are proportional to $x^2$ and therefore vanish in the on--cone limit
\begin{align}
\label{KomD}
\Kle{\D_\alpha,x^2}
&=
2 \, x^2 \, \partial_\alpha
\\
\label{KomX}
\Kle{\X,x^2}
&=
2 \, x^2 \; .
\intertext{The non vanishing commutators of the basis elements
$\Kls{ \tx_\alpha,\tilde\D_\alpha,\tilde\X,\tilde\X_{\kle{\alpha\beta } },\delta_{\alpha\beta}}$ read }
\Kle{\tilde \D_\alpha,\tx_\beta}
&=
\delta_{\alpha\beta} \; \tilde \X + \tilde \X_{\kle{\alpha\beta}}
\\
\Kle{\tilde \X_{\kle{\alpha\beta}},\tx_\mu}
&=
\delta_{\alpha\mu} \, \tx_{\beta} - \delta_{\beta\mu} \, \tx_\alpha
\\
\Kle{\tilde \X_{\kle{\alpha\beta}},\tilde \D_\mu}
&=
\delta_{\alpha\mu} \, \tilde \D_{\beta} - \delta_{\beta\mu} \, \tilde \D_\alpha
\\
\Kle{\tilde \D_\alpha,\tilde \X}
&=
\tilde \D_\alpha
\\
\Kle{\tx_\alpha,\tilde \X}
&=
- \tx_\alpha \; .
\end{align}
All basis elements can be used in an ansatz for $\HPT{\alpha}{\rho}{k}{n}$ but it is reasonable to further reduce the number of possible elements. Since the projector $\HPT{\alpha}{\rho}{k}{n}$ acts on local tensors which are homogeneous polynomials of order $n$, the action of $\tilde \X$ on such a tensor will return the number $n+h-1$ as a result. Such factors can be absorbed into respective coefficients in our ansatz. Moreover, the action of $\tilde \X_{\kle{\alpha\beta}}$ on a homogeneous polynomial can be expressed via $\tx_\alpha$ and $\tilde \D_\alpha$ and it holds
\begin{equation}
\label{XtoxDRel}
\tilde \X_{\kle{\alpha\beta}} \; \OP_n = - \frac{2}{n+h-2} \cdot \tx_{[\alpha} \; \tilde \D_{\beta]} \; \OP_n \; .
\end{equation}
Accordingly, the operators $\tilde \X$ and $\tilde \X_{\kle{\alpha\beta}}$ are not used in an ansatz for $\HPT{\alpha}{\rho}{k}{n}$.
Additionally, $\HPT{\alpha}{\rho}{k}{n}$ obeys the factorization
\begin{equation}
\label{factorization}
\HPT{\alpha}{\rho}{k}{n} = \HP{\alpha}{\beta}{k}{0} \; \HPTP{\beta}{\sigma}{k}{n} \; \HP{\sigma}{\rho}{k}{0} \; ,
\end{equation}
for which the conditions (\ref{operatorConditionDelta}) and (\ref{operatorConditionDeltaSecondVersion}) are already manifest due to the two projectors $\HP{\alpha}{\beta}{k}{0}$ and $\HP{\sigma}{\rho}{k}{0}$. This means that $\HPTP{\beta}{\sigma}{k}{n}$ cannot include terms proportional to $\delta_{\beta_i\beta_j}$ or $\delta^{\sigma_i\sigma_j}$.

A general ansatz for $\HPTP{\beta}{\sigma}{k}{n}$ then contains all terms of dimension zero which can be constructed out of the elements $\tx_{\beta_i}$, $\tx^{\sigma_i}$, $\tilde \D_{\beta_i}$, $\tilde \D^{\sigma_i}$ and $\delta_{\beta_i}^{\sigma_j}$ together with an undetermined coefficient. By construction $\HPT{\alpha}{\rho}{k}{n}$ therefore is an interior differential operator on the complex~cone $\mathbb{K}^{2h}\kln{\C}$ and manifestly fulfills the relations~(\ref{operatorConditionDelta}), (\ref{operatorConditionDeltaSecondVersion}) and~(\ref{ConditionOnConeH}).

An ansatz for $\HPTP{\beta}{\sigma}{k}{n}$ is generated purely by combinatorial rules. The unknown coefficients contained in this ansatz are determined in three steps
\begin{enumerate}
\item{One has to perform the $\kls{\beta,\sigma}$--contractions in (\ref{factorization}). This generates all terms in $\HPT{\alpha}{\rho}{k}{n}$ which are proportional to the metric $\delta_{\alpha_i \alpha_j}$ or $\delta^{\rho_i \rho_j}$.}
\item{One applies the condition of tracelessness (\ref{operatorConditionDivergenceXQ}) to $\HPT{\alpha}{\rho}{k}{n}$. After this operation the resulting object is completely traceless but it still contains undetermined coefficients and is therefore not unique. This is due to the fact that all spin projectors carrying a definite extended Young symmetry must also be traceless. See Section~\ref{SymmetryPart} for the details.}
\item{The coefficient of $\delta_{\alpha_{1}}^{\rho_1} \cdots \delta_{\alpha_{k}}^{\rho_k}$ is set to one and the consistency condition (\ref{generalConsistencyCondition}) is applied to $\HPT{\alpha}{\rho}{k}{n}$; this step requires the knowledge of $\HPT{\alpha}{\rho}{k-1}{n}$. Now the projector $\HPT{\alpha}{\rho}{k}{n}$ is fully determined and unique.}
\end{enumerate} 
The construction of $\HPT{\alpha}{\rho}{k}{n}$ on the complex light~cone $\mathbb{K}^{2h}\kln{\C}$ is now finished and the relations (\ref{operatorConditionDeltaSecondVersion}) to (\ref{operatorConditionDivergenceXQ}) are fulfilled for this projector. However, if we take $\HPT{\alpha}{\rho}{k}{n}$ off the complex~cone, the relations (\ref{operatorConditionLaplace}) and (\ref{operatorConditionDivergence}) are not valid. We have to construct the harmonic extension of this projector.

\subsubsection*{Harmonic extension}

According to Reference \cite{BT77} each homogeneous polynomial $f_n \! \kln{\tx} \in K^{2h}_n$ of degree $n$ on the complex light~cone $\mathbb{K}^{2h}\kln{\C}$ has a unique harmonic extension off the cone given by $f^H_n \! \kln{x} := H_n \, f_n \! \kln{x}$ with $\Delta f^H_n\kln{x} = 0$ and $H_n$ defined as
\begin{equation}
H_n = \sum_{k=0}^{\kle{\frac{n}{2}}} \frac{(-1)^k \, (n+h-2-k)!}{4^k \, k! \, (n+h-2)!} \, \kln{x^2}^k \Delta^k\; .
\end{equation}
The projector $H_n$ therefore establishes an isomorphism between the space $K^{2h}_n$ of homogeneous polynomials on the complex~cone and the space of their harmonic extensions off the cone.

Accordingly, the off--cone projector $\HP{\alpha}{\rho}{k}{n}$ and the on--cone projector $\HPT{\alpha}{\rho}{k}{n}$ are in a one to one correspondence and $\HP{\alpha}{\rho}{k}{n}$ is the unique harmonic extension of $\HPT{\alpha}{\rho}{k}{n}$ given by
\begin{equation}
\label{HarmExtensionH}
\HP{\alpha}{\rho}{k}{n} = H_n \, \HPT{\alpha}{\rho}{k}{n}  \;  .
\end{equation}
Furthermore, we can conclude that the interior derivatives on the complex~cone denoted by $\tilde Q$ and characterized by (\ref{ConditionOnCone}) must correspond to another set of differential operators which preserve the harmonicity off the complex~cone. These operators are denoted by $Q$ and must fulfill the relation
\begin{equation}
\label{harmonicityOffCone}
\Delta \; Q \; f^H_n\kln{x} = 0 \; ,
\end{equation}
where $f^H_n\kln{x}$ is a harmonic polynomial fulfilling the Laplace equation $\Delta \, f^H_n\kln{x} = 0$. To find a suitable basis for the operators $Q$, we simply calculate the harmonic extension of the operators $\Kls{\tx_\alpha,\tilde\D_\alpha,\tilde\X,\tilde\X_{\kle{\alpha\beta}},\delta_{\alpha\beta}}$ and find
\begin{align}
\label{harmonixExtX}
H_{n+1} \, x_\alpha \, \X
&=
\x_\alpha \, H_n
\\
\label{harmonixExtD}
H_{n-1} \, \D_\alpha
&=
\X \, \pd_\alpha \, H_n
\\
H_n \, \X
&=
\X \, H_n
\\
H_n \, \X_{\kle{\alpha\beta}}
&=
\X_{\kle{\alpha\beta}} \, H_n
\intertext{with $\x_\alpha$ defined as}
\label{DefinitionX}
\x_\alpha
&=
x_\alpha \, \X - \frac{1}{2} \, x^2 \, \pd_\alpha  \; .
\end{align}
$\delta_{\alpha\beta}$ off course also commutes with $H_n$. The operators $\kls{\x_\alpha,\pd_\alpha,\X,\X_{\kle{\alpha\beta}},\delta_{\alpha\beta}}$ preserve the harmonicity off the complex~cone, i.e. they fulfill the relation~(\ref{harmonicityOffCone}), since the only non~vanishing commutators with $\Delta$ are proportional to the Laplacian
\begin{eqnarray}
\label{HEKomx}
\Kle{\Delta,\x_\alpha}
&=&
2 \, x_\alpha \, \Delta
\\
\label{HEKomXX}
\Kle{\Delta,\X}
&=&
2 \, \Delta \; .
\end{eqnarray}
The two latter relations~(\ref{HEKomx}) and~(\ref{HEKomXX}) are the off--cone version of the communtators~(\ref{KomD}) and~(\ref{KomX}).

Since $H_n$ establishes an isomorphism between the spaces of homogeneous polynomials on the complex~cone and their harmonic extensions, the on--cone representation of $\mathfrak{so}\kln{2,2h}$ spanned by $\Kls{\tx_\alpha,\tilde\D_\alpha,\tilde\X,\tilde\X_{\kle{\alpha\beta}},\delta_{\alpha\beta}}$ is mapped to an equivalent off--cone representation. The operators $\kls{\x_\alpha,\pd_\alpha,\X,\X_{\kle{\alpha\beta}},\delta_{\alpha\beta}}$ therefore fulfill the commutation relations
\begin{eqnarray}
\Kle{\pd_\alpha,\x_\beta}
&=&
\delta_{\alpha\beta} \; \X + \X_{\kle{\alpha\beta}}
\\
\Kle{\X_{\kle{\alpha\beta}},\x_\mu}
&=&
\delta_{\alpha\mu} \, \x_{\beta} - \delta_{\beta\mu} \, \x_\alpha
\\
\Kle{\X_{\kle{\alpha\beta}},\pd_\mu}
&=&
\delta_{\alpha\mu} \, \pd_{\beta} - \delta_{\beta\mu} \, \pd_\alpha
\\
\Kle{\pd_\alpha,\X}
&=&
\pd_\alpha
\\
\Kle{\x_\alpha,\X}
&=&
- \x_\alpha \; .
\end{eqnarray}
This second off--cone representation will be important for the determination of the complete trace decomposition in Subsection~\ref{CompleteTraceDecomposition}.

Relation (\ref{harmonixExtD}) also gives us the form of the conditions of tracelessness (\ref{operatorConditionLaplace}) and (\ref{operatorConditionDivergence}) on the complex cone. The condition of harmonicity $\Delta \, \HP{\alpha}{\alpha'}{k}{n} = 0 $ vanishes on the cone since the square of the interior derivative $\D_\alpha$ is proportional to $x^2$
\begin{equation}
\D^2 = \frac{1}{4} \; x^2 \, \Delta^2  \;  .
\end{equation}
The condition $\pd^{\alpha_i} \, \HP{\alpha}{\alpha'}{k}{n} = 0 $ on the other hand is turned into
\begin{equation}
\tilde \D^{\alpha_i} \, \HPT{\alpha}{\alpha'}{k}{n} = 0   \qquad  \fo  \quad 1 \les i \les k 
\end{equation}
on the complex cone. Here, the interior derivative $\tilde \D^{\alpha_i}$ must be used to formulate the condition of tracelessness.  

\subsubsection*{Results for $\HP{\alpha}{\rho}{k}{n}$}

We will now list results for the projectors $\HP{\alpha}{\rho}{k}{n}$ for $k=1$ and $k=2$ for all Young symmetries. These results have been calculated in Ref.~\cite{E04} by a \texttt{Java} application which implements all rules explained in the above Sections.

\subsubsection*{Vector}

We start with the vector case where the result for $H_{\alpha_1|n}^{\rho_1}$ reads
\begin{eqnarray}
\label{HVector}
H_{\alpha_1|n}^{\rho_1}
&=&
H_n \kls{ \delta_{\alpha_1}^{\rho_1} - \frac{1}{(n+h-1)(n+2h-3)} \cdot x_{\alpha_1} \, \D^{\rho_1} } \; .
\end{eqnarray}
This result has first been given in Ref.~\cite{GLR99} for $h=2$ and for general space--time dimension $d=2h$ in Ref.~\cite{GLR00}.

\subsubsection*{Antisymmetric second rank tensor}

The result in the antisymmetric second rank tensor case reads
\begin{eqnarray}
\label{HAntisymTensor}
H_{[\alpha_1\alpha_2]|n}^{[\rho_1\rho_2]}
&=&
 H_n \KLSSo{} \delta_{[\alpha_1}^{[\rho_1}\delta_{\alpha_2]}^{\rho_2]}
+ \frac{2}{(n+h-1)(n+2h-4)} \KLNo{} x_{[\alpha_1}^{\phantom{|}} \delta_{\alpha_2]}^{[\rho_1} \D^{\rho_2]}_{\phantom{|}}
\\
\nonumber
&&
\qquad \qquad \qquad \!\!
- \frac{1}{(n+h-2)^2(n+2h-2)} \cdot x_{[\alpha_1}\D_{\alpha_2]} \; x^{[\rho_1}\D^{\rho_2]} \oKLN{} \oKLSS{}
\\
\nonumber
&=&
 H_n \KLSSo{} \delta_{[\alpha_1}^{[\rho_1}\delta_{\alpha_2]}^{\rho_2]}
+ \frac{2}{(n+h-1)(n+2h-4)} \KLNo{} x_{[\alpha_1}^{\phantom{|}} \delta_{\alpha_2]}^{[\rho_1} \D^{\rho_2]}_{\phantom{|}}
\\
\nonumber
&&
\qquad \qquad \qquad \!\!
- \frac{1 }{4(n+2h-2) } \cdot \X_{ [\alpha_1\alpha_2] } \; \X^{ [\rho_1\rho_2] } \oKLN{} \oKLSS{} \;  .
\end{eqnarray}
The second form has been obtained using relation (\ref{XtoxDRel}). The result (\ref{HAntisymTensor}) for $H_{[\alpha_1\alpha_2]|n}^{[\rho_1\rho_2]}$ has also first been given in Ref.~\cite{GLR99} for $h=2$ and for general $h$ in Ref.~\cite{GLR00} in an equivalent form.

\subsubsection*{Symmetric second rank tensor}

In the symmetric second rank tensor case we find
\begin{align}
\label{HNUll}
H_{\alpha_1\alpha_2|0}^{\rho_1\rho_2 }
&=
\delta_{\alpha_1}^{\rho_1}\delta_{\alpha_2}^{\rho_2}
-\frac{1}{2h}\cdot\delta_{\alpha_1\alpha_2}\delta^{\rho_1\rho_2}
\\
\nonumber
H_{(\alpha_1\alpha_2)|n}^{(\rho_1\rho_2)}
&=
H_n \; H_{\alpha_1\alpha_2|0}^{\beta_1\beta_2} \; \tilde H_{(\beta_1\beta_2)|n}^{'(\sigma_1\sigma_2)} \;
H_{\sigma_1\sigma_2|0}^{\rho_1\rho_2}
\\
\label{ResultSymTensor}
&=
H_n \KLSSo{} \tilde H_{(\alpha_1\alpha_2)|n}^{'(\rho_1\rho_2)}
- \frac{1}{2(h-1)} \cdot \delta_{\alpha_1\alpha_2}\delta^{\rho_1\rho_2}
\\
\nonumber
&
\qquad \qquad \qquad
+ \frac{1}{(h-1)(n+h)(n+2h-2)} \KLNo{} x_{(\alpha_1} \D_{\alpha_2)}\, \delta^{\rho_1\rho_2}
+\delta_{\alpha_1\alpha_2} \, \D^{(\rho_1} x^{\rho_2)} \oKLN{}
 \oKLSS{}
\intertext{with $\tilde H_{(\alpha_1\alpha_2)|n}^{'(\rho_1\rho_2)}$ given by}
\tilde H_{(\alpha_1\alpha_2)|n}^{'(\rho_1\rho_2)}
&=
\delta_{(\alpha_1}^{(\rho_1}\delta_{\alpha_2)}^{\rho_2)}
-\frac{1}{(n+h-1)(n+2h-2)}
\KLNNo{} 2 \cdot x_{(\alpha_1}^{\phantom{|}}\delta_{\alpha_2)}^{(\rho_1}\D^{\rho_2)}_{\phantom{|}}
\\
\nonumber
&
+ \frac{1}{n+h-2} \KLEo{}
\frac{2}{(h-1)(n+h)} \cdot x_{(\alpha_1}\D_{\alpha_2)} \; x^{(\rho_1}\D^{\rho_2)}
- \frac{1}{n+2h-3} \cdot x_{\alpha_1}x_{\alpha_2} \, \D^{\rho_1}\D^{\rho_2} \oKLE{} \oKLNN{} \; .
\end{align}
This result for $H_{(\alpha_1\alpha_2)|n}^{(\rho_1\rho_2)}$ has also first been given in an equivalent form in Ref.~\cite{GLR00} for general $h$.

The reader can check that the above results fulfill the relations~(\ref{operatorConditionDivergence}) and~(\ref{operatorConditionDivergenceXQ}) by using the following two contractions
\begin{equation}
\D^{\alpha} \, x_{\alpha} = \pd^{\alpha} \, \x_{\alpha} = \kln{\X+h-1} \kln{\X+1} - \frac{1}{2} \; x^2 \Delta \; .
\end{equation}
All the above results for the projectors $H_{\alpha_1|n}^{\rho_1}$, $H_{[\alpha_1\alpha_2]|n}^{[\rho_1\rho_2]}$ and $H_{(\alpha_1\alpha_2)|n}^{(\rho_1\rho_2)}$ can be easily restricted onto the complex~cone by taking $\tx_{\alpha_i}$ and $\tilde \D_{\alpha_i}$ on the cone and observing that $\tilde H_n = \Id$.

\subsection{The construction of the complete trace decomposition}
\label{CompleteTraceDecomposition}

In this Section we determine the complete trace decomposition of local operators $\OP_{\alpha_1\dots\alpha_k|n}$ which is denoted here by $\TRD_{\alpha_1\dots\alpha_k|n}^{\rho_1\dots\rho_k}$; it's on--cone limit will be denoted by $\widetilde \TRD_{\alpha_1\dots\alpha_k|n}^{\rho_1\dots\rho_k}$. These decompositions determine all traceless parts contained in the local operator under consideration and separate them from the connected pure trace parts. In Ref.~\cite{EGL04} we have already given an extensive discussions for the deduction of the complete trace decomposition in the vector case. There, we made an ansatz for $\TRD_{\alpha_1|n}^{\rho_1}$ and then used a complicated method based on coupled equations containing the unknown coefficients. These equations had to be solved iteratively and the proof of correctness for the determined coefficients was done by induction. The same method has also successfully been applied to the (antisymmetric and symmetric) second rank tensor case.

In both cases the operator structure of the complete trace decomposition can also be deduced directly from the result for $\HP{\alpha}{\rho}{k}{n}$ by performing the replacement
\begin{equation}
\label{ReplacementRule}
\delta_{\alpha_1}^{\rho_1} \cdots \delta_{\alpha_l}^{\rho_l} H_n \ra \HP{\alpha}{\rho}{l}{n} \qquad \text{with} \qquad 1 \les l \les k 
\end{equation}
in each term of $\HP{\alpha}{\rho}{k}{n}$. After this replacement we sandwich each term with the operators $\kln{x^2}^{j-t}$ and $\Delta^{j-t}$ where $t$ counts the number of interior derivatives $\D^{\rho_i}$ and Kronecker deltas $\delta^{\rho_i\rho_j}$. $j$ thereby picks up the number of traces taken of the local operator $\OP_{\alpha_1\dots\alpha_k|n}$.

Each term which is generated by this rule is attached with an undetermined coefficient $f_i\kln{n,h,j}$ being proportional to
\begin{equation}
c\kln{n,h,j} = \frac{(n+h-1-2j)!}{4^j \, j! \, (n+h-1-j)!} \; .
\end{equation}
These unknown coefficients $f_i\kln{n,h,j}$ being contained in the ansatz for $\TRD_{\alpha_1\dots\alpha_k|n}^{\rho_1\dots\rho_k}$ can then simply be determined by the relation
\begin{equation}
\label{ConConDTr}
x^{\alpha_l} \; \TRD_{\alpha_1\dots\alpha_k|n}^{\rho_1\dots\rho_k}
= \TRD_{\alpha_1\dots\alpha_{l-1}\alpha_{l+1}\dots\alpha_k|n+1}^{\rho_1\dots\rho_{l-1}\rho_{l+1}\dots\rho_k} \; x^{\rho_l}
\qquad
\text{for}
\qquad
1 \les l \les k \; ,
\end{equation}
which follows directly from the fact that the complete trace decomposition must a decomposition of the identity in the respective tensor space, i.e. it must hold
\begin{equation}
\label{SimpleIdentity}
\TRD_{\alpha_1\dots\alpha_k|n}^{\rho_1\dots\rho_k}
\overset{!}{=}
\prod_{i=1}^k \delta_{\alpha_i}^{\rho_i} \; .
\end{equation}
A discussion of the replacement rule~(\ref{ReplacementRule}) can be found in Ref.~\cite{E04} Section~6.5. Here, we use this rule as a simple device to deduce the complete trace decompositions for local operators of the respective rank. For the vector and second rank tensor case the following results have also been proven by induction.

\subsubsection*{Results for $\TRD_{\alpha_1\dots\alpha_k|n}^{\rho_1\dots\rho_k}$}

In this Subsection we will give the result for the complete trace decompositions $\TRD_{\alpha_1\dots\alpha_k|n}^{\rho_1\dots\rho_k}$ for $k=1$ and also for $k=2$ on and also off the complex~cone. The following results have been calculated by the \texttt{Java} application developed in Ref.~\cite{E04} implementing the replacement rule (\ref{ReplacementRule}).

\subsubsection*{Vector}

We start with the vector case where the result for $\TRD_{\alpha_1|n}^{\rho_1}$ is 
\begin{eqnarray}
\label{TRDVector}
\TRD_{\alpha_1|n}^{\rho_1}
&=&
\sum_{j=0}^{\kle{\frac{n+1}{2}}}\frac{\kln{n+h-1-2j}!}{4^j j! \kln{n+h-1-j}!} \KLSSo{}
\kln{x^2}^{j}  \, H_{\alpha_1|n-2j}^{\rho_1} \, \Delta^{j} 
\\
\nonumber
&&
\qquad \qquad
+ \frac{4j}{(n+h-j)(n+2h-1-2j)} \cdot \kln{x^2}^{j-1} \x_{\alpha_1} \; H_{n+1-2j} \; \D^{\rho_1} \, \Delta^{j-1} \oKLSS{} \; .
\end{eqnarray}
Strictly speaking, the separation of traceless parts form pure traces proportional to $x_{\alpha_1}$ and/or~$x^2$ is not complete in the above result. According to relation~(\ref{DefinitionX}), the operator $\x_{\alpha_1}$ contains a contribution which is proportional to $x^2 \, \pd_{\alpha_1}$. The condition of tracelessness is unaltered by the action of this derivative. This means that we have to decompose the operator $\x_{\alpha_1}$ according to (\ref{DefinitionX}) into its parts to obtain a complete separation of the traceless contributions from the trace terms. However, in the following we will stick to the form~(\ref{TRDVector}) for the complete trace decomposition since the results are more compact in this form. We will return to this topic in Section~\ref{spindecompositions} where we discuss the complete decompositions into irreducible components. 

Taking the on--cone limit of $\TRD_{\alpha_1|n}^{\rho_1}$ given by~(\ref{TRDVector}) we get
\begin{eqnarray}
\widetilde \TRD_{\alpha_1|n}^{\rho_1} =
\tilde H_{\alpha_1|n}^{\rho_1} + \frac{1}{(n+h-1)(n+2h-3)} \cdot \tx_{\alpha_1} \, \tilde \D^{\rho_1} \; .
\end{eqnarray}
A quick comparison with $\tilde H_{\alpha_1|n}^{\rho_1}$ given by the on--cone limit of (\ref{HVector}) shows that $\widetilde \TRD_{\alpha_1|n}^{\rho_1}$ is a decomposition of $\delta_{\alpha_1}^{\rho_1}$. In contrast to the off--cone decomposition~(\ref{TRDVector}) the separation of traces is always complete in the on--cone limit. This holds for all following on--cone trace decompositions. The above results for $\TRD_{\alpha_1|n}^{\rho_1}$ and $\widetilde \TRD_{\alpha_1|n}^{\rho_1}$ have first been given in Reference~\cite{EGL04} but the following trace decompositions are all new results.

\subsubsection*{Antisymmetric second rank tensor}

In the antisymmetric second rank tensor case we find the result
\begin{eqnarray}
\label{TRDASTensor}
\TRD_{[\alpha_1\alpha_2]|n}^{[\rho_1\rho_2]}
&=&
\sum_{j=0}^{\kle{\frac{n+1}{2}}}\frac{\kln{n+h-1-2j}!}{4^j j! \kln{n+h-1-j}!}
\KLSSo{} \kln{x^2}^{j} H_{[\alpha_1\alpha_2]|n-2j}^{[\rho_1\rho_2]} \; \Delta^{j}
\\
\nonumber
&&
\qquad 
- \frac{8j}{(n+h-j)(n+2h-2-2j)} \cdot \kln{x^2}^{j-1  } \KLNo{} \x_{[\alpha_1}^{\phantom{|}}H_{\alpha_2]|n+1-2j}^{[\rho_1} \; \D^{\rho_2]}_{\phantom{|}}
\\
\nonumber
&&
\qquad \qquad \quad
- \frac{1}{(n+h-2j)(n+2h-2j)} \cdot \x_{[\alpha_1}\partial_{\alpha_2]} \, H_{n+2-2j} \; x^{[\rho_1}\D^{\rho_2]}
\oKLN{} \Delta^{j-1} \oKLSS{} \; .
\end{eqnarray}
On the complex~cone we get
\begin{eqnarray}
\widetilde \TRD_{[\alpha_1\alpha_2]|n}^{[\rho_1\rho_2]}
&=&
\tilde H_{[\alpha_1\alpha_2]|n}^{[\rho_1\rho_2]}
- \frac{2}{(n+h-1)(n+2h-4)} \KLNo{} \tx_{[\alpha_1}^{\phantom{|}} \tilde H_{\alpha_2]|n-1}^{[\rho_1} \; \tilde\D^{\rho_2]}_{\phantom{|}}
\\
\nonumber
&&
\qquad \qquad \qquad \qquad \qquad \qquad
- \frac{1}{(n+h-2)^2(n+2h-2)} \cdot \tx_{[\alpha_1}\tilde\D_{\alpha_2]} \, \tx^{[\rho_1}\tilde\D^{\rho_2]}
\oKLN{}
\\
\nonumber
&=&
\tilde H_{[\alpha_1\alpha_2]|n}^{[\rho_1\rho_2]}
- \frac{2}{(n+h-1)(n+2h-4)} \KLNo{} \tx_{[\alpha_1}^{\phantom{|}} \tilde H_{\alpha_2]|n-1}^{[\rho_1} \; \tilde\D^{\rho_2]}_{\phantom{|}}
\\
\nonumber
&&
\qquad \qquad \qquad \qquad \qquad \qquad
- \frac{1}{4(n+2h-2)} \cdot \tilde \X_{[\alpha_1\alpha_2]} \, \tilde \X^{[\rho_1\rho_2]}
\oKLN{} \; .
\end{eqnarray}
Again, a quick comparison with $\tilde H_{\alpha_1|n}^{\rho_1}$ and $\tilde H_{[\alpha_1\alpha_2]|n}^{[\rho_1\rho_2]}$ given by the on--cone limits of (\ref{HVector}) and (\ref{HAntisymTensor}) shows that $\widetilde \TRD_{[\alpha_1\alpha_2]|n}^{[\rho_1\rho_2]}$ decomposes $\delta_{[\alpha_1}^{[\rho_1}\delta_{\alpha_2]}^{\rho_2]}$.

\subsubsection*{Symmetric second rank tensor}

In the symmetric second rank tensor case the complete trace decomposition $\TRD_{(\alpha_1\alpha_2)|n}^{(\rho_1\rho_2)}$ has the form 
\begin{eqnarray}
\label{TRDSTensor}
&&
\TRD_{(\alpha_1\alpha_2)|n}^{(\rho_1\rho_2)}
\\
\nonumber
&&
= \sum_{j=0}^{\kle{\frac{n+2}{2}}} \, \frac{\kln{n+h-1-2j}!}{4^j \, j! \,\kln{n+h-1-j}!}
\KLSSo{}
\kln{x^2}^j \, H_{(\alpha_1\alpha_2)|n-2j}^{(\rho_1\rho_2)} \, \Delta^j
\\
\nonumber
&&
\qquad 
+ \frac{8j}{\kln{n+h-j}\kln{n+2h-2j}} \cdot
\kln{x^2}^{j-1} \, \x_{(\alpha_1}^{\phantom{|}} H_{\alpha_2)|n+1-2j}^{(\rho_1} \, \D^{\rho_2)}_{\phantom{|}} \; \Delta^{j-1}
\\
\nonumber
&&
\qquad 
+ \frac{4j}{\kln{h-1}\kln{n+h-j}\kln{n+h+2-2j}\kln{n+2h-2j}} \KLNNo{}
\\
\nonumber
&&
\qquad \qquad \quad
2 \cdot \kln{x^2}^{j-1} \x_{(\alpha_1  } \partial_{\alpha_2)} \; H_{n+2-2j} \; x^{(\rho_1} \D^{\rho_2)} \; \Delta^{j-1}
\\
\nonumber
&&
\qquad \qquad
- \kln{n+h-2j} \kln{n+h+1-2j} \cdot \kln{x^2}^{j-1} \KLEo{}
\\
\nonumber
&&
\qquad \qquad \qquad
\x_{(\alpha_1} \partial_{\alpha_2)} \; H_{n+2-2j} \; \delta^{\rho_1\rho_2} 
+ \delta_{\alpha_1\alpha_2} \; H_{n+2-2j} \; \D^{(\rho_1} x^{\rho_2)}
\oKLE{}  \Delta^{j-1}
\\
\nonumber
&&
\qquad \qquad
+ \, \frac{ 4\kln{h-1}\kln{j-1}\kln{n+h-2j} }{ \kln{n+h+1-j}\kln{n+2h+1-2j} } \cdot
\kln{x^2}^{j-2} \; \x_{\alpha_1} \x_{\alpha_2} \; H_{n+2-2j} \; \D^{\rho_1} \D^{\rho_2} \, \Delta^{j-2} \oKLNN{}
\\
\nonumber
&&
\qquad
+ \frac{2j \kln{n+h-2j}\kln{n+h+1-2j}}{\kln{h-1}\kln{n+h-j}} \cdot \kln{x^2}^{j-1} \delta_{\alpha_1\alpha_2} \, H_{n+2-2j} \, \delta^{\rho_1\rho_2}
\Delta^{j-1} \oKLSS{} \; .
\end{eqnarray}
In the on--cone limit this reduces to
\begin{eqnarray}
\label{TRDsecondRankOnCone}
\widetilde \TRD_{(\alpha_1\alpha_2)|n}^{(\rho_1\rho_2)}
&=&
\tilde H_{(\alpha_1\alpha_2)|n}^{(\rho_1\rho_2)} + \frac{1}{2(h-1)} \cdot \delta_{\alpha_1\alpha_2} \, \delta^{\rho_1\rho_2}
\\
\nonumber
&&
+ \frac{2}{\kln{n+h-1}\kln{n+2h-2}} \cdot
\tx_{(\alpha_1}^{\phantom{|}} \tilde H_{\alpha_2)|n-1}^{(\rho_1} \, \tilde \D^{\rho_2)}_{\phantom{|}}
\\
\nonumber
&&
+ \frac{1}{\kln{n+h-1}\kln{n+h-2}\kln{n+2h-3}\kln{n+2h-4}}
\cdot \tx_{\alpha_1} \tx_{\alpha_2} \; \tilde \D^{\rho_1} \tilde \D^{\rho_2}
\\
\nonumber
&&
+ \frac{2}{\kln{h-1}\kln{n+h}\kln{n+h-1}\kln{n+h-2}\kln{n+2h-2}}
 \cdot \tx_{(\alpha_1} \tilde \D_{\alpha_2)} \; \tx^{(\rho_1} \tilde \D^{\rho_2)}
\\
\nonumber
&&
- \frac{1}{\kln{h-1}\kln{n+h}\kln{n+2h-2}}
\KLEo{}
  \tx_{(\alpha_1} \tilde \D_{\alpha_2)} \; \delta^{\rho_1\rho_2}
+ \delta_{\alpha_1\alpha_2} \; \tilde \D^{(\rho_1} \tx^{\rho_2)} \oKLE{} \; .
\end{eqnarray}
Taking the on--cone limit of $\TRD_{(\alpha_1\alpha_2)|n}^{(\rho_1\rho_2)}$ one has to pay attention to additional factors coming from the relations~(\ref{harmonixExtX}) and~(\ref{harmonixExtD}). The operators $\x_\alpha$ and $\pd_\alpha$ must be shifted through the harmonic extension $H_n$ contained in $H_{(\alpha_1\alpha_2)|n}^{(\rho_1\rho_2)}$ and $H_{\alpha_1|n}^{\rho_1}$; see relation~(\ref{HarmExtensionH}).

It is a good exercise to check that $\widetilde \TRD_{(\alpha_1\alpha_2)|n}^{(\rho_1\rho_2)}$ given by (\ref{TRDsecondRankOnCone}) really is a decomposition of $\delta_{(\alpha_1}^{(\rho_1}\delta_{\alpha_2)}^{\rho_2)}$. On has to use $\tilde H_{\alpha_1|n}^{\rho_1}$ and $\tilde H_{(\alpha_1\alpha_2)|n}^{(\rho_1\rho_2)}$ given by the on--cone limits of (\ref{HVector}) and (\ref{ResultSymTensor}).

\newpage

\section{Symmetry Part}
\label{SymmetryPart}

In Section~\ref{TracePart} we have given methods for the construction of a unique projector $\HP{\alpha}{\rho}{k}{n}$ subtracting all traces of the local operator $\OP_{\rho_1\dots\rho_k|n}$ and for the related complete trace decomposition $\TRD_{\alpha_1\dots\alpha_k|n}^{\rho_1\dots\rho_k}$. But since our local operators shall be decomposed into irreducible $SO(2h,\C)$--tensors we have only solved half of the problem up to now. To make a traceless tensor $\OPT_{\alpha_1\dots\alpha_k\kln{\zeta_1\dots\zeta_n}}$ irreducible under this group it must carry a definite Young symmetry in all $m$ indices $\kls{\xi} = \kls{\zeta} \cup \kls{\alpha}$ which determines the behavior of this tensor under index permutations. This requirement of irreducibility for the group $SO(2h,\C)$ is inherited from the $GL(2h,\C)$.

The irreducible representations of the general linear group $GL(2h,\C)$ are uniquely determined by the idempotent Young operators
\begin{equation}
\label{YoungOperators}
\Y_{[\m]} = \frac{f_{(\m)}}{m!} \; \QU_{[\m]} \PRa_{[\m]} \qquad \text{with} \qquad \PRa_{[\m]} = \sum_{p\in H_{[\m]}} p \qquad
\text{and} \qquad \QU_{[\m]} = \sum_{q\in V_{[\m]}} \delta_q \, q \; ,
\end{equation}
where $\delta_q$ is the parity of the permutation $q$. The Young operators are related to corresponding Young tableaux being denoted by $[\m]$. These Young tableaux are obtained from corresponding Young patterns $(\m)$ defined by
\begin{equation}
\label{pattern}
(\m) = \kln{\m_1,\dots,\m_r} \qquad \text{with} \qquad \m_1 \ges \m_2 \ges \dots \ges \m_r \qquad \text{and} \qquad \sum_{i = 1}^r \m_i = m
\end{equation}
by inserting, without repetition, the indices $\xi_1,\dots,\xi_k$. $H_{[\m]}$ and $V_{[\m]}$ in (\ref{YoungOperators}) denote the horizontal and vertical permutations defined by $[\m]$. The coefficient $f_{(\m)}$ in (\ref{YoungOperators}) is given by
\begin{equation}
\label{NormalizationF}
f_{(\m)} = k! \; \frac{\prod_{i<j}\kln{l_i-l_j}}{\prod_{i=1}^{r} l_i !} \qquad \text{with} \qquad l_i = \m_i + r - i \; ,
\qquad \sum_{(m)} f^2_{(\m)} = m!
\end{equation}
and depends only on the given pattern $(\m)$ but not on the related Young tableaux $[\m]$. It gives the number of different, but equivalent, irreducible representations belonging to the same Young pattern $(\m)$ and determines the dimension of these representations. The number of irreducible representations is also given by the number of standard Young tableaux which are obtained by inserting the indices $\xi_1,\dots,\xi_m$ lexicographically into a given Young pattern. This means that the indices are increasing from left to right and from top to bottom.

The construction of irreducible representations of the group $GL(2h,\C)$ and their interrelation with the irreducible representation of the symmetric group $S_k$ is explained in many standard text books. Here, we refer to~\cite{BR77,H62,B55}.

In the following we will denote specific Young tableaux by $[\m]_j$ where $\m$ will denote the different patterns $(\m)$ and $j$ the different tableaux belonging to this pattern. Unfortunately, the form (\ref{YoungOperators}) of the Young projectors $\Y_{[\m]_j}$ does not fully supply us with what we need. For $k \ges 5$ the Young projectors $\Y_{[\m]_j}$ are neither orthogonal nor are they a decomposition of the identity in the respective tensor space. To obtain the desired orthogonal decomposition in terms of symmetry projectors one has to follow the procedure given in Ref.~\cite{B55} Chapter~IV.5 and in Ref.~\cite{E04}. There, an additional operator $q_{[\m]_j}$ is constructed which orthogonalizes the projectors $\Y_{[\m]_j}$ given by~(\ref{YoungOperators}). These new Young operators are defined as
\begin{equation}
\label{OrthoPerformed}
\YH_{[\m]_j} := q_{[\m]_j} \, \Y_{[\m]_j} \; .
\end{equation}
The projectors $\YH_{[\m]_j}$ now fulfill the desired orthogonality condition
\begin{equation}
\label{orthogonality}
\YH_{\kle{\m}_{j}}\YH_{\kle{\m'}_{j'}} = \delta_{\kle{\m}_{j}\kle{\m'}_{j'}} \; \YH_{\kle{\m}_{j}} \qquad \text{and} \qquad \sum_{\kle{\m}_{j}} \, \YH_{\kle{\m}_{j}} = \Id \; .
\end{equation}
Since all projectors $\YH_{[\m]}$ are orthogonal, we can also construct an orthogonal decomposition  on the level of Young patterns $\kln{\m}$. We just have to sum all standard Young projectors $\YH_{[\m]_j}$ belonging to the same pattern
\begin{equation}
\label{PatternProjector}
\YH_{(\m)} := \sum_{j=1}^{f_{(\m)}} \, \YH_{[\m]_j} \; .
\end{equation}
The projectors $\YH_{(\m)}$ then also fulfill an orthogonality relation and decompose the identity
\begin{equation}
\label{othnonormalDec}
\YH_{(\m)}  \YH_{(\m')} = \delta_{(\m)(\m')} \; \YH_{(\m)} \qquad \text{and} \qquad \sum_{\kln{\m}} \, \YH_{\kln{\m}} = \Id \; .
\end{equation}

\subsection{The construction of Extended Young projectors}
\label{ExtendedYoungprojectors}

To obtain the desired irreducible parts of $\OP_{\alpha_1\dots\alpha_k|n}$ we must now construct extended Young projectors operating on $\alpha_1$ to $\alpha_k$ and $\zeta_1$ to $\zeta_n$. For a fixed integer $n$ this can, in principle, be done in a straightforward manner just by building all standard Young projectors $\YH_{\xi_1\dots\xi_{m}|\kle{\m}_j}^{\xi_1'\dots\xi_{m}'}$ where $\kls{\xi}$ is the joint set of the indices given by~(\ref{JoinSets}). The problem is that $n$ is not restricted and that the standard Young projectors $\YH_{\xi_1\dots\xi_{m}|\kle{\m}_j}^{\xi_1'\dots\xi_{m}'}$, in general, cannot be constructed in closed form for $m=n+k$ indices if $n$ is not a fixed integer number.

Fortunately, we do not need to know these full projectors $\YH_{\xi_1\dots\xi_{m}|\kle{\m}_j}^{\xi_1'\dots\xi_{m}'}$ since they are to be contracted with the symmetric tensor $\XS{\zeta}{n}$. It is therefore sufficient to determine the contracted Young projectors
\begin{equation}
\XS{\zeta}{n} \; \YH_{\xi_1\dots\xi_{m}|\kle{\m}_j}^{\xi_1'\dots\xi_{m}'}  \; .
\end{equation}
Figure~\ref{generalYoungTableau} shows such a general Young tableau $\kle{\m}_j$ related to $\YH_{\xi_1\dots\xi_{m}|\kle{\m}_j}^{\xi_1'\dots\xi_{m}'}$.
\begin{figure}[ht]
\unitlength1.0cm
\begin{center}
\begin{picture}(13,6)
\linethickness{0.075mm}
\put(0,0){\framebox(1,1){$\alpha^{1}_{k^{1}}$}}
\put(0,1.0075){\framebox(1,3){$\vdots$}}
\put(0,4.015){\framebox(1,1){$\alpha^{1}_{1}$}}
\put(1.0075,1){\framebox(1,1){$\alpha^{2}_{k^{2}}$}}
\put(1.0075,2.0075){\framebox(1,2){$\vdots$}}
\put(1.0075,4.015){\framebox(1,1){$\alpha^{2}_{1}$}}
\put(3.0225,2){\framebox(1,1){$\alpha^{\mathfrak{a}}_{k^{{\mathfrak{a}}}}$}}
\put(3.0225,3.0075){\framebox(1,1){$\vdots$}}
\put(3.0225,4.015){\framebox(1,1){$\alpha^{\mathfrak{a}}_{1}$}}
\put(0,5.0225){\framebox(1,1){$\zeta_1$}}
\put(1.0075,5.0225){\framebox(1,1){$\zeta_2$}}
\put(2.015,5.0225){\framebox(1,1){$\dots$}}
\put(3.0225,5.0225){\framebox(1,1){$\zeta_\mathfrak{a}$}}
\put(4.03,5.0225){\framebox(1,1){$\dots$}}
\put(5.0375,5.0225){\framebox(1,1){$\zeta_n$}}
\put(6.045,5.0225){\framebox(1,1){$\alpha^0_1$}}
\put(7.0525,5.0225){\framebox(1,1){$\dots$}}
\put(8.06,5.0225){\framebox(1,1){$\alpha^0_{k^0}$}}
\linethickness{0.0mm}
\put(2.015,3.0075){\framebox(1,1){$\dots$}}
\put(4.03,0){\framebox(1,1){$\m_r$}}
\put(4.03,1.015){\framebox(1,1){$\vdots$}}
\put(4.03,2.015){\framebox(1,1){$\vdots$}}
\put(4.03,3.0225){\framebox(1,1){$\vdots$}}
\put(4.03,4.03){\framebox(1,1){$\m_2$}}
\put(9.4,5){\framebox(2,1){$\m_1 = n + k^0$}}
\put(7,3){$\text{with} \quad k^{1} \ges k^{2} \ges \dots \ges k^{\mathfrak{a}} \ges 1$}
\put(7,2){$\text{and}  \quad k = k^0 + \sum\limits_{i=1}^{\mathfrak{a}} \, k^{i}$}
\put(7,1){$\text{and}  \quad n+k = \sum\limits_{i=1}^{r} \,\m_i$}
\end{picture}
\end{center}
\caption{\label{generalYoungTableau} General Young tableau $\kle{\m}_j = \kle{\m_1,\dots,\m_r}_j$ constructed out of $\kls{\alpha}$ and $\kls{\zeta}$} 
\end{figure}

\noindent
To fill this tableaux lexicographically we enumerate the indices $\xi_i$ according to the definition (\ref{JoinSets}). We therefore have $\xi_1 = \zeta_1$, $\xi_n = \zeta_n$ and $\xi_{n+1} = \alpha_1$, $\xi_{n+k} = \alpha_k$. In the following we will develop a \textit{polynomial technique} to determine $\XS{\zeta}{n} \YH_{\xi_1\dots\xi_{m}|\kle{\m}_j}^{\xi_1'\dots\xi_{m}'}$. This is again done by constructing a differential operator $\YH^{\kle{\m}_j|\alpha_1'\dots\alpha_k'}_{\alpha_1\dots\alpha_k|n}$ having the property
\begin{equation}
\label{desiredProperty}
\XS{\zeta}{n} \; \YH_{\xi_1\dots\xi_{m}|\kle{\m}_j}^{\xi_1'\dots\xi_{m}'}
= \YH^{\kle{\m}_j|\alpha_1'\dots\alpha_k'}_{\alpha_1\dots\alpha_k|n} \;\; \XS{\zeta'}{n} \; .
\end{equation}
Here, two cases are to be treated:
\begin{enumerate}
\item{
A subset $\kls{\tilde \alpha} \subset \kls{\alpha}$ is symmetrized together with the internal indices $\zeta_1$ to $\zeta_n$, i.e. all indices contained in $\kls{\tilde \alpha}$ and $\kls{\zeta}$ are put into a completely symmetric Young tableaux $\Kle{m}=\Kle{n+k}$ given by
\\ \\
\begin{picture}(8,1)
\unitlength0.75cm
\linethickness{0.075mm}
\put(0,-0.7){\framebox(1,1){$\zeta_1$}}
\put(1.01,-0.7){\framebox(2,1){$\dots$}}
\put(3.02,-0.7){\framebox(1,1){$\zeta_n$}}
\put(4.03,-0.7){\framebox(1,1){$\tilde \alpha_1$}}
\put(5.04,-0.7){\framebox(2,1){$\dots$}}
\put(7.05,-0.7){\framebox(1,1){$\tilde \alpha_k$}}
\put(8.3,-0.7){.}
\end{picture}
\\ \\ \\
The contracted standard Young projector related to this tableau is given by
\begin{equation}
\label{extendedSymYP}
\XS{\zeta}{n} \; \YH^{\kls{\zeta'}\kls{\tilde \alpha'}}_{\kls{\zeta}\kls{\tilde \alpha}|\kle{\m}} =
\frac{n!}{m!} \cdot
\mathfrak{Y}^{S|\kls{\tilde \alpha'}}_{\kls{\tilde \alpha}} \; \XS{\zeta'}{n}
\end{equation}
with $\mathfrak{Y}^{S|\kls{\tilde \alpha'}}_{\kls{\tilde \alpha}}$ given by
\begin{equation}
\label{DefinitionYHS}
\mathfrak{Y}^{S|\kls{\tilde \alpha'}}_{\kls{\tilde \alpha}}\kln{\pd,x} =
\prod _{i=1}^k \pd_{\tilde \alpha_i}^x \;
\prod _{j=1}^k x^{\tilde \alpha_{j}'}  \; .
\end{equation}
The complete structure of the contracted standard Young projector on the left hand side of (\ref{extendedSymYP}) is reproduced by the differential operator $\mathfrak{Y}^{S|\kls{\tilde \alpha'}}_{\kls{\tilde \alpha}}$ acting on the symmetric tensor~$\XS{\zeta'}{n}$. The normalizing factor in (\ref{extendedSymYP}) is determined by the normalizing factor $f_{(m)}$ defined through (\ref{NormalizationF}).
} 
\item{A subset $\kls{\tilde \alpha} \subset \kls{\alpha}$ of $l$ indices is antisymmetrized with the internal indices $\zeta_1$ to $\zeta_n$,
i.e. all indices contained in $\kls{\zeta}$ and $\kls{\tilde \alpha}$ are inserted into the Young tableaux $\kle{n,1,\dots,1}$
according to
\\ \\ \\ \\ \\ \\ \\ \\
\begin{picture}(5,5)
\unitlength0.75cm
\linethickness{0.075mm}  
\put(0,-0.7){\framebox(1,1){$\tilde \alpha_{l}$}}
\put(0,0.31){\framebox(1,2){$\vdots$}}
\put(0,2.32){\framebox(1,1){$\tilde \alpha_{1}$}}
\put(0,3.33){\framebox(1,1){$\zeta_1$}}
\put(1.01,3.33){\framebox(1,1){$\zeta_2$}}
\put(2.02,3.33){\framebox(2,1){$\dots$}}
\put(4.03,3.33){\framebox(1,1){$\zeta_n$}}
\put(1.3,-0.7){.}
\end{picture}
\\ \\ \\
The contracted Young projector related to this tableau obtains the form
\begin{equation}
\label{extendedASymYP}
\XS{\zeta}{n} \; \YH^{\kls{\zeta'}\kls{\tilde \alpha'}}_{\kls{\zeta}\kls{\tilde \alpha}|\kle{n,1,\dots,1}} =
\frac{1+l}{n+l} \cdot
x^{\sigma} \; 
\mathfrak{Y}^{A|\kls{\tilde \alpha'}}_{\kls{\tilde \alpha} \sigma} \; \XS{\zeta'}{n}
\end{equation}
with $\mathfrak{Y}^{A|\kls{\tilde \alpha'}}_{\kls{\tilde \alpha} \sigma}$ given by 
\begin{eqnarray}
\label{antiAnsatz}
\mathfrak{Y}^{A|\kls{\tilde \alpha'}}_{\kls{\tilde \alpha} \sigma}
&=&
\pd_{[\tilde \alpha_1}^{y_{1}} \cdots \pd_{\tilde \alpha_{l}}^{y_{l}} \; \pd_{\sigma]}^x \; y_1^{\tilde \alpha'_1} \cdots y_{l}^{\tilde \alpha'_{l}}
\\
&=:&
\mathfrak{D}_{\kls{\tilde \alpha} \sigma} \text{Y}^{\kls{\tilde \alpha'}} \; .
\end{eqnarray}
The normalizing factor in (\ref{extendedASymYP}) is determined by the normalizing factor $f_{(n,1,\dots,1)}$ defined through (\ref{NormalizationF}). Again, we reproduce the complete structure of the contracted standard Young projector on the left hand side of (\ref{extendedASymYP}) with the help of the differential operator~$x^{\sigma} \;\mathfrak{Y}^{A|\kls{\tilde \alpha'}}_{\kls{\tilde \alpha} \sigma}$ acting on the tensor~$\XS{\zeta'}{n}$.

A Young pattern $\kln{n,1,\dots,1}$ of course leads to many different lexicographically filled standard tableaux but only the one where all indices $\kls{\zeta}$ are put into the first row survives the symmetrization of $\zeta_1$ to $\zeta_n$ which follows from the definition (\ref{YoungOperators}) of $\Y_{\kle{\m}}$. All other standard tableaux cancel under this symmetrization.
}
\end{enumerate}
With these pre--requisites we can now discuss all standard Young tableaux which can be constructed out of $\kls{\alpha}$ and $\kls{\zeta}$ under the constraint that all indices $\zeta_i$ are symmetric. The contracted Young projector related to the general Young tableaux shown in Figure~\ref{generalYoungTableau} can be obtained with the help of the differential operators $\mathfrak{Y}^{S|\kls{\alpha'}^0}_{\kls{\alpha}^0}$ and $\mathfrak{Y}^{A|\kls{\alpha'}^i}_{\kls{\alpha}^i \sigma}$. First, we perform the symmetrization of $\kls{\alpha}^0$ followed by an antisymmetrization of all subsets $\kls{\alpha}^i$ with $i=1,...,\mathfrak{a}$.

To this purpose we define the following commutative product of $\mathfrak{Y}^{A|\kls{\alpha'}^i}_{\kls{\alpha}^i \sigma_i}$ and $\mathfrak{Y}^{A|\kls{\alpha'}^j}_{\kls{\alpha}^j \sigma_j}$
\begin{eqnarray}
\label{newProduct}
\mathfrak{Y}^{A|\kls{\alpha'}^i}_{\kls{\alpha}^i \sigma_i} \cdot \mathfrak{Y}^{A|\kls{\alpha'}^j}_{\kls{\alpha}^j \sigma_j}
&:=&
\mathfrak{D}_{\kls{\alpha}^i \sigma_i}
\mathfrak{D}_{\kls{\alpha}^j \sigma_j} \cdot
\text{Y}^{\kls{\alpha'}^i} \text{Y}^{\kls{\alpha'}^j}
\end{eqnarray}
and can now perform all antisymmetrizations in Figure~\ref{generalYoungTableau} by successively taking products of $\mathfrak{Y}^{A|\kls{\alpha'}^i}_{\kls{\alpha}^i \sigma_i}$. The key point is that these products automatically perform the remaining symmetrizations in $\m_2$ to $\m_r$. 

To realize the orthogonalization $q_{[\m]_j}$ of $[\m]_j$ we have to apply a standard Young projector $\YH^{\kls{\alpha'}^a}_{\kls{\alpha}^a|\kle{\m_2,\dots,\m_r}_j}$ which we obtain from Figure~\ref{generalYoungTableau} by removing the first row, where $\kls{\alpha}^a$ is given by $\kls{\alpha}^a = \cup_{i=1}^\mathfrak{a} \kls{\alpha}^i$ and is the complement of $\kls{\alpha}^0$. Up to a normalizing factor $f^{\kln{\m}}\kln{n}$, we get
\begin{eqnarray}
\label{TableauxDiffOp}
&&
\XS{\zeta}{n} \;
\YH^{\kls{\zeta'}\kls{\alpha'}}_{\kls{\zeta}\kls{\alpha}|\kle{\m_1,\dots,\m_r}_j}
\\
\nonumber
&&
\quad 
= f^{\kln{\m}}\kln{n} \; \cdot \; 
\YH^{\kls{\beta}^a}_{\kls{\alpha}^a|\kle{\m_2,\dots,\m_r}_j} \; \XS{\sigma}{\mathfrak{a}}
\prod_{i=1}^{\mathfrak{a}} \; \mathfrak{Y}^{A|\kls{\alpha'}^i}_{\kls{\beta}^i \sigma_i} \; \mathfrak{Y}^{S|\kls{\alpha'}^0}_{\kls{\alpha}^0} \;\; \XS{\zeta'}{n}
\end{eqnarray}
The action of $q_{[\m]_j}$ can be reduced to the rows $\m_2$ to $\m_r$ due to the symmetrization of the indices $\zeta_1$ to $\zeta_n$. We refer to the References~\cite{B55,E04} for further details about this operator. The operator $\YH^{\kle{\m}_j|\alpha_1'\dots\alpha_k'}_{\alpha_1\dots\alpha_k|n}$ can now be read off from~(\ref{TableauxDiffOp}).

To obtain an extended Young projector for the Young pattern $\kln{\m}=\kln{\m_1,\dots,\m_r}$ we have to evaluate two sums. First, we have to sum over all possible subsets $\kls{\alpha}^0$ and second, over all lexicographically filled standard tableaux $\kle{\m_2,\dots,\m_r}_j$ according to relation (\ref{PatternProjector}). The extended Young projector for the pattern $\kln{\m}=\kln{\m_1,\dots,\m_r}$ then obtains the form
\begin{eqnarray}
\label{heuristicConstruction}
\YH^{\kln{\m}|\kls{\alpha'}}_{\kls{\alpha}|n}
= f^{\kln{\m}}\kln{n} \; \cdot \; \sum_{\kls{\alpha}^0} \;
\YH^{\kls{\beta}^a}_{\kls{\alpha}^a|\kln{\m_2,\dots,\m_r}} \;
\XS{\sigma}{\mathfrak{a}} \prod_{i=1}^{\mathfrak{a}} \; \mathfrak{Y}^{A|\kls{\alpha'}^i}_{\kls{\beta}^i \sigma_i} \;
\mathfrak{Y}^{S|\kls{\alpha'}^0}_{\kls{\alpha}^0} \; .
\end{eqnarray}
In the above construction $f^{\kln{\m}}\kln{n}$ is an undetermined normalizing coefficient which will be determined by the projector condition $\YH^{\kln{\m}|\kls{\sigma}}_{\kls{\alpha}|n} \; \YH^{\kln{\m}|\kls{\alpha'}}_{\kls{\sigma}|n} = \YH^{\kln{\m}|\kls{\alpha'}}_{\kls{\alpha}|n}$. By taking the sum over all Young~tableaux we have summed over all equivalent irreducible representations belonging to that pattern. Due to relation~(\ref{othnonormalDec}) we obtain the identity, if the summation over all extended Young patterns~$(\m)$ is performed
\begin{eqnarray}
\label{orthonormal}
\sum_{(\m)} \, \YH^{\kln{\m}|\rho_1 \dots \rho_k}_{\alpha_1 \dots \alpha_k|n} = \Id \; .
\end{eqnarray}
The symmetry of the indices $\alpha_1$ to $\alpha_k$ will be determined by the application of a standard Young projector.

\subsubsection*{Results for the extended Young projectors}

We will now give the results for extended Young projectors for $k=1$ and for $k=2$. The results will be presented in two versions. The first version gives the results in the form~(\ref{heuristicConstruction}) and in a second form we will perform all $\sigma$--contractions. This second version of the results is very useful to check that the extended Young projectors are an orthogonal decomposition of the identity.

\subsubsection*{Vector}

In the vector case we get the results
\begin{align}
\intertext{(I)
\begin{picture}(5,0)
\unitlength0.5cm
\linethickness{0.075mm}
\put(1.000,0){\framebox(1,1){$\zeta_1$}}
\put(2.015,0){\framebox(3,1){$\cdots$}}
\put(5.030,0){\framebox(1,1){$\zeta_n$}}
\put(6.045,0){\framebox(1,1){$\alpha_1$}}
\end{picture} }
\label{VectorI}
\YH_{\alpha_1|n}^{(n+1)|\rho_1}
&=
\frac{n!}{(n+1)!} \cdot \mathfrak{Y}^{S|\rho_1}_{\alpha_1}
\qquad
\fo \; n
\\
&=
\frac{1}{n+1} \cdot \partial_{\alpha_1 } \, x^{\rho_1}
\intertext{(II)
\begin{picture}(5,2)
\unitlength0.5cm
\linethickness{0.075mm}
\put(1.000,0){\framebox(1,1){$\zeta_1$}}
\put(2.015,0){\framebox(3,1){$\cdots$}}
\put(5.030,0){\framebox(1,1){$\zeta_n$}}
\put(1.000,-1.015){\framebox(1,1){$\alpha_1$}}
\end{picture}}
\label{VectorII}
\YH_{\alpha_1|n}^{(n,1)|\rho_1 }
&=
\frac{2}{n+1} \cdot x^{\sigma_1} \; \mathfrak{Y}^{A|\rho_1}_{\alpha_1\sigma_1}
\qquad
\fo \; n \ges 1
\\
&=
\frac{2}{n+1} \cdot x^{\sigma_1} \; \delta_{[\alpha_1}^{\rho_1}\partial_{\sigma_1]}^{\phantom{|}}
\\
&=
\delta_{\alpha_1}^{\rho_1} - \frac{1}{n+1} \cdot \partial_{\alpha_1} \, x^{\rho_1} \; .
\end{align}

Due to the differentiation in~(\ref{VectorII}) the extended Young projector $\YH_{\alpha_1|n}^{(n,1)|\rho_1}$ annihilates a constant tensor with $n=0$.

To perform the $\sigma$--contraction, we assumed that these projectors act on a homogeneous polynomial $\OP_n$ of the order $n$. It is obvious that (\ref{VectorI}) and (\ref{VectorII}) are orthonormal projectors and decompose $\delta_{\alpha_1}^{\rho_1}$.

\subsubsection*{Antisymmetric second rank tensor}

In the antisymmetric second rank tensor case we find the results
\begin{align}
\intertext{(II)
\begin{picture}(5,2)
\unitlength0.5cm
\linethickness{0.075mm}
\put(1.000,0){\framebox(1,1){$\zeta_1$}}
\put(2.015,0){\framebox(3,1){$\cdots$}}
\put(5.030,0){\framebox(1,1){$\zeta_n$}}
\put(6.045,0){\framebox(1,1){$\alpha_2$} }
\put(1.000,-1.015){\framebox(1,1){$\alpha_1$}}
\end{picture}}
\label{YoungATI}
\YH_{[\alpha_1\alpha_2]|n}^{(n+1,1)|[\rho_1\rho_2]}
&=
\frac{4}{n(n+2) } \cdot \delta_{[\alpha_1}^{[\beta_1} \delta_{\alpha_2]}^{\beta_2]} \; 
x^{\sigma_1} \, \mathfrak{Y}^{A|[\rho_1}_{\beta_1\sigma_1} \, \mathfrak{Y}^{S|\rho_2]}_{\beta_2}
\qquad
\fo \; n \ges 1
\\
&=
\frac{4}{n(n+2)} \cdot \delta_{[\alpha_1}^{[\beta_1} \delta_{\alpha_2]}^{\beta_2]} \; 
x^{\sigma_1} \, \delta_{[\beta_1}^{[\rho_1}  \partial_{\sigma_1]}^{\phantom{|}} \, \partial_{\beta_2}^{\phantom{|}} x^{\rho_2]}_{\phantom{|}}
\\
&=
- \frac{2}{n+2} \cdot \partial_{[\alpha_1}^{\phantom{|}} \delta_{\alpha_2]}^{[\rho_1} \, x^{\rho_2]}_{\phantom{|}}
\intertext{(III)
\begin{picture}(5,2)
\unitlength0.5cm
\linethickness{0.075mm}
\put(1.000,0){\framebox(1,1){$\zeta_1$}}
\put(2.015,0){\framebox(3,1){$\cdots$}}
\put(5.030,0){\framebox(1,1){$\zeta_n$}}
\put(1.000,-1.015){\framebox(1,1){$\alpha_1$}}
\put(1.000,-2.030){\framebox(1,1){$\alpha_2$}}
\end{picture}}
\label{YoungATII}
\YH_{[\alpha_1\alpha_2]|n}^{(n+1,1)|[\rho_1\rho_2]}
&=
\frac{3}{n+2} \cdot
x^{\sigma_1} \, \mathfrak{Y}^{A|\rho_1\rho_2}_{\alpha_1\alpha_2\sigma_1 }
\qquad
\fo \; n \ges 1
\\
&=
\frac{3}{n+2} \cdot
x^{\sigma_1} \, \delta_{[\alpha_1}^{[\rho_1} \delta_{\alpha_2\phantom{|}}^{\rho_2]} \partial_{\sigma_1]}^{\phantom{|}}
\\
&=
\delta_{[\alpha_1}^{[\rho_1} \delta_{\alpha_2]}^{\rho_2]} +
\frac{2}{n+2} \cdot \partial_{[\alpha_1}^{\phantom{|}} \delta_{\alpha_2]}^{[\rho_1} \, x^{\rho_2]}_{\phantom{|}} \; .
\end{align}
Again, the orthogonal decomposition of $\delta_{[\alpha_1}^{[\rho_1} \delta_{\alpha_2]}^{\rho_2]}$ is obvious.

\subsubsection*{Symmetric second rank tensor}

For the symmetric tensor case we obtain
\begin{align}
\intertext{(I)
\begin{picture}(5,2)
\unitlength0.5cm
\linethickness{0.075mm}
\put(1.000,0){\framebox(1,1){$\zeta_1$}}
\put(2.015,0){\framebox(3,1){$\cdots$}}
\put(5.030,0){\framebox(1,1){$\zeta_n$}}
\put(6.045,0){\framebox(1,1){$\alpha_1$}}
\put(7.060,0){\framebox(1,1){$\alpha_2$}}
\end{picture}}
\label{YoungSI}
\YH_{(\alpha_1\alpha_2)|n}^{(n+2)|(\rho_1\rho_2)}
&=
\frac{n!}{(n+2)!} \cdot \mathfrak{Y}^{S|\rho_1\rho_2}_{\alpha_1\alpha_2}
\qquad
\fo \; n 
\\
&=
\frac{1}{(n+2)(n+1)}\cdot \partial_{\alpha_1}\partial_{\alpha_2}x^{\rho_1}x^{\rho_2}
\intertext{(II)
\begin{picture}(5,2)
\unitlength0.5cm
\linethickness{0.075mm}
\put(1.000,0){\framebox(1,1){$\zeta_1$}}
\put(2.015,0){\framebox(3,1){$\cdots$}}
\put(5.030,0){\framebox(1,1){$\zeta_n$}}
\put(6.045,0){\framebox(1,1){$\alpha_2$}}
\put(1.000,-1.015){\framebox(1,1){$\alpha_1$}}
\end{picture}}
\label{YoungSII}
\YH_{(\alpha_1\alpha_2)|n}^{(n+1,1)|(\rho_1\rho_2)}
&=
\frac{4}{n(n+2)}\cdot \delta_{(\alpha_1}^{(\beta_1}\delta_{\alpha_2)}^{\beta_2)} \; x^{\sigma_1} \, \mathfrak{Y}^{A|(\rho_1}_{\beta_1\sigma_1} \, \mathfrak{Y}^{S|\rho_2)}_{\beta_2}
\qquad
\fo \; n \ges 1
\\
&=
\frac{4}{n(n+2)}\cdot \delta_{(\alpha_1}^{(\beta_1}\delta_{\alpha_2)}^{\beta_2)} \; x^{\sigma_1} \, \delta_{[\beta_1}^{(\rho_1}\partial_{\sigma_1]}^{\phantom{|}}\partial_{\beta_2}^{\phantom{|}} x^{\rho_2)}_{\phantom{|}}
\\
&=
\frac{2}{n}\kln{ \partial_{(\alpha_1}^{\phantom{|}}\delta_{\alpha_2)}^{(\rho_1}x^{\rho_2)}_{\phantom{|}}
-\frac{1}{n+2}\cdot \partial_{\alpha_1}\partial_{\alpha_2}x^{\rho_1}x^{\rho_2} }
\intertext{(IV)
\begin{picture}(5,2)
\unitlength0.5cm
\linethickness{0.075mm}
\put(1.000,0){\framebox(1,1){$\zeta_1$}}
\put(2.015,0){\framebox(1,1){$\zeta_2$}}
\put(3.030,0){\framebox(3,1){$\cdots$}}
\put(6.045,0){\framebox(1,1){$\zeta_n$}}
\put(1.000,-1.015){\framebox(1,1){$\alpha_1$}}
\put(2.015,-1.015){\framebox(1,1){$\alpha_2$}}
\end{picture}}
\label{YoungSIII}
\YH_{(\alpha_1\alpha_2)|n}^{(n,2)|(\rho_1\rho_2)}
&=
\frac{4}{n(n+1)}\cdot \delta_{(\alpha_1}^{(\beta_1}\delta_{\alpha_2)}^{\beta_2)} \; x^{\sigma_1}x^{\sigma_2} \; \mathfrak{Y}^{A|(\rho_1}_{\beta_1\sigma_1} \, \mathfrak{Y}^{A|\rho_2)}_{\beta_2\sigma_2}
\qquad
\fo \; n \ges 2
\\
&=
\frac{4}{n(n+1)}\cdot \delta_{(\alpha_1}^{(\beta_1}\delta_{\alpha_2)}^{\beta_2)} \; x^{\sigma_1}x^{\sigma_2} \, \delta_{[\beta_1}^{(\rho_1}\partial_{\sigma_1]}^{\phantom{|}}
\delta_{[\beta_2}^{\rho_2)}\partial_{\sigma_2]}^{\phantom{|}}
\\
&=
\delta_{(\alpha_1}^{(\rho_1}\delta_{\alpha_2)}^{\rho_2)} - \frac{2}{n}\kln{ \partial_{(\alpha_1}^{\phantom{|}} \delta_{\alpha_2)}^{(\rho_1} x^{\rho_2)}_{\phantom{|}}
-\frac{1}{2(n+1)} \cdot \pd_{\alpha_1} \pd_{\alpha_2} x^{\rho_1} x^{\rho_2} } \; .
\end{align}
Here, one only has to sum the coefficients of the term $\pd_{\alpha_1} \pd_{\alpha_2} x^{\rho_1} x^{\rho_2}$ in $\YH_{(\alpha_1\alpha_2)|n}^{(n+2)|(\rho_1\rho_2)}$, $\YH_{(\alpha_1\alpha_2)|n}^{(n+1,1)|(\rho_1\rho_2)}$ and $\YH_{(\alpha_1\alpha_2)|n}^{(n,2)|(\rho_1\rho_2)}$ to see that these three extended Young projectors decompose $\delta_{(\alpha_1}^{(\rho_1}\delta_{\alpha_2)}^{\rho_2)}$.

\subsection{The construction of spin projectors}
\label{SpinProjectors}

In this Section we will construct all spin projectors for $k=1$ and $k=2$ free indices and will thereby make use of the previous results obtained for the extended Young projectors $\EYPN{\alpha}{\rho}{\m}{n}$ and for the projector onto traceless tensor polynomials $\HP{\alpha}{\rho}{k}{n}$. Here, we remark that the name of the spin projectors is borrowed from the notion of Lorentz spin which is defined in four dimensional space--time only.

Irreducible tensor representations of the group $SO(2h;\C)$ are realized in the space of traceless tensors carrying a proper Young symmetry. In the Sections~\ref{ExtendedYoungprojectors} and~\ref{SubtractionOfAllTraces} we have given an algorithm for the explicit determination of all extended Young projectors and for the determination of $\HP{\alpha}{\rho}{k}{n}$ including all possible Young symmetrizations of these objects. We can therefore dispose of all pre--requisites for the construction of projectors onto irreducible tensor representations of the group $SO(2h;\C)$.

To obtain an irreducible tensor we first apply the projector onto traceless tensor polynomials $\HP{\alpha}{\rho}{k}{n}$
followed by a projection onto extended Young symmetry with the help of the extended Young projectors
$\EYPN{\alpha}{\rho}{\m}{n}$. The product of these two projectors is then called a spin projector defined as
\begin{align}
\label{twistProjector}
\SP^{\kln{\m}|\rho_1\dots\rho_k}_{\alpha_1\dots\alpha_k|n}
&:=
\EYPN{\alpha}{\beta}{\m}{n} \; \HP{\beta}{\rho}{k}{n}
\intertext{projecting onto a local irreducible tensor given by}
\OP_{\alpha_1\dots\alpha_k|n}^{\kln{\m}}
&:=
\SP^{\kln{\m}|\rho_1\dots\rho_k}_{\alpha_1\dots\alpha_k|n} \; \OP_{\rho_1\dots\rho_k|n} \; .
\end{align}

\subsubsection*{Results for the spin projectors}

To obtain the results for the spin projectors we have to apply all Young projectors that were found in the respective tensor case to the related projector onto traceless tensor polynomials.

\subsubsection*{Vector}

In the vector case we have found two relevant Young patterns (I) and (II). The extended Young projectors (\ref{VectorI}) and (\ref{VectorII}) belonging to these patterns lead to two spin projectors. For the first Young symmetry $(n+1)$ we get
\begin{align}
\label{VectorSpinI}
\SP_{\alpha_1|n}^{(n+1)|\rho_1}
&=
\YH_{\alpha_1|n}^{(n+1)|\beta_1} \; H_{\beta_1|n}^{\rho_1}
\\
&=
\frac{1}{(n+1)(n+h-1)} \cdot  H_n \; \D_{\alpha_1} \, x^{\rho_1}
\intertext{and for the second Young symmetry $(n,1)$ we obtain}
\label{VectorSpinII}
\SP_{\alpha_1|n}^{(n,1)|\rho_1}
&=
\YH_{\alpha_1|n}^{(n,1)|\beta_1} \; H_{\beta_1|n}^{\rho_1}
\\
&=
H_n \kln{\delta_{\alpha_1}^{\rho_1} - \frac{1}{n+h-1}
\kls{ \frac{1}{n+2h-3} \cdot x_{\alpha_1} \, \D^{\rho_1} + \frac{1}{n+1} \cdot \D_{\alpha_1} \, x^{\rho_1} } } \; .
\end{align}

\subsubsection*{Antisymmetric Tensor}

In this case we have to apply the extended Young projectors (\ref{YoungATI}) and (\ref{YoungATII}) belonging to the pattern (II) and (III). Again we get two spin projectors. For the Young symmetry $(n+1,1)$ we find
\begin{align}
\label{ASTensorSpinI}
\SP_{[\alpha_1\alpha_2]|n}^{(n+1,1)|[\rho_1\rho_2]}
&=
\YH_{[\alpha_1\alpha_2]|n}^{(n+1,1)|[\beta_1\beta_2]} \;
H_{[\beta_1\beta_2]|n}^{[\rho_1\rho_2]}
\\
\nonumber
&=
- \frac{2}{(n+2)(n+h-1)} \cdot H_n \KLSo{}
\D_{[\alpha_1}^{\phantom{|}}\delta_{\alpha_2]}^{[\rho_1} x^{\rho_2]}_{\phantom{|}}
\\
\nonumber
&
\qquad \qquad \;
- \frac{n+h}{(n+h-2)^2(n+2h-2)} \cdot x_{[\alpha_1}\D_{\alpha_2]} x^{[\rho_1}\D^{\rho_2]} \oKLS{} \; .
\intertext{For the Young symmetry $(n,1,1)$ we get}
\label{ASTensorSpinII}
\SP_{[\alpha_1\alpha_2]|n}^{(n,1,1)|[\rho_1\rho_2]}
&=
\YH_{[\alpha_1\alpha_2]|n}^{(n,1,1)|[\beta_1\beta_2]} \; H_{[\beta_1\beta_2]|n}^{[\rho_1\rho_2]}
\\
\nonumber
&=
H_n \KLSo{} \delta_{[\alpha_1}^{[\rho_1}\delta_{\alpha_2]}^{\rho_2]}
+ \frac{2}{n+h-1} \KLEo{} \frac{1}{n+2h-4} \cdot x_{[\alpha_1}^{\phantom{|}}\delta_{\alpha_2]}^{[\rho_1}\D^{\rho_2]}_{\phantom{|}}
+ \frac{1}{n+2} \cdot \D_{[\alpha_1}^{\phantom{|}}\delta_{\alpha_2]}^{[\rho_1}x^{\rho_2]}_{\phantom{|}} \oKLE{}
\\
\nonumber
&
\qquad \qquad \;
- \frac{2}{(n+2)(n+h-2)^2(n+2h-4)} \cdot x_{[\alpha_1}\D_{\alpha_2]}x^{[\rho_1}\D^{\rho_2]} \oKLS{} \; .
\end{align}

\subsubsection*{Symmetric Tensor}

Here, we apply the extended Young projectors (\ref{YoungSI}), (\ref{YoungSII}) and (\ref{YoungSIII}) belonging to the pattern (I), (II) and (IV); we obtain three spin projectors. For the Young pattern $(n+2)$ we get
\begin{align}
\label{STensorSpinI}
\SP_{(\alpha_1\alpha_2)|n}^{(n+2)|(\rho_1\rho_2)}
&=
\YH_{(\alpha_1\alpha_2)|n}^{(n+2)|(\beta_1\beta_2)} \; H_{(\beta_1\beta_2)|n}^{(\rho_1\rho_2)}
\\
\nonumber
&=
\frac{1}{(n+1)(n+2)(n+h-1)(n+h)} \cdot H_n \; \D_{\alpha_1}\D_{\alpha_2} \, x^{\rho_1} x^{\rho_2}
\intertext{and for the pattern $(n+1,1)$ we obtain}
\label{STensorSpinII}
\SP_{(\alpha_1\alpha_2)|n}^{(n+1,1)|(\rho_1\rho_2)}
&=
\YH_{(\alpha_1\alpha_2)|n}^{(n+1,1)|(\beta_1\beta_2)} \; H_{(\beta_1\beta_2)|n}^{(\rho_1\rho_2)}
\\
\nonumber
&=
H_n \, H_{\alpha_1\alpha_2|0}^{\beta_1\beta_2} \; \tilde \SP_{(\beta_1\beta_2)|n}^{'(n+1,1)|(\sigma_1\sigma_2)} \;
H_{\sigma_1\sigma_2|0}^{\rho_1\rho_2}
\\
\nonumber
&=
H_n \KLSSo{} \tilde \SP_{(\alpha_1\alpha_2)|n}^{'(n+1,1)|(\rho_1\rho_2)}
- \frac{2}{n(n+h)(n+2h-2)} \KLEo{} x_{(\alpha_1}\D_{\alpha_2)} \, \delta^{\rho_1\rho_2}
+\delta_{\alpha_1\alpha_2} \, \D^{(\rho_1}x^{\rho_2)} \oKLE{} \oKLSS{}
\intertext{with}
\tilde \SP_{(\alpha_1\alpha_2)|n}^{'(n+1,1)|(\rho_1\rho_2)}
&=
\frac{2}{n(n+h-1)} \cdot
\\
&
\KLNNo{} \D_{(\alpha_1}^{\phantom{|}}\delta_{\alpha_2)}^{(\rho_1}x^{\rho_2)}_{\phantom{|}}
- \frac{1}{n+h} \KLEo{}
\frac{1}{n+2h-2} \cdot x_{(\alpha_1}\D_{\alpha_2)}x^{(\rho_1}\D^{\rho_2)}
+ \frac{1}{n+2} \cdot \D_{\alpha_1}\D_{\alpha_2} \, x^{\rho_1}x^{\rho_2} \oKLE{} \oKLNN{}
\end{align}
and $H_{\alpha_1\alpha_2|0}^{\beta_1\beta_2}$ given by (\ref{HNUll}). For the third and last Young pattern $(n,2)$ we find
\begin{align}
\SP_{(\alpha_1\alpha_2)|n}^{(n,2)|(\rho_1\rho_2)}
\label{STensorSpinIII}
&=
\YH_{(\alpha_1\alpha_2)|n}^{(n,2)|(\beta_1\beta_2)} \; H_{(\beta_1\beta_2)|n}^{(\rho_1\rho_2)}\\
\nonumber
&=
H_n \, H_{\alpha_1\alpha_2|0}^{\beta_1\beta_2} \; \tilde \SP_{(\beta_1\beta_2)|n}^{'(n,2)|(\sigma_1\sigma_2)} \;
H_{\sigma_1\sigma_2|0}^{\rho_1\rho_2}
\\
\nonumber
&=
H_n \KLSSo{} \tilde \SP_{(\alpha_1\alpha_2)|n}^{'(n,2)|(\rho_1\rho_2)}
- \frac{1}{2\kln{h-1}} \cdot \delta_{\alpha_1\alpha_2}\delta^{\rho_1\rho_2}
\\
\nonumber
&
\qquad \quad
 + \frac{1}{n(h-1)(n+h)} \KLEo{} x_{(\alpha_1}\D_{\alpha_2)} \; \delta^{\rho_1\rho_2}
 +\delta_{\alpha_1\alpha_2} \; \D^{(\rho_1}x^{\rho_2)} \oKLE{}\oKLSS{}
\intertext{with}
\tilde \SP_{\alpha_1\alpha_2|n}^{'(n,2)|\rho_1\rho_2}
&=
\delta_{(\alpha_1}^{(\rho_1}\delta_{\alpha_2)}^{\rho_2)}
-\frac{2}{n+h-1}
\KLEo{} \frac{1}{n+2h-2} \cdot x_{(\alpha_1}^{\phantom{|}} \delta_{\alpha_2)}^{(\rho_1} \D^{\rho_2)}_{\phantom{|}}
+ \frac{1}{n} \cdot \D_{(\alpha_1}^{\phantom{|}} \delta_{\alpha_2)}^{(\rho_1} x^{\rho_2)}_{\phantom{|}} \oKLE{}
\\
\nonumber
&
\quad
+ \frac{1}{(n+h-2)(n+2h-2)} \KLEo{}
\frac{2(h-2)}{n(h-1)(n+h)} \cdot x_{(\alpha_1} \D_{\alpha_2)} \, x^{(\rho_1} \D^{\rho_2)}
\\
\nonumber
&
\qquad\qquad\qquad\qquad\qquad\qquad\qquad
+ \frac{1}{(n+h-1)(n+2h-3)} \cdot x_{\alpha_1} x_{\alpha_2} \, \D^{\rho_1} \D^{\rho_2} \oKLE{}
\\
\nonumber
&
\quad
+ \frac{1}{n(n+1)(n+h)(n+h-1)} \cdot \D_{\alpha_1} \D_{\alpha_2} \, x^{\rho_1} x^{\rho_2}  \;  .
\end{align}
In this form, the spin projectors (\ref{ASTensorSpinI}), (\ref{ASTensorSpinII}) and (\ref{STensorSpinI}) to (\ref{STensorSpinIII}) are given here for the first time. Different representations in local and nonlocal form can be found in the Refs.~\cite{GLR99,GLR00,GL01}. The key point is, that we have consequently used the interior derivative $\D_\alpha$ in the above results which shows the deep relations to the conformal Lie algebra $\mathfrak{so}(2,2h)$ and makes them more compact.

Since any spin projector $\SP^{\kln{\m}|\rho_1\dots\rho_k}_{\alpha_1\dots\alpha_k|n}$ projects onto local traceless tensor polynomials, it fulfills the factorization (\ref{factorization}). In the above results this holds especially for $\SP_{(\alpha_1\alpha_2)|n}^{(n+1,1)|(\rho_1\rho_2)}$ and $\SP_{(\alpha_1\alpha_2)|n}^{(n,2)|(\rho_1\rho_2)}$.

The on--cone limit of the spin projectors can be obtained by $\tilde H_n = \Id$ and taking $\tx_\alpha$ and $\tilde \D_\alpha$ on the complex cone.

\subsection{The construction of complete spin decompositions}
\label{spindecompositions}

We are now able to deduce our main results, namely the complete decomposition of local tensors $\OP_{\alpha_1\dots\alpha_k\kln{\zeta_1\dots\zeta_n}}$ into irreducible components, from the complete trace decompositions obtained in Section~\ref{CompleteTraceDecomposition}. There, we isolated all traceless contributions from the trace terms $x^2$, $x_{\alpha_i}$ and $\delta_{\alpha_i\alpha_j}$ and can now transform these traceless parts into irreducible tensors by replacing the projectors onto traceless polynomials $\HP{\alpha}{\rho}{k}{n}$ by their representations in terms of spin projectors.

According to relation~(\ref{orthonormal}) and Section~\ref{SpinProjectors} we find the following representations for $H_n$, $H_{\alpha_1|n}^{\rho_1}$, $H_{[\alpha_1\alpha_2]|n}^{[\rho_1\rho_2]}$ and $H_{(\alpha_1\alpha_2)|n}^{(\rho_1\rho_2)}$
\begin{eqnarray}
\label{repI}
H_n
&=&
\SP_{n}^{(n)}
\\
\label{repII}
H_{\alpha_1|n}^{\rho_1}
&=&
\SP_{\alpha_1|n}^{(n+1)|\rho_1} + \SP_{\alpha_1|n}^{(n,1)|\rho_1}
\\
\label{repIII}
H_{[\alpha_1\alpha_2]|n}^{[\rho_1\rho_2]}
&=&
\SP_{[\alpha_1\alpha_2]|n}^{(n+1,1)|[\rho_1\rho_2]} + \SP_{[\alpha_1\alpha_2]|n}^{(n,1,1)|[\rho_1\rho_2]}
\\
\label{repIV}
H_{(\alpha_1\alpha_2)|n}^{(\rho_1\rho_2)}
&=&
\SP_{(\alpha_1\alpha_2)|n}^{(n+2)|(\rho_1\rho_2)} + \SP_{(\alpha_1\alpha_2)|n}^{(n+1,1)|(\rho_1\rho_2)} + \SP_{(\alpha_1\alpha_2)|n}^{(n,2)|(\rho_1\rho_2)} \; .
\end{eqnarray}
These representations can now be inserted into the complete trace decompositions (\ref{TRDVector}), (\ref{TRDASTensor}) and (\ref{TRDSTensor}) to obtain complete decompositions into irreducible components.

\subsubsection*{Scalar}

To begin the list of complete decompositions into irreducible components, we reformulate the result of Bargmann and Todorov~\cite{BT77} for the scalar case
\begin{equation}
\label{ScalarDec}
\SPD_n = \sum_{j=0}^{\kle{\frac{n}{2}}}\frac{\kln{n+h-1-2j}!}{4^j j! \kln{n+h-1-j}!} \cdot \kln{x^2}^j  \;  \SO_{n-2j}^{(n-2j)}(x,j)
\end{equation}
with 
\begin{equation}
\SO_{n-2j}^{(n-2j)}(x,j) := \SP_{n-2j}^{(n-2j)} \; \Delta^j \; .
\end{equation}
On the complex cone the scalar decomposition reduces to the identity which means that any fully contracted scalar local tensor on the cone is irreducible under $SO(2h;\C)$.

\subsubsection*{Vector}

In the vector case we insert the relations (\ref{repI}) and (\ref{repII}) into the trace decomposition (\ref{TRDVector}) and find
\begin{eqnarray}
\nonumber
\SPD_{\alpha_1|n}^{\rho_1} 
&=&
\sum_{j=0}^{\kle{\frac{n+1}{2}}}\frac{\kln{n+h-1-2j}!}{4^j j! \kln{n+h-1-j}!} \KLSSo{}
\kln{x^2}^j \KLE{ \SO_{\alpha_1|n-2j}^{2|(n-2j+1)|\rho_1}\kln{x,j} + \SO_{\alpha_1|n-2j}^{3|(n-2j,1)|\rho_1}\kln{x,j} }
\\
\label{SPDVectorI}
&&
\qquad \;
+ \, \frac{4j}{(n+h-j)(n+2h-1-2j)} \cdot \kln{x^2}^{j-1}
\x_{\alpha_1} \, \SO_{n+1-2j}^{1|(n-2j+1)|\rho_1}\kln{x,j} \oKLSS{}
\end{eqnarray}
The operators $\SO_{\alpha_1|n}^{i|(\m)|\rho_1}\kln{x,j}$ are irreducible under the group $SO(2h,\C)$. For the vector case they are given by one scalar operator
\begin{align}
\label{VectorOPSI}
\SO_{n+1-2j}^{1|(n-2j+1)|\rho_1}(x,j)
&:=
\SP_{n+1-2j}^{(n-2j+1)}\;\D^{\rho_1} \; \Delta^{j-1}
\intertext{and two vector operators}
\SO_{\alpha_1|n-2j}^{2|(n-2j+1)|\rho_1}(x,j)
&:=
\SP_{\alpha_1|n-2j}^{(n-2j+1)|\rho_1} \;\; \Delta^{j}
\\
\label{VectorOPSIII}
\SO_{\alpha_1|n-2j}^{3|(n-2j,1)|\rho_1}(x,j)
&:=
\SP_{\alpha_1|n-2j}^{(n-2j,1)|\rho_1} \;\; \Delta^{j} \; .
\end{align}
According to (\ref{DefinitionX}) the operator $\x_{\alpha_1}$ still contains a derivative $\pd_{\alpha_1}$ which can be applied in (\ref{SPDVectorI}) to the following operator $\SO_{n+1-2j}^{1|(n-2j+1)|\rho_1}$ without destroying its irreducibility. Decomposing $\x_{\alpha_1}$ we get the complete decomposition
\begin{eqnarray}
\nonumber
\SPD_{\alpha_1|n}^{\rho_1}
&=&
\sum_{j=0}^{\kle{\frac{n+1}{2}}}\frac{\kln{n+h-1-2j}!}{4^j j! \kln{n+h-1-j}!} \KLSSo{}
\kln{x^2}^j \KLE{ \SO_{\alpha_1|n-2j}^{2|(n-2j+1)|\rho_1}\kln{x,j} + \SO_{\alpha_1|n-2j}^{3|(n-2j,1)|\rho_1}\kln{x,j} }
\\ 
\label{SPDVectorII}
&&
\qquad
- \, \frac{2j}{(n+h-j)(n+2h-1-2j)} \cdot \kln{x^2}^{j-1}
\\
\nonumber
&&
\qquad \qquad
\times \KLEo{} x^2 \, \SO_{\alpha_1|n-2j}^{1|(n-2j+1)|\rho_1}\kln{x,j} - 2 \kln{n+h-2j} \cdot x_{\alpha_1} \, \SO_{n+1-2j}^{1|(n-2j+1)|\rho_1}\kln{x,j} \oKLE{}  \oKLSS{}  
\end{eqnarray}
with $\SO_{\alpha_1|n-2j}^{1|(n-2j+1)|\rho_1}$ given by
\begin{eqnarray}
\label{VectorOPSIV}
\SO_{\alpha_1|n-2j}^{1|(n-2j+1)|\rho_1}\kln{x,j}
&:=&
\pd_{\alpha_1}^{\phantom{|}} \, \SO_{n+1-2j}^{1|(n-2j+1)|\rho_1}\kln{x,j}   \; .
\end{eqnarray}
For $h=2$ this complete decomposition (\ref{SPDVectorII}) has first been given in Ref.~\cite{EGL04} for the quark--antiquark vector operator; see also Section~\ref{ApplicationQCD} for the twist decomposition of this operator. Taking the on--cone limit of (\ref{SPDVectorII}) we get
\begin{eqnarray}
\label{SPDVectorOnCone}
\widetilde \SPD_{\alpha_1|n}^{\rho_1}
&=&
\tilde \SO_{\alpha_1|n}^{2|(n+1)|\rho_1}(\tilde x,0) + \tilde \SO_{\alpha_1|n}^{3|(n,1)|\rho_1}(\tilde x,0)
\\
\nonumber
&&
+ \frac{1}{(n+h-1)(n+2h-3)}\cdot\tilde x_{\alpha_1} \; \tilde \SO_{n-1}^{1|(n-1)|\rho_1}(\tilde x,1)  \;  .
\end{eqnarray}
If we contract the results (\ref{SPDVectorI}) or (\ref{SPDVectorII}) with $x^{\alpha_1}$ we arrive at the scalar decomposition (\ref{ScalarDec}) with $n \ra n+1$ after some calculation. This also holds for the respective on--cone decompositions.

\subsubsection*{Antisymmetric tensor}

To obtain the complete decomposition into irreducible tensors in the antisymmetric tensor case, we insert the representations (\ref{repI}) to (\ref{repIII}) into the complete trace decomposition~(\ref{TRDASTensor}) and find the result
\begin{eqnarray}
\label{SPDASTensorI}
\SPD_{[\alpha_1\alpha_2]|n}^{[\rho_1\rho_2]}
&=&
\sum_{j=0}^{\kle{\frac{n+2}{2}}} \frac{(n+h-1-2j)!}{4^j \, j! \, (n+h-1-j)!} \KLSSo{}
\\
\nonumber
&&
\quad
\kln{x^2}^j \KLE{ \SO_{[\alpha_1\alpha_2]|n-2j}^{3|(n-2j+1,1)|[\rho_1\rho_2]}\kln{x,j} + \SO_{[\alpha_1\alpha_2]|n-2j}^{4|(n-2j,1,1)|[\rho_1\rho_2]}\kln{x,j} }
\\
\nonumber
&&
- \frac{8j}{(n+h-j)(n+2h-2-2j)} \cdot \kln{x^2}^{j-1} \; \x_{[\alpha_1}^{\phantom{|}} \SO_{\alpha_2]|n+1-2j}^{2|(n-2j+1,1)|[\rho_1\rho_2]}\kln{x,j} 
\\
\nonumber
&&
- \frac{8j(n+h+1-2j)}{(n+h-j)(n+h-2j)(n+2h-2j)} \cdot \kln{x^2}^{j-1} \; \x_{[\alpha_1}^{\phantom{|}} \SO_{\alpha_2]|n+1-2j}^{1|(n-2j+2)|[\rho_1\rho_2]}\kln{x,j} \oKLSS{}
\end{eqnarray}
with two vector operators
\begin{align}
\SO_{\alpha_2|n+1-2j}^{1|(n-2j+2)|[\rho_1\rho_2]}\kln{x,j}
&:=
\SP_{\alpha_2|n+1-2j}^{(n-2j+2)|[\rho_1} \, \D^{\rho_2]}_{\phantom{|}} \, \Delta^{j-1}
\\
\SO_{\alpha_2|n+1-2j}^{2|(n-2j+1,1)|[\rho_1\rho_2]}\kln{x,j}
&:=
\SP_{\alpha_2|n+1-2j}^{(n-2j+1,1)|[\rho_1} \, \D^{\rho_2]}_{\phantom{|}} \, \Delta^{j-1}
\intertext{and two antisymmetric tensor operators}
\SO_{[\alpha_1\alpha_2]|n-2j}^{3|(n-2j+1,1)|[\rho_1\rho_2]}\kln{x,j}
&:=
\SP_{[\alpha_1\alpha_2]|n-2j}^{(n-2j+1,1)|[\rho_1\rho_2]} \, \Delta^{j}
\\
\SO_{[\alpha_1\alpha_2]|n-2j}^{4|(n-2j,1,1)|[\rho_1\rho_2]}\kln{x,j}
&:=
\SP_{[\alpha_1\alpha_2]|n-2j}^{(n-2j,1,1)|[\rho_1\rho_2]} \, \Delta^{j} \; .
\end{align}
The last term in the decomposition (\ref{SPDASTensorI}) receives two contributions. Again, we decompose the operator $\x_{\alpha_1}$ to expand the above result and get a decomposition with five contributions
\begin{eqnarray}
\label{SPDASTensorII}
&&
\SPD_{[\alpha_1\alpha_2]|n}^{[\rho_1\rho_2]} =
\\
\nonumber
&&
\sum_{j=0}^{\kle{\frac{n+2}{2}}} \frac{(n+h-1-2j)!}{4^j \, j! \, (n+h-1-j)!} \KLSSo{}
\\
\nonumber
&&
\qquad
\kln{x^2}^j \KLE{ \SO_{[\alpha_1\alpha_2]|n-2j}^{3|(n-2j+1,1)|[\rho_1\rho_2]}\kln{x,j} + \SO_{[\alpha_1\alpha_2]|n-2j}^{4|(n-2j,1,1)|[\rho_1\rho_2]}\kln{x,j} }
\\
\nonumber
&&
\quad
+ \frac{4j}{(n+h-j)(n+2h-2-2j)} \cdot 
\kln{x^2}^{j-1} \KLEo{} x^2 \, \SO_{[\alpha_1\alpha_2]|n-2j}^{2|(n-2j+1,1)|[\rho_1\rho_2]}\kln{x,j}
\\
\nonumber
&&
\qquad\qquad\qquad\qquad\qquad\qquad\qquad\qquad\qquad\quad\;
- 2(n+h-2j) \cdot x_{[\alpha_1}^{\phantom{|}} \SO_{\alpha_2]|n+1-2j}^{2|(n-2j+1,1)|[\rho_1\rho_2]}\kln{x,j}  \oKLE{}
\\
\nonumber
&&
\quad
- \frac{8j(n+h+1-2j)}{(n+h-j)(n+2h-2j)} \cdot \kln{x^2}^{j-1} \; x_{[\alpha_1}^{\phantom{|}} \SO_{\alpha_2]|n+1-2j}^{1|(n-2j+2)|[\rho_1\rho_2]}\kln{x,j} \oKLSS{}
\end{eqnarray}
with
\begin{eqnarray}
\SO_{[\alpha_1\alpha_2]|n-2j}^{2|(n-2j+1,1)|[\rho_1\rho_2]}\kln{x,j}
&:=&
\pd_{[\alpha_1}^{\phantom{|}} \SO_{\alpha_2]|n+1-2j}^{2|(n-2j+1,1)|[\rho_1\rho_2]}\kln{x,j} \; .
\end{eqnarray}
On the complex cone we find the result
\begin{eqnarray}
\label{SPDASTensorOnCone}
\widetilde \SPD_{[\alpha_1\alpha_2]|n}^{[\rho_1\rho_2]}
&=&
 \tilde \SO_{[\alpha_1\alpha_2]|n}^{3|(n+1,1)|[\rho_1\rho_2]}(\tilde x,0)
+\tilde \SO_{[\alpha_1\alpha_2]|n}^{4|(n,1,1)|[\rho_1\rho_2]}(\tilde x,0)
\\
\nonumber
&&
-\frac{2}{(n+h-1)(n+2h-4)} \cdot
\tilde x_{[\alpha_1}^{\phantom{|}}\tilde \SO_{\alpha_2]|n-1}^{2|(n-1,1)|[\rho_1\rho_2]}(\tilde x,1)
\\
\nonumber
&&
-\frac{2}{(n+h-2)(n+2h-2)} \cdot 
\tilde x_{[\alpha_1}^{\phantom{|}}\tilde \SO_{\alpha_2]|n-1}^{1|(n)|[\rho_1\rho_2]}(\tilde x,1)  \;  .
\end{eqnarray}

\subsubsection*{Symmetric tensor}

To obtain the final spin decomposition in the symmetric tensor case we insert the representations (\ref{repI}), (\ref{repII}) and (\ref{repIV}) into the complete trace decomposition (\ref{TRDSTensor}). After some calculation the result is
\begin{eqnarray}
\label{SPDSTensorI}
&&
\SPD_{(\alpha_1\alpha_2)|n}^{(\rho_1\rho_2)} =
\\
\nonumber
&&
 \sum_{j=0}^{\kle{\frac{n+2}{2}}}\frac{\kln{n+h-1-2j}!}{4^j j! \kln{n+h-1-j}!} \KLSSo{}
\\
&&
\nonumber
\qquad 
  \kln{x^2}^j \KLE{ 
  \SO_{(\alpha_1\alpha_2)|n-2j}^{5|(n-2j+2)|(\rho_1\rho_2)}
+ \SO_{(\alpha_1\alpha_2)|n-2j}^{6|(n-2j+1,1)|(\rho_1\rho_2)}
+ \SO_{(\alpha_1\alpha_2)|n-2j}^{7|(n-2j,2)|(\rho_1\rho_2)} }
\\
&&
\nonumber
\quad 
+ \frac{8j}{(n+h-j)(n+2h-2j)} \cdot \kln{x^2}^{j-1} \x_{(\alpha_1}^{\phantom{|}} \SO_{\alpha_2)|n+1-2j}^{4|(n-2j+1,1)|(\rho_1\rho_2)}
\\
&&
\nonumber
\quad 
- \frac{4j(n+h+1-2j)}{(h-1)(n+h-j)(n+h+2-2j)(n+2h-2j)} \cdot \kln{x^2}^{j-1} \KLEo{}
\\
&&
\nonumber
\qquad
(n+h-2j) \kln{ \x_{(\alpha_1}^{\phantom{|}} \SO_{\alpha_2)|n+1-2j}^{1|(n-2j+2)|(\rho_1\rho_2)}
+ \delta_{\alpha_1\alpha_2} \, \SO_{n+2-2j}^{0|(n-2j+2)|(\rho_1\rho_2)} }
- 2h \cdot \x_{(\alpha_1}^{\phantom{|}} \SO_{\alpha_2)|n+1-2j}^{3|(n-2j+2)|(\rho_1\rho_2)} \oKLE{}
\\
\nonumber
&&
\quad 
+ \frac{2j(n+h-2j)(n+h+1-2j)}{(h-1)(n+h-j)} \cdot \kln{x^2}^{j-1} \; \delta_{\alpha_1\alpha_2} \, \SO_{n+2-2j}^{1|(n-2j+2)|(\rho_1\rho_2)}
\\
\nonumber
&&
\quad 
+ \frac{16j(j-1)(n+h-2j)(n+h-j)^{-1}(n+2h-2j)^{-1}}{(n+h+1-j)(n+h+2-2j)(n+2h+1-2j)} \cdot \kln{x^2}^{j-2} \x_{\alpha_1} \x_{\alpha_2} \, \SO_{n+2-2j}^{2|(n-2j+2)|(\rho_1\rho_2)} \oKLSS{}
\end{eqnarray}
with three scalar operators
\begin{align}
\SO_{n+2-2j}^{0|(n-2j+2)|(\rho_1\rho_2)}(x,j)
&:=
\SP_{n+2-2j}^{(n-2j+2)}\;\D^{(\rho_1}x^{\rho_2)} \; \Delta^{j-1}
\\
\SO_{n+2-2j}^{1|(n-2j+2)|(\rho_1\rho_2)}(x,j)
&:=
\SP_{n+2-2j}^{(n-2j+2)}\;\delta^{\rho_1\rho_2} \; \Delta^{j-1}
\\
\SO_{n+2-2j}^{2|(n-2j+2)|(\rho_1\rho_2)}(x,j)
&:=
\SP_{n+2-2j}^{(n-2j+2)}\;\D^{\rho_1}\D^{\rho_2} \; \Delta^{j-2} \; ,
\intertext{three vector operators}
\SO_{\alpha_1|n+1-2j}^{1|(n-2j+2)|(\rho_1\rho_2)}(x,j)
&:=
\partial_{\alpha_1\phantom{|}}^{\phantom{|}} \!
\SO_{n+2-2j}^{1|(n-2j+2)|(\rho_1\rho_2)}(x,j)
\\
\SO_{\alpha_1|n+1-2j}^{3|(n-2j+2)|(\rho_1\rho_2)}(x,j)
&:=
\SP_{\alpha_1|n+1-2j}^{(n-2j+2)|(\rho_1}\;\D^{\rho_2)}_{\phantom{|}} \; \Delta^{j-1}
\\
\SO_{\alpha_1|n+1-2j}^{4|(n-2j+1,1)|(\rho_1\rho_2)}(x,j)
&:=
\SP_{\alpha_1|n+1-2j}^{(n-2j+1,1)|(\rho_1}\;\D^{\rho_2)}_{\phantom{|}} \; \Delta^{j-1}
\intertext{and three tensor operators}
\SO_{(\alpha_1\alpha_2)|n-2j}^{5|(n-2j+2)|(\rho_1\rho_2)}(x,j)
&:=
\SP_{(\alpha_1\alpha_2)|n-2j}^{(n-2j+2)|(\rho_1\rho_2)} \;\; \Delta^{j}
\\
\SO_{(\alpha_1\alpha_2)|n-2j}^{6|(n-2j+1,1)|(\rho_1\rho_2)}(x,j)
&:=
\SP_{(\alpha_1\alpha_2)|n-2j}^{(n-2j+1,1)|(\rho_1\rho_2)} \;\; \Delta^{j}
\\
\SO_{(\alpha_1\alpha_2)|n-2j}^{7|(n-2j,2)|(\rho_1\rho_2)}(x,j)
&:=
\SP_{(\alpha_1\alpha_2)|n-2j}^{(n-2j,2)|(\rho_1\rho_2)} \;\; \Delta^{j} \; .
\end{align}
Expanding all operators $\x_{\alpha}$ in the latter decomposition (\ref{SPDSTensorI}) we get
\begin{eqnarray}
\label{SPDSTensorII}
&&
\SPD_{(\alpha_1\alpha_2)|n}^{(\rho_1\rho_2)} =
\\
\nonumber
&&
\sum_{j=0}^{\kle{\frac{n+2}{2}}}\frac{\kln{n+h-1-2j}!}{4^j j! \kln{n+h-1-j}!} \KLSSo{}
\\
\nonumber
&&
\quad 
  \kln{x^2}^j \KLE{ 
  \SO_{(\alpha_1\alpha_2)|n-2j}^{5|(n-2j+2)|(\rho_1\rho_2)}
+ \SO_{(\alpha_1\alpha_2)|n-2j}^{6|(n-2j+1,1)|(\rho_1\rho_2)}
+ \SO_{(\alpha_1\alpha_2)|n-2j}^{7|(n-2j,2)|(\rho_1\rho_2)} }
\\
\nonumber
&&
\quad 
- \frac{4j}{(n+h-j)(n+2h-2j)} \cdot \kln{x^2}^{j-1} \KLEo{}
\\
\nonumber
&&
\qquad \quad \; \;
  x^2 \, \SO_{(\alpha_1\alpha_2)|n-2j}^{4|(n-2j+1,1)|(\rho_1\rho_2)}
- 2\kln{n+h-2j} \cdot x_{(\alpha_1}^{\phantom{|}}\SO_{\alpha_2)|n+1-2j}^{4|(n-2j+1,1)|(\rho_1\rho_2)} \oKLE{}
\\
\nonumber
&&
\quad 
+ \frac{2j(n+h+1-2j)}{(h-1)(n+h-j)(n+h+2-2j)(n+2h-2j)} \cdot \kln{x^2}^{j-1} \KLEo{}
\\
\nonumber
&&
\qquad \quad \; \;
(n+h-2j) \cdot x^2 \, \SO_{(\alpha_1\alpha_2)|n-2j}^{1|(n-2j+2)|(\rho_1\rho_2)}
\\
\nonumber
&&
\qquad \quad
 -2(n+h-2j) \kln{ 
\kln{n+h-2j} \cdot x_{(\alpha_1}^{\phantom{|}}\SO_{\alpha_2)|n+1-2j}^{1|(n-2j+2)|(\rho_1\rho_2)}
+ \delta_{\alpha_1\alpha_2} \, \SO_{n+2-2j}^{0|(n-2j+2)|(\rho_1\rho_2)}}
\\
\nonumber
&&
\qquad \quad
- 2h \kln{ x^2 \; \SO_{(\alpha_1\alpha_2)|n-2j}^{3|(n-2j+2)|(\rho_1\rho_2)}
- 2  \kln{n+h-2j} \cdot x_{(\alpha_1}^{\phantom{|}}\SO_{\alpha_2)|n+1-2j}^{3|(n-2j+2)|(\rho_1\rho_2)} } \oKLE{}
\\
\nonumber
&&
\quad 
+ \frac{2j(n+h-2j)(n+h+1-2j)}{(h-1)(n+h-j)} \cdot \kln{x^2}^{j-1} \, \delta_{\alpha_1\alpha_2} \, \SO_{n+2-2j}^{1|(n-2j+2)|(\rho_1\rho_2)}
\\
\nonumber
&&
\quad 
+ \frac{4j(j-1)(n+h-2j)(n+h-j)^{-1}(n+2h-2j)^{-1}}{(n+h+1-j)(n+h+2-2j)(n+2h+1-2j)} \cdot \kln{x^2}^{j-2 } \KLEo{}
\\
\nonumber
&&
\qquad \quad \; \;
\kln{x^2}^2 \; \SO_{(\alpha_1\alpha_2)|n-2j}^{2|(n-2j+2)|(\rho_1\rho_2)}
\\
\nonumber
&&
\qquad \quad
- 2\kln{n+h+1-2j} \cdot x^2 \, \kln{ 2 \cdot x_{(\alpha_1}^{\phantom{|}}\SO_{\alpha_2)|n+1-2j}^{2|(n-2j+2)|(\rho_1\rho_2)}
+ \, \delta_{\alpha_1\alpha_2}\SO_{n+2-2j}^{2|(n-2j+2)|(\rho_1\rho_2)} }
\\
\nonumber
&&
\qquad \quad
+ 4\kln{n+h+1-2j}\kln{n+h+2-2j} \cdot x_{\alpha_1}x_{\alpha_2}\SO_{n+2-2j}^{2|(n-2j+2)|(\rho_1\rho_2)} \oKLE{} \oKLSS{}
\end{eqnarray}
with the additional differentiated operators
\begin{eqnarray}
\SO_{\alpha_2|n+1-2j}^{2|(n-2j+2)|(\rho_1\rho_2)}(x,j)
&:=&
\partial_{\alpha_2\phantom{|}}^{\phantom{|}} \!
\SO_{n+2-2j}^{2|(n-2j+2)|(\rho_1\rho_2)}(x,j)
\\
\SO_{(\alpha_1\alpha_2)|n-2j}^{1-3|(n-2j+2)|(\rho_1\rho_2)}(x,j)
&:=&
\partial_{(\alpha_1\phantom{|}}^{\phantom{|}} \!
\SO_{\alpha_2)|n+1-2j}^{1-3|(n-2j+2)|(\rho_1\rho_2)}(x,j)
\\
\SO_{(\alpha_1\alpha_2)|n-2j}^{4|(n-2j+1,1)|(\rho_1\rho_2)}(x,j)
&:=&
\partial_{(\alpha_1\phantom{|}}^{\phantom{|}} \!
\SO_{\alpha_2)|n+1-2j}^{4|(n-2j+1,1)|(\rho_1\rho_2)}(x,j) \; .
\end{eqnarray}
Taking the on--cone limit of the result (\ref{SPDSTensorII}) we obtain the complete spin decomposition on the complex cone for symmetric second rank tensors
\begin{eqnarray}
\label{SPDSTensorOnCone}
&&
\widetilde \SPD_{(\alpha_1\alpha_2)|n}^{(\rho_1\rho_2)} =
\tilde \SO_{(\alpha_1\alpha_2)|n}^{5|(n+2)|(\rho_1\rho_2)}(\tilde x,0)
+\tilde \SO_{(\alpha_1\alpha_2)|n}^{6|(n+1,1)|(\rho_1\rho_2)}(\tilde x,0)
+\tilde \SO_{(\alpha_1\alpha_2)|n}^{7|(n,2)|(\rho_1\rho_2)}(\tilde x,0)
\\
\nonumber
&&
\qquad 
+ \frac{2}{(n+h-1)(n+2h-2)} \cdot \tilde x_{(\alpha_1}^{\phantom{|}}\tilde \SO_{\alpha_2)|n-1}^{4|(n-1,1)|(\rho_1\rho_2)}(\tilde x,1)
\\
\nonumber
&&
\qquad 
- \frac{1}{(h-1)(n+h)(n+2h-2)} \KLNo{} \kln{n+h-2} \cdot \tilde x_{(\alpha_1}^{\phantom{|}}\tilde \SO_{\alpha_2)|n-1}^{1|(n)|(\rho_1\rho_2)}(\tilde x,1)
\\
\nonumber
&&
\qquad \qquad \qquad \qquad \qquad \qquad \qquad \qquad
+ \, \delta_{\alpha_1\alpha_2} \tilde \SO_{n}^{0|(n)|(\rho_1\rho_2)}(\tilde x,1)
- 2h \cdot \tilde x_{(\alpha_1}^{\phantom{|}}\tilde \SO_{\alpha_2)|n-1}^{3|(n)|(\rho_1\rho_2)}(\tilde x,1) \oKLN{}
\\
\nonumber
&&
\qquad 
+\frac{1}{2\kln{h-1}} \cdot \delta_{\alpha_1\alpha_2} \tilde \SO_{n}^{1|(n)|(\rho_1\rho_2)}(\tilde x,1)
\\
\nonumber
&&
\qquad 
+\frac{1}{(n+h-1)(n+h-2)(n+2h-3)(n+2h-4)} \cdot
\tilde x_{\alpha_1}^{\phantom{|}}\tilde x_{\alpha_2}^{\phantom{|}}\tilde \SO_{n-2}^{2|(n-2)|(\rho_1\rho_2)}(\tilde x,2) \; .
\end{eqnarray}
If we sum the antisymmetric decomposition $\SPD_{[\alpha_1\alpha_2]|n}^{[\rho_1\rho_2]}$ given by (\ref{SPDASTensorII}) and the symmetric decomposition $\SPD_{(\alpha_1\alpha_2)|n}^{(\rho_1\rho_2)}$ given by (\ref{SPDSTensorII}) and contract with $x^{\alpha_2}$ we reobtain the vector decomposition (\ref{SPDVectorII}) with $n \ra n+1$ after some calculation. Contacting the symmetric decomposition (\ref{SPDSTensorII}) with $\delta^{\alpha_1\alpha_2}$ we arrive at the scalar decomposition~(\ref{ScalarDec}). This holds also for the respective on--cone decompositions (\ref{SPDASTensorOnCone}), (\ref{SPDSTensorOnCone}) and (\ref{SPDVectorOnCone}).

The application of our algorithm is not limited by $k=2$ indices. The complete decomposition into irreducible components for the generic local tensor $\OP_{\alpha_1\alpha_2\alpha_3(\zeta_1\dots\zeta_n)}$ can be found in Ref.~\cite{E04}. There, all Young symmetries $\kln{1,1,1}$, $\kln{2,1}$ and $\kln{3}$ of the indices $\alpha_1$ to $\alpha_3$ are treated.

\section{Application to QCD operators}
\label{ApplicationQCD}

In our final Section we will return to the initial task of constructing complete twist decompositions for the different QCD operators given in Section~\ref{Compton}. We will thereby suppress the flavour and axial structures of these operators since they are not relevant for the respective decompositions. Furthermore, we will restrict our considerations to the first part of the centered operators. The second part can be obtained by substituting $\kappa$ by $-\kappa$.

The equivalent group theoretical problem of decomposing these local operators into $SO(1,3;\R)$--irreducible components has been solved in the two preceeding Sections. According to Ref.~\cite{BR77} Chapter 8 any complex--analytic irreducible representation of $SO(2h;\C)$ subduces a real irreducible representation of any real form of this group. Since the Lorentz group $SO(1,2h-1;\R)$ is such a real form of the complex orthogonal group we can straightforwardly apply all results of the preceeding sections to obtain complete decompositions into $SO(1,2h-1;\R)$--irreducible tensor components.

The change of the metric from $\delta_{\alpha_i\alpha_j}$ to $g_{\alpha_i\alpha_j}$ also changes the Laplacian $\Delta$ to the d'Alembert operator $\Box$. In a second step we reduce to four space--time dimensions, i.e. we set $h=2$. 

\subsubsection*{Scalar}

As a first and simple application of complete decompositions into irreducible components we will decompose the contracted local quark--antiquark operator $O_{n+1}$ given by
\begin{align}
\label{OScalarExample}
O_{n+1}
&=
x^{\alpha_1} \, X^{\zeta_1\cdots\zeta_n} \; \,
\bar \psi\kln{y} \; \gamma_{(\alpha_1} \; \DA_{\zeta_1} \cdots \DA_{\zeta_n)} \psi\kln{y} \PIpe{y=0}
\intertext{into its twist parts. To obtain the respective nonlocal operator $O(-\kappa\,x,\kappa\,x)$ we have to sum according to (\ref{SummationToNonlocalOperators}) and obtain}
O(-\kappa\,x,\kappa\,x)
&=
\bar\psi\kln{-\kappa \, x} \; \kln{x\gamma} \; U\kln{-\kappa \, x,\kappa \, x} \psi\kln{\kappa \, x} \; .
\intertext{In four space--time dimensions the complete twist decomposition off the light~cone reads}
\label{OScalarExampleDec}
O_{n+1} 
&=
\sum_{j=0}^{\kle{\frac{n+1}{2}}}\frac{\kln{n+2-2j}!}{4^j j! \kln{n+2-j}!} \cdot \kln{x^2}^j \, O_{n+1-2j}^{\tw(2+2j)}  
\intertext{with}
O_{n+1-2j}^{\tw(2+2j)}
&:=
\SO_{n+1-2j}^{(n-2j+1)} \, O_{n+1}
\\
&=
H_{n+1-2j} \; \Box^j \; O_{n+1} \; .
\end{align}
According to the definition~(\ref{DefinitionOfTwist}) of geometric twist as canonical dimension minus Lorentz spin the twist of $O_{n+1-2j}^{\tw(2+2j)}$ is calculated as $\tau = d_n - j_n = (n+3) - (n-2j+1) = 2+2j$.

The decomposition~(\ref{OScalarExample}) is a direct consequence of the results of Bargmann and Todorov~\cite{BT77} and has first been formulated in Ref.~\cite{GLR01} in this form. The local operator $N_{n} = X^{\zeta_1\cdots\zeta_n} \; \,
\bar \psi\kln{y} \; \DA_{(\zeta_1} \cdots \DA_{\zeta_n)} \psi\kln{y} \PIpe{y=0}$ possesses a twist decomposition which is completely analogous to (\ref{OScalarExampleDec}). All twists are to be shifted by plus one since the order of the operator is reduced by one. The result is
\begin{align}
\label{NScalarExampleDec}
N_{n} 
&=
\sum_{j=0}^{\kle{\frac{n}{2}}}\frac{\kln{n+1-2j}!}{4^j j! \kln{n+1-j}!} \cdot \kln{x^2}^j \, N_{n-2j}^{\tw(3+2j)}  
\intertext{with}
\label{NDef}
N_{n-2j}^{\tw(3+2j)}
&:=
\SO_{n-2j}^{(n-2j)} \, N_{n}
\\
&=
H_{n-2j} \; \Box^j \; N_{n} \; .
\end{align}
This decomposition for $N_n$ can also be found in Ref.~\cite{GLR01}. A final example for the scalar twist decomposition is the local operator $G_{n} = X^{\zeta_1\cdots\zeta_n} \; \, F^{\sigma\rho}\kln{y} \; \DA_{(\zeta_1} \cdots \DA_{\zeta_n)} F_{\sigma\rho}\kln{y} \PIpe{y=0}$. Due to the canonical dimension of $n+4$ of $G_n$ all twists are to be shifted by plus one in (\ref{NScalarExampleDec}) and (\ref{NDef}) if $N$ is replaced by $G$.

The complete (infinite) twist decompositions of the operators $O(-\kappa\,x,\kappa\,x)$ and $N(-\kappa\,x,\kappa\,x)$ are given in nonlocal form in Ref.~\cite{GLR01}.

\subsubsection*{Vector}

As an example for the vector decomposition we will now apply the results of the Section~\ref{spindecompositions} to the local quark--antiquark operator $O_{\alpha_1|n}$ given by
\begin{align}
\label{OExample}
O_{\alpha_1|n}
&=
X^{\zeta_1\cdots\zeta_n} \; \,
\bar \psi\kln{y} \; \gamma_{\alpha_1} \; \DA_{(\zeta_1} \cdots \DA_{\zeta_n)} \psi\kln{y} \PIpe{y=0} \; .
\intertext{This local operator generates, when summed according to (\ref{SummationToNonlocalOperators}), the nonlocal QCD operator
\newline
$O_{\alpha_1}\kln{-\kappa\,x,\kappa\,x}$ given by}
O_{\alpha_1}\kln{-\kappa\,x,\kappa\,x}
&= 
\bar\psi\kln{-\kappa \, x} \; \gamma_{\alpha_1} \; U\kln{-\kappa \, x,\kappa \, x} \psi\kln{\kappa \, x} \; .
\end{align}
To obtain the complete twist decomposition for $O_{\alpha_1|n}$ we have to restrict all relevant results of Section~\ref{spindecompositions} to four space--time dimensions and define local operators of definite geometric twist with the help of the spin operators $\SPO_{\alpha_1\dots\alpha_l|n}^{i|(\m)|\rho_1}$ with $0\les l \les 1$.

If we observe that the canonical dimension of the local operator (\ref{OExample}) (without the $x^{\zeta_i}$'s) is given by $n+3$ and read off the Lorentz spin of the Young pattern
$(\m)$ according to (\ref{SpinsI}) to (\ref{SpinsIII}) we can immediately define local operators of definite geometric
twist off the light cone. This is done by applying the operators given $\SPO$ by (\ref{VectorOPSI}) to (\ref{VectorOPSIII}) and (\ref{VectorOPSIV}) to $O_{\alpha_1|n}$. Accordingly, we get one scalar part
\begin{align}
\label{ODefTwistOpI}
O^{1|\tw\kln{2+2j}}_{n+1-2j}
&:=
\SPO_{n+1-2j}^{1|(n-2j+1)|\rho_1} \; O_{\rho_1|n}
\intertext{and three vector parts}
O^{1|\tw\kln{2+2j}}_{\alpha_1|n-2j}
&:=
\SPO_{\alpha_1|n-2j}^{1|(n-2j+1)|\rho_1} \; O_{\rho_1|n}
\\
O^{2|\tw\kln{2+2j}}_{\alpha_1|n-2j}
&:=
\SPO_{\alpha_1|n-2j}^{2|(n-2j+1)|\rho_1} \; O_{\rho_1|n}
\\
\label{ODefTwistOpIV}
O^{3|\tw\kln{3+2j}}_{\alpha_1|n-2j}
&:=
\SPO_{\alpha_1|n-2j}^{3|(n-2j,1)|\rho_1} \; O_{\rho_1|n} \; .
\end{align}
Here, one has to replace the Laplacian $\Delta$ by the d'Alembert operator $\Box$ and set $h=2$ in the spin operators $\SPO_{\alpha_1\dots\alpha_l|n}^{i|(\m)|\rho_1}$. In explicit form the above four operators read
\begin{eqnarray}
O^{1|\tw\kln{2+2j}}_{n+1-2j}
&=&
H_{n+1-2j} \; \D^{\rho_1} \; \Box^{j-1} \, O_{\rho_1|n}
\\
O^{1|\tw\kln{2+2j}}_{\alpha_1|n-2j}
&=&
\pd_{\alpha_1}^{\phantom{|}} \; H_{n+1-2j} \; \D^{\rho_1} \; \Box^{j-1} \, O_{\rho_1|n}
\\
O^{2|\tw\kln{2+2j}}_{\alpha_1|n-2j}
&=&
\frac{1}{(n+1-2j)^2} \cdot  H_{n-2j} \; \D_{\alpha_1} \, x^{\rho_1} \; \Box^{j} \, O_{\rho_1|n}
\\
O^{3|\tw\kln{3+2j}}_{\alpha_1|n-2j}
&=&
H_{n-2j} \kln{\delta_{\alpha_1}^{\rho_1} - \frac{1}{(n+1-2j)^2}
\Kls{ x_{\alpha_1} \, \D^{\rho_1} + \D_{\alpha_1} \, x^{\rho_1} } } \; \Box^{j} \, O_{\rho_1|n}
\end{eqnarray}
The complete twist decomposition of $O_{\alpha_1|n}$ can now be directly obtained from (\ref{SPDVectorII}) for $h=2$ and reads
\begin{eqnarray}
\label{OIIEx}
O_{\alpha_1|n}
&=&
\sum_{j=0}^{\kle{\frac{n+1}{2}}}\frac{\kln{n+1-2j}!}{4^j j! \kln{n+1-j}!} \KLSSo{}
\kln{x^2}^j \KLE{ O^{2|\tw\kln{2+2j}}_{\alpha_1|n-2j} + O^{3|\tw\kln{3+2j}}_{\alpha_1|n-2j} }
\\ 
\nonumber
&&
- \, \frac{2j}{(n+2-j)(n+3-2j)} \cdot \kln{x^2}^{j-1 }
\KLEo{} x^2 \, O^{1|\tw\kln{2+2j}}_{\alpha_1|n-2j} - 2 \kln{n+2-2j} \cdot x_{\alpha_1} \, O^{1|\tw\kln{2+2j}}_{n+1-2j} \oKLE{}  \oKLSS{}.
\end{eqnarray}
This complete twist decomposition off the light cone for the local quark--antiquark operator $O_{\alpha_1|n}$ has first been given in an equivalent form by the relation (84) in Ref.~\cite{EGL04}. There, also a summation to a nonlocal form is performed. Contracting the vector decomposition (\ref{OIIEx}) with $x^{\alpha_1}$ we, of course, arrive at the scalar decomposition of $O_{n+1}$ given by (\ref{OScalarExampleDec}) after some calculation.  

On the light cone we define operators of definite geometric twist in complete analogy to (\ref{ODefTwistOpI}) to (\ref{ODefTwistOpIV}).
For $h=2$ we can read off from (\ref{SPDVectorOnCone}) the related complete twist decomposition on the light cone
\begin{align}
\label{SPDOOnCone}
\tilde O_{\alpha_1|n}
&=
\tilde O_{\alpha_1|n}^{2|\tw(2)} + \tilde O_{\alpha_1|n}^{3|\tw(3)}
+ \frac{1}{(n+1)^2} \cdot \tilde x_{\alpha_1} \; \tilde O_{n-1}^{1|\tw(4)}
\intertext{with the operators}
\tilde O^{1|\tw\kln{4}}_{n-1}
&=
\tilde \D^{\rho_1} \, \tilde O_{\rho_1|n}
\\
\tilde O^{2|\tw\kln{2}}_{\alpha_1|n}
&=
\frac{1}{(n+1)^2} \cdot  \tilde \D_{\alpha_1} \, \tx^{\rho_1} \, \tilde O_{\rho_1|n}
\\
\tilde O^{3|\tw\kln{3}}_{\alpha_1|n}
&=
\kln{\delta_{\alpha_1}^{\rho_1} - \frac{1}{(n+1)^2}
\Kls{ \tx_{\alpha_1} \, \tilde \D^{\rho_1} + \tilde \D_{\alpha_1} \, \tx^{\rho_1} } } \, \tilde O_{\rho_1|n} \; .
\end{align}
The on--cone decomposition (\ref{SPDOOnCone}) of $\tilde O_{\alpha_1|n}$ was first given in Ref.~\cite{GL01} in this form. A preliminary form not making use of the interior derivative $\D_{\alpha_1}$ can be found in Ref.~\cite{GLR99}.

The complete decompositions off and on the light~cone for the partly contracted local Gluon operator
\begin{equation}
G_{\alpha_1|n+1} = \delta^{(\beta_1}_{\alpha_1} x^{\beta_2)} \, X^{\zeta_1\cdots\zeta_n} \; \, {F^{\sigma}}_{\beta_1}\kln{y} \; \DA_{(\zeta_1} \cdots \DA_{\zeta_n)} F_{\beta_2\sigma}\kln{y} \PIpe{y=0}
\end{equation}
can be read off from the respective decompositions for the quark--antiquark operator given by (\ref{OIIEx}) and (\ref{SPDOOnCone}) with $n \ra n+1$. The twists of the different operators are unchanged since both the canonical dimension and the spins are shifted by plus one if $O_{\alpha_1|n}$ is replaced by $G_{\alpha_1|n+1}$.

As a third example for the complete vector decomposition (\ref{SPDVectorII}) we will apply it to the local operator $M_{\alpha_1|n+1}$ given by
\begin{align}
\label{MExample}
M_{\alpha_1|n+1}
&=
x^{\alpha_2} \; X^{\zeta_1\cdots\zeta_n} \; \,
\bar \psi\kln{y} \; \sigma_{[\alpha_1(\alpha_2]} \; \DA_{\zeta_1\phantom{|}} \cdots \DA_{\zeta_n)} \psi\kln{y} \PIpe{y=0} \; .
\intertext{This local operator generates, when summed up, the nonlocal QCD operator $M_{\alpha_1}\kln{-\kappa\,x,\kappa\,x}$ given by}
M_{\alpha_1}\kln{-\kappa\,x,\kappa\,x}
&= 
x^{\alpha_2} \; \bar\psi\kln{-\kappa \, x} \; \sigma_{[\alpha_1\alpha_2]} \; U\kln{-\kappa \, x,\kappa \, x} \psi\kln{\kappa \, x} \; .
\end{align}
The canonical dimension of the local operator (\ref{MExample}) is again given by $n+3$. We read off the Lorentz spin of the involved Young patterns and define in complete analogy to (\ref{ODefTwistOpI}) to (\ref{ODefTwistOpIV}) local operators of definite geometric twist off the light cone by
\begin{align}
\label{MDefTwistOpI}
M^{1|\tw\kln{1+2j}}_{n+2-2j}
&:=
\SPO_{n+2-2j}^{1|(n-2j+2)|\rho_1} \; M_{\rho_1|n+1}
\\
M^{1|\tw\kln{1+2j}}_{\alpha_1|n+1-2j}
&:=
\SPO_{\alpha_1|n+1-2j}^{1|(n-2j+2)|\rho_1} \; M_{\rho_1|n+1}
\\
M^{2|\tw\kln{1+2j}}_{\alpha_1|n+1-2j}
&:=
\SPO_{\alpha_1|n+1-2j}^{2|(n-2j+2)|\rho_1} \; M_{\rho_1|n+1}
\\
\label{MDefTwistOpIV}
M^{3|\tw\kln{2+2j}}_{\alpha_1|n+1-2j}
&:=
\SPO_{\alpha_1|n+1-2j}^{3|(n-2j+1,1)|\rho_1} \; M_{\rho_1|n+1} \; .
\end{align}
Comparing with (\ref{ODefTwistOpI}) to (\ref{ODefTwistOpIV}) all spins are shifted by plus one which results from the additional symmetrized index $\alpha_2$. Accordingly, all twists in (\ref{MDefTwistOpI}) to (\ref{MDefTwistOpIV}) are to be shifted by minus one since the canonical dimensions of the local operators $O_{\alpha_1|n}$ and $M_{\alpha_1|n+1}$ are equal.

Due to the internal antisymmetry of $M_{\alpha_1|n+1}$ it holds $x^{\alpha_1} M_{\alpha_1|n+1} = 0$. As a consequence, the operators $M^{1|\tw\kln{1+2j}}_{\alpha_1|n+1-2j}$ and $M^{2|\tw\kln{1+2j}}_{\alpha_1|n+1-2j}$ are equal up to a factor
\begin{equation}
\label{MRelASym}
M^{2|\tw\kln{1+2j}}_{\alpha_1|n+1-2j}
=
- \frac{2j}{(n+3-j)(n+2-2j)} \cdot M^{1|\tw\kln{1+2j}}_{\alpha_1|n+1-2j} \; .
\end{equation}
With respect to the latter relation (\ref{MRelASym}) the complete twist decomposition of the local operator $M_{\alpha_1|n+1}$ is obtained from the generic decomposition (\ref{SPDVectorII}) for $n \ra n+1$ and $h=2$ and reads
\begin{eqnarray}
\label{MDec}
M_{\alpha_1|n+1}
&=&
\sum_{j=0}^{\kle{\frac{n+2}{2}}}\frac{\kln{n+2-2j}!}{4^j j! \kln{n+2-j}!} \KLSSo{}
\kln{x^2}^j \, M^{3|\tw\kln{2+2j}}_{\alpha_1|n+1-2j} 
\\ 
\nonumber
&&
- \frac{4j\kln{n+3-2j}}{\kln{n+3-j}\kln{n+4-2j}} \cdot \kln{x^2}^{j-1 }
\KLEo{} \frac{1}{n+2-2j} \cdot x^2 \,  M^{1|\tw\kln{1+2j}}_{\alpha_1|n+1-2j} - \, x_{\alpha_1} \, M^{1|\tw\kln{1+2j}}_{n+2-2j} \oKLE{}  \oKLSS{}
\end{eqnarray}
This complete twist decomposition off the light cone for $M_{\alpha_1|n+1}$ has first been given in an equivalent form by the relation (120) in Ref.~\cite{EGL04} together with a summation to nonlocal operators. Differences to this reference appear due to the usage of the interior derivative $\D^{\rho_1}$ in the operators $M^{1|\tw\kln{1+2j}}_{n+2-2j}$ and $M^{1|\tw\kln{1+2j}}_{\alpha_1|n+1-2j}$.

On the light cone the complete decomposition into operators of definite geometric twist is easily obtained from the off--cone decomposition (\ref{MDec}) or the generic form (\ref{SPDVectorOnCone}) and reads
\begin{align}
\label{SPDMOnCone}
\tilde M_{\alpha_1|n+1}
&=
\tilde M_{\alpha_1|n+1}^{3|\tw(2)}
+ \frac{1}{(n+2)^2} \cdot \tilde x_{\alpha_1} \; \tilde M_{n}^{1|\tw(3)}
\intertext{with the on--cone operators}
\tilde M^{1|\tw\kln{3}}_{n}
&=
\tilde \D^{\rho_1} \, \tilde M_{\rho_1|n+1}
\\
\tilde M^{3|\tw\kln{2}}_{\alpha_1|n+1}
&=
\kln{\delta_{\alpha_1}^{\rho_1} - \frac{1}{(n+2)^2} \cdot \tx_{\alpha_1} \, \tilde \D^{\rho_1} } \, \tilde M_{\rho_1|n+1 } \; .
\end{align}
In this form, the decomposition of $\tilde M_{\alpha_1|n+1}$ on the light cone was first given in Ref.~\cite{GL01}. A preliminary form can be found in Ref.~\cite{GLR99}.

\subsubsection*{Antisymmetric second rank tensor}

We will now apply the generic decomposition in the antisymmetric second rank tensor case given by (\ref{SPDASTensorII}) to the local quark mass operator $M_{[\alpha_1\alpha_2]}$ given by
\begin{align}
\label{MExampleAS}
M_{[\alpha_1\alpha_2]|n}
&=
X^{\zeta_1\cdots\zeta_n} \; \,
\bar \psi\kln{y} \; \sigma_{[\alpha_1\alpha_2]} \; \DA_{(\zeta_1} \cdots \DA_{\zeta_n)} \psi\kln{y} \PIpe{y=0} \; .
\intertext{This local operator is the uncontracted form of $M_{\alpha_1|n+1}$ given by (\ref{MExample}) and generates, when summed up, the nonlocal QCD operator $M_{[\alpha_1\alpha_2]}\kln{-\kappa\,x,\kappa\,x}$ given by}
M_{[\alpha_1\alpha_2]}\kln{-\kappa\,x,\kappa\,x}
&= 
\bar\psi\kln{-\kappa \, x} \; \sigma_{[\alpha_1\alpha_2]} \; U\kln{-\kappa \, x,\kappa \, x} \psi\kln{\kappa \, x} \; .
\end{align}
We define two vector operators of definite geometric twist
\begin{align}
M_{\alpha_2|n+1-2j}^{1|\tw(1+2j)} 
&:=
\SO_{\alpha_2|n+1-2j}^{1|(n-2j+2)|[\rho_1\rho_2]} \, M_{[\rho_1\rho_2]|n}
\\
M_{\alpha_2|n+1-2j}^{2|\tw(2+2j)}
&:=
\SO_{\alpha_2|n+1-2j}^{2|(n-2j+1,1)|[\rho_1\rho_2]} \, M_{[\rho_1\rho_2]|n}
\intertext{and three antisymmetric tensor operators}
M_{[\alpha_1\alpha_2]|n-2j}^{2|\tw(2+2j)}
&:=
\SO_{[\alpha_1\alpha_2]|n-2j}^{2|(n-2j+1,1)|[\rho_1\rho_2]} \, M_{[\rho_1\rho_2]|n}
\\
M_{[\alpha_1\alpha_2]|n-2j}^{3|\tw(2+2j)}´
&:=
\SO_{[\alpha_1\alpha_2]|n-2j}^{3|(n-2j+1,1)|[\rho_1\rho_2]} \, M_{[\rho_1\rho_2]|n}
\\
M_{[\alpha_1\alpha_2]|n-2j}^{4|\tw(3+2j)}
&:=
\SO_{[\alpha_1\alpha_2]|n-2j}^{4|(n-2j,1,1)|[\rho_1\rho_2]}  \, M_{[\rho_1\rho_2]|n} \; .
\end{align}
In explicit form the two vector operators read
\begin{align}
M_{\alpha_2|n+1-2j}^{1|\tw(1+2j)} 
&=
\frac{1}{(n+2-2j)^2} \cdot H_{n+1-2j} \; \D_{\alpha_2} \, x^{[\rho_1} \D^{\rho_2]} \, \Box^{j-1} \, M_{[\rho_1 \rho_2]|n}
\\
M_{\alpha_2|n+1-2j}^{2|\tw(2+2j)}
&=
H_{n+1-2j} \kln{\delta_{\alpha_2}^{[\rho_1} - \frac{1}{(n+2-2j)^2} \cdot
\D_{\alpha_2 } \, x^{[\rho_1}  }  \D^{\rho_2]} \, \Box^{j-1} \, M_{[\rho_1 \rho_2]|n} \; ,
\intertext{whereas the three tensor operators are of the form}
M_{[\alpha_1\alpha_2]|n-2j}^{2|\tw(2+2j)}
&=
\pd_{[\alpha_1}^{\phantom{|}} H_{n+1-2j} \kln{\delta_{\alpha_2]}^{[\rho_1} - \frac{1}{(n+2-2j)^2} \cdot
\D_{\alpha_2]}^{\phantom{|}} \, x^{[\rho_1} }  \D^{\rho_2]} \, \Box^{j-1} \, M_{[\rho_1 \rho_2]|n}
\\
M_{[\alpha_1\alpha_2]|n-2j}^{3|\tw(2+2j)}´
&=
- \frac{2}{(n+2-2j) (n+1-2j) } \cdot H_{n-2j} \KLSo{}
\D_{[\alpha_1}^{\phantom{|}}\delta_{\alpha_2]}^{[\rho_1} x^{\rho_2]}_{\phantom{|}}
\\
\nonumber
&
\qquad \qquad \qquad \qquad \qquad
- \frac{1}{(n-2j)^2} \cdot x_{[\alpha_1}\D_{\alpha_2]} x^{[\rho_1}\D^{\rho_2]} \oKLS{} \; \Box^j \, M_{[\rho_1\rho_2]|n}
\\
M_{[\alpha_1\alpha_2]|n-2j}^{4|\tw(3+2j)}
&=
H_{n-2j} \KLSo{} \delta_{[\alpha_1}^{[\rho_1}\delta_{\alpha_2]}^{\rho_2]}
+ \frac{2}{n+1-2j} \KLEo{} \frac{1}{n-2j} \cdot x_{[\alpha_1}^{\phantom{|}}\delta_{\alpha_2]}^{[\rho_1}\D^{\rho_2]}_{\phantom{|}}
+ \frac{1}{n+2-2j} \cdot \D_{[\alpha_1}^{\phantom{|}}\delta_{\alpha_2]}^{[\rho_1}x^{\rho_2]}_{\phantom{|}} \oKLE{}
\\
\nonumber
&
\qquad \qquad \;\;\,
- \frac{2}{(n+2-2j)(n-2j)^3} \cdot x_{[\alpha_1}\D_{\alpha_2]}x^{[\rho_1}\D^{\rho_2]} \oKLS{} \; \Box^j \, M_{[\rho_1 \rho_2]|n} .
\end{align}
The final complete twist decomposition of the local operator $M_{[\alpha_1\alpha_2]|n}$ can now be read off for $h=2$ from (\ref{SPDASTensorII})
\begin{eqnarray}
\label{SPDMExample}
M_{[\alpha_1\alpha_2]|n}
&=&
\sum_{j=0}^{\kle{\frac{n+2}{2}}} \frac{(n+1-2j)!}{4^j \, j! \, (n+1-j)!} \KLSSo{}
\\
\nonumber
&&
\qquad
\kln{x^2}^j \KLE{ M_{[\alpha_1\alpha_2]|n-2j}^{3|\tw(2+2j)} + M_{[\alpha_1\alpha_2]|n-2j}^{4|\tw(3+2j)} }  
\\
\nonumber
&&
\quad
+ \frac{8j}{n+2-j} \cdot 
\kln{x^2}^{j-1} \KLEo{} \frac{1}{2(n+2-2j)} \cdot x^2 \, M_{[\alpha_1\alpha_2]|n-2j}^{2|\tw(2+2j)}
- \, x_{[\alpha_1}^{\phantom{|}} M_{\alpha_2]|n+1-2j}^{2|\tw(2+2j)} 
\\
\nonumber
&&
\qquad\qquad\qquad\qquad\qquad\quad
- \frac{n+3-2j}{n+4-2j} \cdot x_{[\alpha_1}^{\phantom{|}} M_{\alpha_2]|n+1-2j}^{1|\tw(1+2j)} \oKLE{} \oKLSS{} \; .
\end{eqnarray}
On the light cone we deduce the result from the generic on--cone decomposition (\ref{SPDASTensorOnCone}) or take the on--cone limit of (\ref{SPDMExample}) and get
\begin{align}
\label{SPDMExampleOnCone}
\tilde M_{[\alpha_1\alpha_2]|n}
&=
 \tilde M_{[\alpha_1\alpha_2]|n}^{3|\tw(2)}
+\tilde M_{[\alpha_1\alpha_2]|n}^{4|\tw(3)}
-\frac{2}{n} \kle{ \frac{1}{n+1} \cdot
\tilde x_{[\alpha_1}^{\phantom{|}}\tilde M_{\alpha_2]|n-1}^{2|\tw(4)}
+\frac{1}{n+2} \cdot 
\tilde x_{[\alpha_1}^{\phantom{|}}\tilde M_{\alpha_2]|n-1}^{1|\tw(3)} }
\intertext{with}
\nonumber
\tilde M_{\alpha_2|n-1}^{1|\tw(3)} 
&=
\frac{1}{n^2} \cdot \tilde \D_{\alpha_2} \, \tx^{[\rho_1} \, \tilde \D^{\rho_2]} \; \tilde M_{[\rho_1 \rho_2]|n}
\\
\nonumber
\tilde M_{\alpha_2|n-1}^{2|\tw(4)}
&=
\kln{\delta_{\alpha_2}^{[\rho_1} - \frac{1}{n^2} \cdot
\tilde \D_{\alpha_2} \, \tx^{[\rho_1} }  \tilde \D^{\rho_2]} \; \tilde M_{[\rho_1 \rho_2]|n}
\intertext{and}
\tilde M_{[\alpha_1\alpha_2]|n}^{3|\tw(2)}
&=
- \frac{2}{(n+2) (n+1) } \KLSo{}
\tilde \D_{[\alpha_1}^{\phantom{|}}\delta_{\alpha_2]}^{[\rho_1} \tx^{\rho_2]}_{\phantom{|}}
- \frac{1}{n^2} \cdot \tx_{[\alpha_1} \tilde\D_{\alpha_2]} \tx^{[\rho_1} \tilde\D^{\rho_2]} \oKLS{} \; \tilde M_{[\rho_1\rho_2]|n}
\\
\tilde M_{[\alpha_1\alpha_2]|n}^{4|\tw(3)}
&=
\KLSo{} \delta_{[\alpha_1}^{[\rho_1}\delta_{\alpha_2]}^{\rho_2]}
+ \frac{2}{n+1} \KLEo{} \frac{1}{n} \cdot \tx_{[\alpha_1}^{\phantom{|}} \delta_{\alpha_2]}^{[\rho_1} \tilde\D^{\rho_2]}_{\phantom{|}}
+ \frac{1}{n+2} \cdot \tilde \D_{[\alpha_1}^{\phantom{|}} \delta_{\alpha_2]}^{[\rho_1} \tx^{\rho_2]}_{\phantom{|}} \oKLE{}
\\
\nonumber
&
\qquad \qquad \qquad \qquad \qquad \quad
- \frac{2}{n^3(n+2)} \cdot \tx_{[\alpha_1} \tilde\D_{\alpha_2]} \tx^{[\rho_1} \tilde\D^{\rho_2]} \oKLS{} \; \tilde M_{[\rho_1 \rho_2]|n} .
\end{align}
The antisymmetric part of the local gluon operator $G_{\alpha_1\alpha_2|n}$ given by
\begin{equation}
\label{GExampleAnti}
G_{\kle{\alpha_1\alpha_2}|n}
=
\delta_{[\alpha_1}^{[\beta_1} \delta_{\alpha_2]}^{\beta_2]} \; \,
X^{\zeta_1\cdots\zeta_n} \;
{F^\sigma}_{\!\!\beta_1} \kln{y} \; \DA_{(\zeta_1} \cdots \DA_{\zeta_n)} F_{\beta_2\sigma} \kln{y} \PIpe{y=0} \; .
\end{equation}
possesses a twist decomposition which is completely analogous to (\ref{SPDMExample}) and (\ref{SPDMExampleOnCone}).
One only has to observe that the canonical dimension of $G_{\alpha_1\alpha_2|n}$ is $n+4$.
Thereby, all twists in the decompositions (\ref{SPDMExample}) and (\ref{SPDMExampleOnCone}) are shifted by plus one if $M$ is replaced by $G$.

The on--cone twist decomposition (\ref{SPDMExampleOnCone}) has already been obtained in Ref.~\cite{GLR99,L02} together with the related
summation to nonlocal operators. The new result (\ref{SPDMExample}) extends this known twist decomposition of $\tilde M_{[\alpha_1\alpha_2]|n}$ on the
light cone to the complete twist decomposition of the local operators $M_{\kle{\alpha_1\alpha_2}|n}$ and $G_{[\alpha_1\alpha_2]|n}$ off the light cone.

\subsubsection*{Symmetric second rank tensor}

In our final example we will apply the complete decomposition into irreducible tensors given by (\ref{SPDSTensorII}) in the symmetric second rank tensor case to the symmetric part of the gluon operator given by
\begin{align}
\label{GExampleSym}
G_{\kln{\alpha_1\alpha_2}|n}
&=
\delta_{(\alpha_1}^{(\beta_1} \delta_{\alpha_2)}^{\beta_2)} \; \,
X^{\zeta_1\cdots\zeta_n} \;
{F^\sigma}_{\!\!\beta_1} \kln{y} \; \DA_{(\zeta_1} \cdots \DA_{\zeta_n)} F_{\beta_2\sigma} \kln{y} \PIpe{y=0} \; .
\intertext{After the summation this operator generates the nonlocal QCD operator}
G_{\kln{\alpha_1\alpha_2}}\kln{- \kappa \, x, \kappa \, x}
&= 
{F^\sigma}_{\!\!(\alpha_1} \kln{-\kappa \, x} \; U(-\kappa \, x, \kappa \, x) \; F_{\alpha_2)\sigma}\kln{\kappa \, x} \; .
\end{align}
Again, we restrict all relevant results of Section \ref{spindecompositions} to four space--time dimensions and define local operators of definite geometric twist with the help of the spin operators $\SPO_{\alpha_1\dots\alpha_l|n}^{i|(\m)|(\rho_1\rho_2)}$ with $0\les l \les 2$. Since the canonical dimension of the local Gluon operator (\ref{GExampleSym}) is given by $n+4$ and the Lorentz spin of the Young patterns $(\m)$ is defined by (\ref{SpinsI}) to (\ref{SpinsIV}) we can immediately define local operators of definite geometric twist off the light cone.

First, we define three scalar operators
\begin{align}
\label{GOpI}
G_{n+2-2j}^{0|\tw(2+2j)}
&:=
\SO_{n+2-2j}^{0|(n-2j+2)|(\rho_1\rho_2)} \; G_{(\rho_1\rho_2)|n}
\\
G_{n+2-2j}^{1|\tw(2+2j)}
&:=
\SO_{n+2-2j}^{1|(n-2j+2)|(\rho_1\rho_2)} \; G_{(\rho_1\rho_2)|n}
\\
G_{n+2-2j}^{2|\tw(2+2j)}
&:=
\SO_{n+2-2j}^{2|(n-2j+2)|(\rho_1\rho_2)} \; G_{(\rho_1\rho_2)|n} \; ,
\intertext{two vector operators}
G_{\alpha_1|n+1-2j}^{3|\tw(2+2j)}
&:=
\SO_{\alpha_1|n+1-2j}^{3|(n-2j+2)|(\rho_1\rho_2)} \; G_{(\rho_1\rho_2)|n}
\\
G_{\alpha_1|n+1-2j}^{4|\tw(3+2j)}
&:=
\SO_{\alpha_1|n+1-2j}^{4|(n-2j+1,1)|(\rho_1\rho_2)} \; G_{(\rho_1\rho_2)|n}
\intertext{and three tensor operators}
G_{ (\alpha_1\alpha_2) |n-2j}^{5|\tw(2+2j)}
&:=
\SO_{(\alpha_1\alpha_2)|n-2j}^{5|(n-2j+2)|(\rho_1\rho_2)} \; G_{(\rho_1\rho_2)|n}
\\
G_{ (\alpha_1\alpha_2) |n-2j}^{6|\tw(3+2j)}
&:=
\SO_{(\alpha_1\alpha_2)|n-2j}^{6|(n-2j+1,1)|(\rho_1\rho_2)} \; G_{(\rho_1\rho_2)|n}
\\
\label{GOpEicht}
G_{ (\alpha_1\alpha_2) |n-2j}^{7|\tw(4+2j)}
&:=
\SO_{(\alpha_1\alpha_2)|n-2j}^{7|(n-2j,2)|(\rho_1\rho_2)} \; G_{(\rho_1\rho_2)|n}
\intertext{of definite geometric twist. Six additional differentiated operators are given by}
\label{GDiffI}
G_{\alpha_2|n+1-2j}^{1-2|\tw(2+2j)}
&:=
\SO_{\alpha_2|n+1-2j}^{1-2|(n-2j+2)|(\rho_1\rho_2)} \; G_{(\rho_1\rho_2)|n}
\\
G_{ (\alpha_1\alpha_2) |n-2j}^{1-3|\tw(2+2j)}
&:=
\SO_{(\alpha_1\alpha_2)|n-2j}^{1-3|(n-2j+2)|(\rho_1\rho_2)} \; G_{(\rho_1\rho_2)|n}
\\
\label{GDiffIII}
G_{ (\alpha_1\alpha_2) |n-2j}^{4|\tw(3+2j)}
&:=
\SO_{(\alpha_1\alpha_2)|n-2j}^{4|(n-2j+1,1)|(\rho_1\rho_2)} \; G_{(\rho_1\rho_2)|n} \; .
\end{align}
In the following we will give the eight operators (\ref{GOpI}) to (\ref{GOpEicht}) in explicit form. The remaining ones (\ref{GDiffI}) to (\ref{GDiffIII}) are obtained by differentiation.

The three scalar operator are given by
\begin{align}
G_{n+2-2j}^{0|\tw(2+2j)}
&=
H_{n+2-2j} \;\, \D^{(\rho_1} x^{\rho_2) } \; \Box^{j-1} \; G_{(\rho_1\rho_2)|n}
\\
G_{n+2-2j}^{1|\tw(2+2j)}
&=
H_{n+2-2j} \;\, g^{\rho_1\rho_2 } \; \Box^{j-1} \; G_{(\rho_1\rho_2)|n}
\\
G_{n+2-2j}^{2|\tw(2+2j)}
&=
H_{n+2-2j} \;\, \D^{\rho_1} \D^{\rho_2 } \; \Box^{j-2} \; G_{(\rho_1\rho_2)|n} \; .
\intertext{The two vector operators read}
\nonumber
G_{\alpha_2|n+1-2j}^{3|\tw(2+2j)} 
&=
\frac{1}{(n+2-2j)^2} \cdot H_{n+1-2j} \;\, \D_{\alpha_2} \, x^{(\rho_1} \D^{\rho_2)} \; \Box^{j-1} \, G_{(\rho_1 \rho_2)|n}
\\
\nonumber
G_{\alpha_2|n+1-2j}^{4|\tw(3+2j)}
&=
H_{n+1-2j} \kln{\delta_{\alpha_2}^{(\rho_1} - \frac{1}{(n+2-2j)^2}
\KLs{ x_{\alpha_2} \, \D^{(\rho_1} + \D_{\alpha_2} \, x^{(\rho_1} } }  \D^{\rho_2)} \; \Box^{j-1} \, G_{(\rho_1 \rho_2)|n}
\intertext{and the three tensor operators are of the form}
\nonumber
G_{(\alpha_1\alpha_2)|n-2j}^{5|\tw(2+2j)} 
&=
\frac{1}{(n+1-2j)^2(n+2-2j)^2} \cdot H_{n-2j} \;\, \D_{\alpha_1} \D_{\alpha_2} \, x^{\rho_1} x^{\rho_2} \; \Box^{j} \, G_{(\rho_1 \rho_2)|n}
\\
\nonumber
G_{(\alpha_1\alpha_2)|n-2j}^{6|\tw(3+2j)} 
&=
\frac{1}{n-2j} \cdot H_{n-2j} \KLSSo{} \frac{2}{n+1-2j} \KLNNo{} \D_{(\alpha_1}^{\phantom{|}}\delta_{\alpha_2)}^{(\rho_1}x^{\rho_2)}_{\phantom{|}}
\\
\nonumber
&
\qquad \qquad \qquad \quad
- \frac{1}{(n+2-2j)^2} \KLEo{}
x_{(\alpha_1}\D_{\alpha_2)}x^{(\rho_1}\D^{\rho_2)}
+ \D_{\alpha_1}\D_{\alpha_2} \, x^{\rho_1}x^{\rho_2} \oKLE{} \oKLNN{}
\\
\nonumber
&
\qquad \qquad \qquad \quad
- \frac{2}{(n+2-2j)^2} \KLEo{} x_{(\alpha_1}\D_{\alpha_2)} \, g^{\rho_1\rho_2}
+ g_{\alpha_1\alpha_2} \, \D^{(\rho_1}x^{\rho_2)} \oKLE{} \oKLSS{} \, \Box^j \, G_{(\rho_1 \rho_2)|n}
\\
\nonumber
G_{(\alpha_1\alpha_2)|n-2j}^{7|\tw(4+2j)} 
&=
H_{n-2j} \KLSSo{} \delta_{(\alpha_1}^{(\rho_1} \delta_{\alpha_2)}^{\rho_2)}
-\frac{2}{n+1-2j}
\KLEo{} \frac{1}{n+2-2j} \cdot x_{(\alpha_1}^{\phantom{|}} \delta_{\alpha_2)}^{(\rho_1} \D^{\rho_2)}_{\phantom{|}}
+ \frac{1}{n-2j} \cdot \D_{(\alpha_1}^{\phantom{|}} \delta_{\alpha_2)}^{(\rho_1} x^{\rho_2)}_{\phantom{|}} \oKLE{}
\\
\nonumber
&
\quad
+ \frac{1}{(n-2j )(n+1-2j)^2(n+2-2j)} \KLE{ x_{\alpha_1} x_{\alpha_2} \, \D^{\rho_1} \D^{\rho_2}
+ \D_{\alpha_1} \D_{\alpha_2} \, x^{\rho_1} x^{\rho_2} }
\\
\nonumber
&
\quad
- \frac{1}{2} \cdot g_{ \alpha_1\alpha_2} \, g^{\rho_1\rho_2}
\\
\nonumber
&
\quad
 + \frac{1}{(n-2j)(n+2-2j)} \KLEo{} x_{(\alpha_1}\D_{\alpha_2)} \; g^{\rho_1\rho_2}
 + g_{\alpha_1\alpha_2} \; \D^{(\rho_1}x^{\rho_2)} \oKLE{}\oKLSS{} \, \Box^j \, G_{(\rho_1 \rho_2 )|n}
\end{align}

The final decomposition off the light cone reads
\begin{eqnarray}
\label{SPDSTensorIIG}
&&
G_{(\alpha_1\alpha_2)|n} =
\\
\nonumber
&&
\sum_{j=0}^{\kle{\frac{n+2}{2}}}\frac{\kln{n+1-2j}!}{4^j j! \kln{n+1-j}!} \KLSSo{}
\\
\nonumber
&&
\qquad 
  \kln{x^2}^j \KLE{ 
  G_{ (\alpha_1\alpha_2) |n-2j}^{5|\tw(2+2j)}
+ G_{ (\alpha_1\alpha_2) |n-2j}^{6|\tw(3+2j)}
+ G_{ (\alpha_1\alpha_2) |n-2j}^{7|\tw(4+2j)} }
\\
\nonumber
&&
\quad 
- \frac{4j}{(n+2-j)(n+4-2j)} \cdot \kln{x^2}^{j-1} \KLEo{}
\\
\nonumber
&&
\qquad \quad \; \;
  x^2 \, G_{ (\alpha_1\alpha_2) |n-2j}^{4|\tw(3+2j)}
- 2\kln{n+2-2j} \cdot x_{(\alpha_1}^{\phantom{|}} G_{ \alpha_2) |n+1-2j}^{4|\tw(3+2j)} \oKLE{}
\\
\nonumber
&&
\quad 
+ \frac{2j(n+3-2j)}{(n+2-j)(n+4-2j)^2} \cdot \kln{x^2}^{j-1} \KLEo{}
\\
\nonumber
&&
\qquad \quad \; \;
(n+2-2j) \cdot x^2 \, G_{ (\alpha_1\alpha_2) | n-2j}^{1|\tw(2+2j)}
\\
\nonumber
&&
\qquad \quad
 -2(n+2-2j) \kln{ 
\kln{n+2-2j} \cdot x_{(\alpha_1}^{\phantom{|}} G_{\alpha_2)|n+1-2j}^{1|\tw(2+2j)} 
+ g_{\alpha_1\alpha_2} \, G_{n+2-2j}^{0|\tw(2+2j)} }
\\
\nonumber
&&
\qquad \quad
- 4 \kln{ x^2 \; G_{ (\alpha_1\alpha_2) |n-2j}^{3|\tw(2+2j)}
- 2 \kln{n+2-2j} \cdot x_{(\alpha_1}^{\phantom{|}} G_{ \alpha_2) |n+1-2j}^{3|\tw(2+2j)} } \oKLE{}
\\
\nonumber
&&
\quad 
+ \frac{2j(n+2-2j)(n+3-2j)}{(n+2-j)} \cdot \kln{x^2}^{j-1} \, g_{\alpha_1\alpha_2} \, G_{n+2-2j}^{1|\tw(2+2j)} 
\\
\nonumber
&&
\quad 
+ \frac{4j(j-1)(n+2-2j)}{(n+2-j)(n+3-j)(n+4-2j)^2(n+5-2j)} \cdot \kln{x^2}^{j-2} \KLEo{}
\\
\nonumber
&&
\qquad \quad \; \;
\kln{x^2}^2 \; G_{(\alpha_1\alpha_2)|n-2j}^{2|\tw(2+2j)} 
\\
\nonumber
&&
\qquad \quad
- 2\kln{n+3-2j} \cdot x^2 \, \kln{ 2 \cdot x_{(\alpha_1}^{\phantom{|}} \, G_{\alpha_2)|n+1-2j}^{2|\tw(2+2j)} 
+ \, g_{\alpha_1\alpha_2} \, G_{n+2-2j}^{2|\tw(2+2j)} }
\\
\nonumber
&&
\qquad \quad
+ 4\kln{n+3-2j}\kln{n+4-2j} \cdot x_{\alpha_1}x_{\alpha_2} \, G_{n+2-2j}^{2|\tw(2+2j)} \oKLE{} \oKLSS{}
\end{eqnarray}
Taking the on--cone limit of (\ref{SPDSTensorIIG}) we find the result
\begin{align}
\label{SPDSTensorOnConeEx}
\tilde G_{(\alpha_1\alpha_2)|n}
&=
\tilde G_{ (\alpha_1\alpha_2) |n}^{5|\tw(2)}
+\tilde G_{ (\alpha_1\alpha_2) |n}^{6|\tw(3)}
+\tilde G_{ (\alpha_1\alpha_2) |n}^{7|\tw(4)}
\\
\nonumber
&
\quad
+ \frac{2}{(n+1)(n+2)} \cdot \tilde x_{(\alpha_1}^{\phantom{|}}\tilde G_{\alpha_2)|n-1}^{4|\tw(5)}
\\
\nonumber
&
\quad
- \frac{1}{(n+2)^2} \KLNo{} n \cdot \tilde x_{(\alpha_1}^{\phantom{|}} \tilde G_{\alpha_2)|n-1}^{1|\tw(4)}
+ \, g_{\alpha_1\alpha_2} \, \tilde G_{n}^{0|\tw(4)}
- 4 \cdot \tilde x_{(\alpha_1}^{\phantom{|}} \tilde G_{\alpha_2)|n-1}^{3|\tw(4)} \oKLN{}
\\
\nonumber
&
\quad
+\frac{1}{2} \cdot g_{\alpha_1\alpha_2} \, \tilde G_{n}^{1|\tw(4)}
\\
\nonumber
&
\quad
+\frac{1}{n^2(n+1)^2} \cdot
\tilde x_{\alpha_1}^{\phantom{|}}\tilde x_{\alpha_2}^{\phantom{|}} \tilde G_{n-2}^{2|\tw(6)}
\intertext{with the three scalar operators}
\tilde G_{n}^{0|\tw(4)}
&=
\tilde \D^{(\rho_1} \tx^{\rho_2)} \; \tilde G_{(\rho_1\rho_2)|n}  
\\
\tilde G_{n}^{1|\tw(4)}
&=
g^{\rho_1\rho_2 } \; \tilde G_{(\rho_1\rho_2)|n}  
\\
\tilde G_{n-2}^{2|\tw(6)}
&=
\tilde \D^{\rho_1} \tilde \D^{\rho_2 } \; \tilde G_{(\rho_1\rho_2)|n}  
\intertext{three vector operators}
\tilde G_{\alpha_2|n-1}^{1|\tw(4)}
&=
\frac{1}{n } \cdot \tilde \D_{\alpha_2} \;\,  g^{\rho_1\rho_2 } \; \tilde G_{(\rho_1\rho_2)|n}  
\\
\tilde G_{\alpha_2|n-1}^{3|\tw(4)}
&=
\frac{1}{n^2} \cdot \tilde \D_{\alpha_2} \, \tx^{(\rho_1} \tilde \D^{\rho_2)} \; \tilde G_{(\rho_1 \rho_2)|n}
\\
\tilde G_{\alpha_2|n-1}^{4|\tw(5)}
&=
\kln{\delta_{\alpha_2}^{(\rho_1} - \frac{1}{n^2}
\KLs{ \tx_{\alpha_2} \, \tilde \D^{(\rho_1} + \tilde \D_{\alpha_2} \, \tx^{(\rho_1} } }  \tilde \D^{\rho_2)} \; \tilde G_{(\rho_1 \rho_2)|n}
\intertext{and three tensor operators}
\tilde G_{(\alpha_1\alpha_2)|n}^{5|\tw(2)} 
&=
\frac{1}{(n+1)^2(n+2)^2} \cdot \tilde \D_{\alpha_1} \tilde \D_{\alpha_2} \, \tx^{\rho_1} \tx^{\rho_2} \; \tilde G_{(\rho_1 \rho_2)|n}
\\
\nonumber
\tilde G_{(\alpha_1\alpha_2)|n}^{6|\tw(3)}
&=
\frac{1}{n} \KLSSo{} \frac{2}{n+1} \KLNNo{} \tilde \D_{(\alpha_1}^{\phantom{|}}\delta_{\alpha_2)}^{(\rho_1} \, \tx^{\rho_2)}_{\phantom{|}}
- \frac{1}{(n+2)^2} \KLEo{}
\tx_{(\alpha_1} \tilde \D_{\alpha_2)} \tx^{(\rho_1} \tilde \D^{\rho_2)}
+ \tilde \D_{\alpha_1} \tilde \D_{\alpha_2} \, \tx^{\rho_1} \tx^{\rho_2} \oKLE{} \oKLNN{}
\\
&
\qquad \quad
- \frac{2}{(n+2)^2} \KLEo{} \tx_{(\alpha_1} \tilde \D_{\alpha_2)} \, g^{\rho_1\rho_2}
+ g_{\alpha_1\alpha_2} \, \tilde \D^{(\rho_1} \tx^{\rho_2)} \oKLE{} \oKLSS{} \; \tilde G_{(\rho_1 \rho_2)|n}
\\
\tilde G_{(\alpha_1\alpha_2)|n}^{7|\tw(4)}
&=
\KLSSo{} \delta_{(\alpha_1}^{(\rho_1} \delta_{\alpha_2)}^{\rho_2)}
-\frac{2}{n+1}
\KLEo{} \frac{1}{n+2} \cdot \tx_{(\alpha_1}^{\phantom{|}} \delta_{\alpha_2)}^{(\rho_1} \, \tilde \D^{\rho_2)}_{\phantom{|}}
+ \frac{1}{n} \cdot \tilde \D_{(\alpha_1}^{\phantom{|}} \delta_{\alpha_2)}^{(\rho_1} \, \tx^{\rho_2)}_{\phantom{|}} \oKLE{}
\\
\nonumber
&
\qquad \quad
+ \frac{1}{n(n+1)^2(n+2)} \KLe{ \tx_{\alpha_1} \tx_{\alpha_2} \, \tilde\D^{\rho_1} \tilde\D^{\rho_2}
+ \tilde\D_{\alpha_1} \tilde\D_{\alpha_2} \, \tx^{\rho_1} \tx^{\rho_2} }
\\
\nonumber
&
\qquad \quad
 - \frac{1}{2} \cdot g_{ \alpha_1\alpha_2} \; g^{\rho_1\rho_2}
 + \frac{1}{n(n+2)} \KLEo{} \tx_{(\alpha_1} \tilde\D_{\alpha_2)} \; g^{\rho_1\rho_2}
 + g_{\alpha_1\alpha_2} \; \tilde\D^{(\rho_1} \tx^{\rho_2)} \oKLE{}\oKLSS{} \, \tilde G_{(\rho_1 \rho_2 )|n} \; .
\end{align}
In nonlocal representation the complete twist decomposition of the operator $\tilde G_{(\alpha_1\alpha_2)|n}$ on the light--cone can be found in Ref.~\cite{GLR00}. Here, however, we have consequently used the interior derivative $\D_{\alpha}$ to find compact representations for the operators of pure geometric twist. The complete twist decomposition of $G_{(\alpha_1\alpha_2)|n}$ off the light--cone given by (\ref{SPDSTensorIIG}) is a new result extending the respective on--cone decomposition.

\section{Conclusions}

The main focus of this work is the complete twist decomposition of local and
nonlocal QCD operators. To this purpose we have introduced a unique group theoretical procedure which is essentially based
on the complete decomposition of (generic) local operators $\OP_{\alpha_1\dots\alpha_k\kln{\zeta_1\dots\zeta_n}}$ into components
carrying an irreducible tensor representation of the Lie group $SO(2h;\C)$.
These irreducible tensor representations remain irreducible when restricted to the real subgroup $SO(1,3;\R)$ of $SO(4;\C)$
and can therefore be labeled by a definite spin and twist according to the original definition
of Gross and Treiman: twist $\tau$ = canonical dimension $d_n$ minus Lorentz spin $j_n$. 
The central tool to obtain the decompositions of $\OP_{\alpha_1\dots\alpha_k\kln{\zeta_1\dots\zeta_n}}$
into irreducible tensors is the
polynomial technique and our procedure of twist decomposition thereby has to be viewed as an extension of the procedure used
in Refs.~\cite{GLR99,GLR00,L02}. Here, we have to emphasize two points. First, we have introduced an off--cone representation of the
conformal algebra $\mathfrak{so}\kln{2,2h}$ which turned out to be very useful for the formulation of traceless as well as
irreducible local and nonlocal tensor operators off the light~cone.
In Ref.~\cite{EGL04} this representation has been used for the first time in the course of twist decomposition.
The second crucial point is the deduction of the complete
trace decomposition from the projectors onto traceless tensor polynomials $H_{\alpha_1\dots\alpha_k|n}^{\rho_1\dots\rho_k}$
thereby making use of both representations
(on--cone as well as off--cone) of the conformal algebra. From this trace decomposition we deduced the complete twist decomposition
of the (generic) local operators $\OP_{\alpha_1\dots\alpha_k|n}\kln{x}$ on the light~cone as well as off the light~cone.
Furthermore, the decomposition into irreducible tensors is formulated for general space--time dimensions $2h$ for the
Lie group $SO(2h;\C)$.

Since we have formulated the twist decomposition for generic operators, it can be applied to many different QCD operators. In the present work, we obtained the respective twist decompositions for the QCD quark operators $O_{\mu|n}$ and $M_{\mu|n+1}$ as well as the complete twist decompositions for the local quark tensor operator $M_{[\mu\nu]|n}\kln{x}$ and the local gluon operator $G_{\mu\nu|n}\kln{x}$. The application to more complicated multilocal QCD operators containing many quark and gluon fields, e.g., Shuryak--Vainshtein~\cite{SV82} type operators, is straightforward and makes use of the same generic twist decomposition related to the respective rank and symmetry of the (multilocal) operator. One only has to adopt the Taylor expansion for multilocal operators given in Refs.~\cite{GLR00,L02}.

\end{document}